\documentclass[12pt]{article} 
\pdfoutput=1

\usepackage[utf8]{inputenc}
\usepackage[T1]{fontenc} % codage moderne des caractères sous Latex
\usepackage{lmodern}
\usepackage[english]{babel}           % style anglais
\usepackage{cancel}
\usepackage[pdftex]{graphicx}
\usepackage{epsfig}
\usepackage{graphicx}
\usepackage{comment}
\usepackage{latexsym}
\usepackage{hyperref}
\usepackage{amsmath}
\usepackage{mathrsfs}
\usepackage[usenames, dvipsnames]{color}
\usepackage{amsbsy}
\usepackage{amssymb}
\usepackage{amsthm}
\usepackage{amsfonts}
\usepackage{cite}
\usepackage{enumitem}
\usepackage{xcolor}
\usepackage{color}
\usepackage{mathtools}
\usepackage{caption} 
\usepackage{subcaption} %sous-figures 
\usepackage{euscript} % FOR HAND CALLIGRAPHIC 
\usepackage{float}
\usepackage{diagbox}
\usepackage[title]{appendix}
\usepackage{tabularx}
\usepackage{tikz}
\usetikzlibrary{intersections}
\usetikzlibrary{calc}
\usepackage{pgfplots} %package to plot graphs
\usepgfplotslibrary{fillbetween}
\pgfplotsset{compat=1.17} %same

\pgfdeclarelayer{pre main}
\pgfdeclarelayer{background}
\pgfdeclarelayer{foreground}
\pgfsetlayers{background, pre main, main, foreground}

\usepackage{tcolorbox}
\usepackage{soul}

\usetikzlibrary{decorations.pathmorphing,arrows.meta}%\usepackage{ulem}

\usepackage{amsmath}
\allowdisplaybreaks
\usepackage{amsfonts}
\usepackage{amssymb}
\usepackage{bbm}
\usepackage{verbatim}
\usepackage{dsfont}
\usepackage{booktabs}
\usepackage{slashed}
\usepackage{braket}
\usepackage{comment}

\usepackage{tikz}
\usepackage{pgfplots}
\usetikzlibrary{intersections, pgfplots.fillbetween}

\renewcommand\d{{\rm d}}

\newcommand{\F}{{\cal F}}

\newcommand{\J}{{\cal J}}

\renewcommand{\S}{{\cal S}}

\newcommand{\be}{\begin{equation}}
\newcommand{\ee}{\end{equation}}
\newcommand{\dis}{\displaystyle}
\renewcommand{\thefootnote}{\fnsymbol{footnote}}

\newcommand{\Eq}[1]{Eq.~\eqref{#1}}
\newcommand{\Eqs}[1]{Eqs.~\eqref{#1}}

\newcommand{\Sect}[1]{Section~\ref{#1}}

\newcommand{\Appendix}[1]{Appendix~\ref{#1}}
\newcommand{\Fig}[1]{Figure~\ref{#1}}
\newcommand{\Figs}[1]{Figures~\ref{#1}}

\newcommand{\R}{\mathbb{R}}

\newcommand{\Le}{{\rm L}}
\newcommand{\E}{{\rm E}}
\newcommand{\Ri}{{\rm R}}

\newcommand{\ie}{{\it i.e.} }
\newcommand{\eg}{{\it e.g.} }

\renewcommand{\and}{\mbox{and}}

\newcommand{\esp}{\phantom{\!\!\overset{\displaystyle |}{|}}}

\newcommand{\bm}{\boldmath} 
\catcode`\@=11
\def\marginnote#1{}
\newcount\hour
\newcount\minute
\newtoks\amorpm
\hour=\time\divide\hour by60 \minute=\time{\multiply\hour by60
\global\advance\minute by-\hour}
\edef\standardtime{{\ifnum\hour<12 \global\amorpm={am}%
        \else\global\amorpm={pm}\advance\hour by-12 \fi
        \ifnum\hour=0 \hour=12 \fi
        \number\hour:\ifnum\minute<10 0\fi\number\minute\the\amorpm}}
\edef\militarytime{\number\hour:\ifnum\minute<10 0\fi\number\minute}
\def\draftlabel#1{{\@bsphack\if@filesw {\let\thepage\relax
   \xdef\@gtempa{\write\@auxout{\string
      \newlabel{#1}{{\@currentlabel}{\thepage}}}}}\@gtempa
   \if@nobreak \ifvmode\nobreak\fi\fi\fi\@esphack}
        \gdef\@eqnlabel{#1}}
\def\@eqnlabel{}
\def\@vacuum{}
\def\draftmarginnote#1{\marginpar{\raggedright\scriptsize\tt#1}}
\def\draft{\oddsidemargin -.2truein
        \def\@oddfoot{\sl preliminary draft \hfil
        \rm\thepage\hfil\sl\today\quad\militarytime}
        \let\@evenfoot\@oddfoot \overfullrule 3pt
        \let\label=\draftlabel
        \let\marginnote=\draftmarginnote
   \def\@eqnnum{(\theequation)\rlap{\kern\marginparsep\tt\@eqnlabel}%
\global\let\@eqnlabel\@vacuum}  }
\def\thebibliography#1{
\vskip 0.5cm \centerline{\bf \Large References}
\list{
[\arabic{enumi}]}{\settowidth\labelwidth{[#1]}
\leftmargin\labelwidth
\advance\leftmargin\labelsep
\usecounter{enumi}}
\def\newblock{\hskip .11em plus .33em minus .07em}
\sloppy\clubpenalty4000\widowpenalty4000
\sfcode`\.=1000\relax}

\renewcommand{\theequation}{\arabic{section}.\arabic{equation}}
\renewcommand{\section}{\setcounter{equation}{0}\@startsection
{section}{1}{0mm}{-\baselineskip}{0.5\baselineskip} {\normalfont\Large\bfseries}}
\renewcommand{\subsection}{\@startsection
{subsection}{2}{0mm}{-\baselineskip}{0.5\baselineskip} {\normalfont\large\bfseries}}
\renewcommand{\subsubsection}{\@startsection
{subsubsection}{3}{0mm}{-\baselineskip}{0.5\baselineskip}
{\normalfont\normalsize\slshape}}

\begin{document}

%%%%%

\begin{titlepage}
\begin{flushright}
%CPHT-RRxxx.xx2022, December 2022
%\vspace{0.0cm}
\end{flushright}
\begin{centering}
{\bm\bf \Large 
Holography for de Sitter bubble geometries}

\vspace{6mm}

 {\bf Anastasios Irakleous$^1$\footnote{anastasios.irakleous@ucy.ac.cy}, Fran\c{c}ois Rondeau$^2$\footnote{francois.rondeau@ens-lyon.fr} and Nicolaos Toumbas$^1$\footnote{nick@ucy.ac.cy}}

 \vspace{3mm}

$^1$ {\em Department of Physics, University of Cyprus, Nicosia 1678, Cyprus}

$^2$ {\em ENS de Lyon, Laboratoire de Physique, CNRS UMR 5672, Lyon 69007, France}

\end{centering}
\vspace{0.5cm}
$~$\\
\centerline{\bf\large Abstract}\vspace{0.2cm}

\vspace{-0.6cm}
\begin{quote}

We generalize the de Sitter static patch holographic proposal and the bilayer holographic entanglement entropy prescription to de Sitter geometries containing a bubble of smaller positive or vanishing cosmological constant. When the causal patch of an observer at the center of the bubble overlaps with the ``parent'' de Sitter region, we propose that the full spacetime can be holographically encoded on two holographic screens. The leading geometrical entanglement entropy between the two screens, which can be constant or time-dependent in some cases, never exceeds the Gibbons-Hawking entropy associated with the ``parent'' de Sitter space. When the causal patch of the observer at the center of the bubble is causally disconnected from the ``parent'' de Sitter region, the holographic proposal no longer applies, and more than two holographic screens are required to encode the whole spacetime.

\end{quote}

\end{titlepage}
\newpage
\setcounter{footnote}{0}
\renewcommand{\thefootnote}{\arabic{footnote}}
 \setlength{\baselineskip}{.7cm} \setlength{\parskip}{.2cm}

\setcounter{section}{0}

%%%%%%%%%%%%%%%%%%%%%%%%%%%%%%%%%%%%%%%%
\newpage
\tableofcontents
\newpage
%%%%%%%%%%%%%%%%%%%%%%%%%%%%%%%%%%%%%%%%%%%%%%
%%%%%%%%%%%%%%%%%%%%%%%%%%%%%%%%%%%%%%%%%%%%%%
%input the txt files for the sections
\section{Introduction}

In our effort to obtain a non-perturbative description of quantum gravity, holography plays a crucial role. Indeed, the holographic principle \cite{tHooft:1993dmi,Susskind:1994vu} is crucial in the description of string theory and quantum gravity in asymptotically anti-de Sitter backgrounds (AdS) in terms of a conformal theory (CFT) on the boundary \cite{Maldacena:1997re,Gubser:1998bc,Witten:1998qj}. Holography can also be seen to be realized in the description of M-theory on flat backgrounds in terms of a maximally supersymmetric matrix model \cite{Banks:1996vh}. Due to the presence of ``asymptotically cold conditions'' \cite{Susskind:2007pv,Seiberg:2006wf} in these examples, the bulk fields decouple as the boundary of space is approached, allowing us to define precise observables and establish the holographic dictionary. 

There are also a number of interesting proposals concerning the holographic description of cosmological backgrounds and, in particular, de Sitter (dS) space. These include \cite{Banks:2000fe, Witten:2001kn}, the dS/CFT correspondence \cite{Strominger:2001pn,Strominger:2001gp}, 
the static patch holography for dS space \cite{Dyson:2002nt,Dyson:2002pf,Goheer:2002vf,Susskind:2021omt}, as well as generalizations of the latter proposal for closed FRW cosmologies in arbitrary dimensions \cite{Bousso:1999cb,Bak:1999hd,Nomura:2016ikr,Franken:2023jas}. However, the absence of ``asymptotic coldness'' in the cosmological examples renders the construction of the precise holographic dictionary more difficult. Some generic properties of the dual holographic system can be extracted on the basis of symmetries and the Bousso covariant entropy conjecture \cite{Bousso:1999xy,Bousso:2002ju}, which allows us to obtain an estimate of the number of the holographic degrees of freedom.  

\begin{figure}[!h]

    \centering
\begin{tikzpicture}[scale=0.9]
\begin{scope}[transparency group]
\begin{scope}[blend mode=multiply]
\path
       +(3,3)  coordinate (IItopright)
       +(-3,3) coordinate (IItopleft)
       +(3,-3) coordinate (IIbotright)
       +(-3,-3) coordinate(IIbotleft)
      
       ;
\draw (IItopleft) -- (IItopright) -- (IIbotright) -- (IIbotleft) --(IItopleft) -- cycle;

\node at (-0.55,0.55) [circle, fill, inner sep=1.5 pt, label = below:$\S_\Le$
]{};
\node at (1,1) [circle, fill, inner sep=1.5 pt, label = below:$\S_\Ri$
]{};

\draw[domain=-3:3, red, smooth, variable=\x] plot ({\x}, {sin(deg((\x/2-1)))+1.5});
%\draw[domain=-0.55:1, smooth, variable=\x, black,dashed] plot ({\x}, {sin(deg((\x/2-1)))+1.5});
%\draw[domain=1:3, smooth, variable=\x] plot ({\x}, {sin(deg((\x/2-1)))+1.5});

\node at (-1.9,1.4) [label=below:$\color{red} \Sigma$]{};

%\fill[fill=red!40] (-0.55,0.55) -- (0.45,1.55) -- (1,1) -- (0,0) -- cycle;

%\fill[fill=blue!50] (-0.55,0.55) -- (0.45,1.55) -- (1,1) -- (0,0) -- cycle;

%\fill [fill=blue!50] (-3,3) -- (-0.55,0.55) -- (-3,-1.95) --  cycle;

%\fill [fill=red!50] (1,1) -- (3,3) -- (3,-1) --  cycle;

\draw (IItopleft) -- (IIbotright)
              (IItopright) -- (IIbotleft) ;

\end{scope}
\end{scope}
\end{tikzpicture}
    \caption{\footnotesize Penrose diagram for the de Sitter spacetime. Spacelike slices have the topology of a sphere and the worldlines of two antipodal observers follow the left and right edges of the diagram. Holographic degrees of freedom lie on the two cosmological horizons depicted by the diagonal lines. The black dots depict the intersections of the cosmological horizons with a bulk Cauchy slice $\rm{\Sigma}$, on which the holographic screens $\S_\Le$ and $\S_\Ri$ are located.
    \label{fig:intro_dS}
    }
\end{figure}
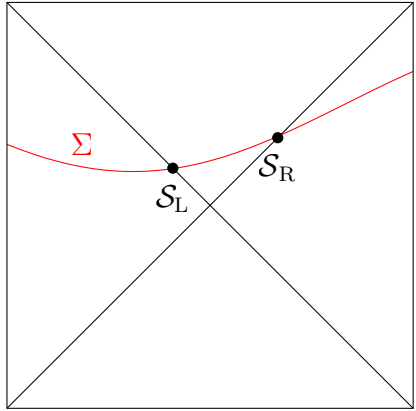

More pertinent to this work is the static patch holographic proposal, which states that the causal patch of a comoving observer in de Sitter space can be described by a holographic dual theory on the boundary of the causal patch, namely, the cosmological horizon \cite{Susskind:2021omt}. The causal patch associated with the comoving observer at the antipodal point of the spatial sphere can also be described by a second copy of the holographic system on the cosmological horizon bounding it. More precisely, given a complete, SO$(n)$-symmetric bulk Cauchy slice $\Sigma$, where $n$ is the number of spatial dimensions ($n\ge 2$), we anchor two holographic screens, $\S_\Le$ and $\S_\Ri$, at the intersections of $\Sigma$ with the cosmological horizons bounding the two static patches. See \Fig{fig:intro_dS}, 
where the Penrose diagram of de Sitter space is shown, together with the slice $\Sigma$ and the two holographic screens. The number of degrees of freedom on each screen is equal to the area of the screen in Planck units. According to the Bousso covariant entropy conjecture, these degrees of freedom suffice to describe the states on $\Sigma_\Le$ and $\Sigma_\Ri$, the parts of $\Sigma$ in the two static patches, respectively \cite{Bousso:1999xy,Bousso:2002ju,Susskind:2021omt,Franken:2023pni}. Moreover, since the two static patches are causally disconnected, the two holographic systems do not interact (except when the screens coincide at the bifurcate horizon), but they are quantum mechanically entangled \cite{Susskind:2021omt}.  

Indeed, it has been argued that the presence of an observer in a spatially closed universe such as de Sitter space is crucial in order to have a non-trivial algebra of observables \cite{Chandrasekaran:2022cip,Witten:2023qsv,Witten:2023xze,Gomez:2023upk} and ``grow'' a non-trivial Hilbert space \cite{McNamara:2020uza,Balasubramanian:2023xyd,Mirbabayi:2023vgl,Harlow:2025pvj}.\footnote{See also the recent work \cite{Liu:2025cml}.} Moreover, a well-defined prescription has been developed for holographic entanglement entropy computations, regarding the subsystems of the two-screen system, which allows us to gain further insight into the nature of holographic dual theory \cite{Susskind:2021esx,Shaghoulian:2021cef,Shaghoulian:2022fop,Franken:2023pni}. This prescription, called the bilayer proposal, amounts to a generalization to de Sitter space of the Ryu-Takayanagi (RT) and the Hubeny-Rangamani-Takayanagi (HRT) prescriptions for holographic entanglement entropy computations in asymptotically AdS spaces \cite{Ryu:2006bv,Ryu:2006ef, Hubeny:2007xt}.\footnote{See also \cite{VanRaamsdonk:2009ar,VanRaamsdonk:2010pw,Lewkowycz:2013nqa,Maldacena:2013xja,Faulkner:2013ana,Wall:2012uf,Engelhardt:2014gca,Freedman:2016zud,Dong:2016hjy,Dong:2016eik,Headrick:2022nbe,Engelhardt:2023xer,Penington:2019npb,Almheiri:2019psf,Almheiri:2019hni,Penington:2019kki,Almheiri:2019qdq,Almheiri:2020cfm,Nomura:2017fyh,Murdia:2022giv} for further important developments and applications to black hole physics.} It involves extremizing a generalized entropy with the geometrical and semiclassical contributions arising from the two static patches and the region between them \cite{Susskind:2021esx,Shaghoulian:2021cef,Shaghoulian:2022fop,Franken:2023pni}. Interesting patterns of entanglement and phase transitions can be uncovered involving the holographic degrees of freedom. More importantly, the entanglement wedge of the two-screen system comprises complete bulk Cauchy slices \cite{Franken:2023pni}. Assuming entanglement wedge reconstruction \cite{Dong:2016eik}, this implies that the full de Sitter spacetime can be reconstructed from the holographic data \cite{Franken:2023pni}. The region between the static patches can be seen as a bridge between them and is built as a result of quantum entanglement between the two single-screen subsystems. This result is a direct manifestation of the ``ER=EPR'' relation \cite{Maldacena:2013xja}. For other applications of the generalized entropy formula to cosmology, see \eg \cite{Hartman:2020khs,Chen:2020tes,VanRaamsdonk:2020tlr,Balasubramanian:2020coy,Balasubramanian:2020xqf,Manu:2020tty,Kames-King:2021etp,Aguilar-Gutierrez:2021bns,Azarnia:2021uch,Goswami:2021ksw,Bousso:2022gth,Ben-Dayan:2022nmb,Kawamoto:2023nki,Franken:2023ugu,Hao:2024nhd,Yadav:2024ray,Noumi:2025cup,Geng:2021wcq,Langhoff:2021uct}.  

The static patch holographic proposal and the bilayer prescription for de Sitter space can be generalized for a large class of closed FRW cosmologies in arbitrary dimensions, supported by perfect fluid sources \cite{Franken:2023jas}. In these examples, the two holographic screens can be anchored on the apparent horizons delimiting the regions of trapped and anti-trapped spheres. Both nonsingular, bouncing cosmologies and Big Bang/Big Crunch examples are included in this class. When the perfect fluid index satisfies $-1<w<2/n-1$ for closed bouncing cosmologies and $2/n-1<w< 4/n-1$ for the Big Bang/Big Crunch ones\footnote{For $w=4/n-1$, the apparent horizons are lightlike and, as in the de Sitter case ($w=-1$), they coincide with the cosmological horizons bounding the two causal patches, respectively. The holographic construction can be generalized for this case as well \cite{Franken:2023jas}.}, the apparent horizons are timelike and lie within the causal patches of the two antipodal observers, respectively. It is this geometrical feature that allows the generalization of the de Sitter holographic construction and the bilayer proposal for holographic entropy computations to these cases as well \cite{Franken:2023jas}.

However, eternal de Sitter space is not a realistic cosmological model for our universe. Observational data supports a very rich phenomenological model with early inflationary and high temperature eras, symmetry breaking phase transitions, and large amounts of dark matter and dark energy dominating the evolution at latter times. If the dark energy persists arbitrarily long in time, our Universe would be asymptotically de Sitter at best. However, evidence based on string theoretic considerations seems to suggest that the current accelerating phase of the Universe is a transient one. Indeed, it is widely believed that string theory has a large landscape of metastable de Sitter vacua \cite{Bousso:2000xa,Susskind:2003kw,Kachru:2003aw}. Due to the large number of choices for fluxes and wrapped branes through non-trivial cycles of the string compactification manifold, the set is discrete and closely spaced, so that vacua of arbitrarily small positive cosmological constants can be found \cite{Bousso:2000xa,Susskind:2003kw,Kachru:2003aw}. The lifetime of the de Sitter vacua can be very large, but is bounded by the Poincare recurrence time (which is set by the exponential of the de Sitter entropy) \cite{Dyson:2002nt,Dyson:2002pf,Goheer:2002vf,Kachru:2003aw,Frey:2003dm}. Therefore, even if we start with a de Sitter vacuum with a small, positive cosmological constant, tunneling transitions will occur into neighboring vacuum states with a smaller positive or zero cosmological constant. In the context of cosmology, bubbles of space with a smaller cosmological constant spontaneously nucleate in a ``parent'' de Sitter background with a higher cosmological constant. The bubbles expand exponentially but at a slower rate than the background. So, they do not get to coalesce and cover the whole space. Moreover, new bubbles are generated continuously within the older ones. Attempts to provide a holographic framework for such an eternally inflating bubble cosmology, at least in the simpler cases where a single bubble appears, can be found in \cite{Susskind:2007pv,Freivogel:2004rd,Freivogel:2005vv,Freivogel:2006xu,Sekino:2009kv}.

The goal of this paper is to generalize the static patch holographic proposal for de Sitter space and the bilayer prescription for holographic entanglement entropy computations to simple bubble geometries in arbitrary dimensions. In particular, we focus on a class of bounce cosmological solutions which are invariant under time-reversal symmetry and involve a single bubble inside a ``parent'' de Sitter spacetime with a larger cosmological constant. The bubble contracts from infinite size in the far past to a finite size and then expands again to infinite size in the far future. By analytically continuing to Lorentzian signature certain Coleman-De Luccia instanton solutions \cite{Coleman:1980aw} arising in models of gravity coupled to a scalar field, such bounce solutions can be obtained. See \eg \cite{Freivogel:2004rd}. The scalar potential must admit different local minima at positive energy densities. The bubble region is surrounded by an $n$-dimensional domain wall which breaks the original SO$(1,n+1)$ symmetry group of dS$_{n+1}$ to SO$(1,n)$. We neglect the thickness of the wall, working in the thin-wall approximation\footnote{Studies of bubble geometries and vacuum decay beyond the thin-wall approximation can be found \eg in \cite{Banks:2002nm,Dong:2011gx,Espinosa:2018hue,Espinosa:2018voj}.}, following \cite{Fabinger:2003gp}. 

\begin{figure}[h!]
 \begin{subfigure}[t]{0.3\linewidth}
    \centering
\begin{tikzpicture}[scale=0.7]
\begin{scope}[transparency group]
\begin{scope}[blend mode=multiply]

%wall
\draw[name path=A, line width= 1 pt,brown] plot[variable=\t,samples=91,domain=-45:45] ({3*sec(\t)-4.23},{3*tan(\t)});
\draw[name path=B, line width= 1 pt,brown] plot[variable=\t,samples=91,domain=-45:45] ({2*sec(\t)-2.83},{3*tan(\t)});
\tikzfillbetween[of=A and B]{gray, opacity=0.3}; 

\draw[name path=C](0,-3) -- (-3,-3) -- (-3,3) -- (0,3);
\draw[name path=D](0,-3) -- (3,-3) -- (3,3) -- (0,3);

\tikzfillbetween[of=A and C]{blue, opacity=0.1};
\tikzfillbetween[of=B and D]{red, opacity=0.1};

\draw[dashed] (-3,3)--(-1.05,1.05);
\draw[dashed] (-3,-3)--(-1.05,-1.05);
\draw[dashed] (-0.5,1.6)--(1.1,0);
\draw[dashed] (-0.5,-1.6)--(1.1,0);
\draw[dashed] (3,3)--(0,0);
\draw[dashed] (3,-3)--(0,0);
%\draw[dotted] (0,0)--(-0.7,0.7);
%\draw[dotted] (0,0)--(-0.7,-0.7);

%Bousso's wedges
%\draw (0.55,1.8) -- (0.75,2) -- (0.95,1.8);
%\draw (-0.2,0.5) -- (-0.0,0.7) -- (0.2,0.5);
%\draw (-2.2,0.2) -- (-2,0) -- (-2.2,-0.2);
%\draw (-0.6,0.2) -- (-0.4,0) -- (-0.6,-0.2);
%\draw (-1.5,2.2) -- (-1.3,2.4) -- (-1.1,2.2);
%\draw (-1.5,-2.2) -- (-1.3,-2.4) -- (-1.1,-2.2);
%\draw (2.2,0.2) -- (2,0) -- (2.2,-0.2);
%\draw (0.6,0.2) -- (0.4,0) -- (0.6,-0.2);
%\draw (0.55,-1.8) -- (0.75,-2) -- (0.95,-1.8);
%\draw (-0.2,-0.5) -- (0,-0.7) -- (0.2,-0.5);

\end{scope}
\end{scope}
\end{tikzpicture}
    \caption{\footnotesize Bubble geometry with $(\varepsilon_L,\varepsilon_R)=(+1,+1)$.}
    \label{fig:Penrose_diag_intro_++}
    \end{subfigure}
\quad \
\begin{subfigure}[t]{0.3\linewidth}
 \centering
\begin{tikzpicture}[scale=0.7]
\begin{scope}[transparency group]
\begin{scope}[blend mode=multiply]

%wall
\draw[name path=A, line width= 1 pt,brown] plot[variable=\t,samples=91,domain=-45:45] ({3*sec(\t)-4.23},{3*tan(\t)});
\draw[name path=B, line width= 1 pt,brown] plot[variable=\t,samples=91,domain=-45:45] ({-2*sec(\t)+2.83},{3*tan(\t)});
\tikzfillbetween[of=A and B]{gray, opacity=0.3}; 

\draw[name path=C](0,-3) -- (-3,-3) -- (-3,3) -- (0,3);
\draw[name path=D](0,-3) -- (3,-3) -- (3,3) -- (0,3);

\tikzfillbetween[of=A and C]{blue, opacity=0.1};
\tikzfillbetween[of=B and D]{red, opacity=0.1};

\draw[dashed] (-3,3)--(-1.05,1.05);
\draw[dashed] (-3,-3)--(-1.05,-1.05);
\draw[dashed] (0.55,1.7)--(2.25,0);
\draw[dashed] (0.55,-1.7)--(2.25,0);
\draw[dashed] (3,3)--(0.8,0.8);
\draw[dashed] (3,-3)--(0.8,-0.8);

\draw[dashed] (-1.8,0)--(-1.1,0.7);
\draw[dashed] (-1.8,0)--(-1.1,-0.7);

%Bousso's wedges
%\draw (1.5,2.2) -- (1.3,2.4) -- (1.1,2.2);
%\draw (-2.7,0.2) -- (-2.5,0) -- (-2.7,-0.2);
%\draw (-1.5,0.2) -- (-1.3,0) -- (-1.5,-0.2);
%\draw (-1.5,2.2) -- (-1.3,2.4) -- (-1.1,2.2);
%\draw (-1.5,-2.2) -- (-1.3,-2.4) -- (-1.1,-2.2);
%\draw (2.7,0.2) -- (2.5,0) -- (2.7,-0.2);
%\draw (1.35,0.2) -- (1.15,0) -- (1.35,-0.2);
%\draw (1.5,-2.2) -- (1.3,-2.4) -- (1.1,-2.2);

\end{scope}
\end{scope}
\end{tikzpicture}
    \caption{\footnotesize Bubble geometry with $(\varepsilon_L,\varepsilon_R)=(+1,-1)$.}
    \label{fig:Penrose_diag_intro_+-}
\end{subfigure}
\quad \
\begin{subfigure}[t]{0.3\linewidth}
    \centering
\begin{tikzpicture}[scale=0.7]
\begin{scope}[transparency group]
\begin{scope}[blend mode=multiply]

%wall
\draw[name path=B, line width= 1 pt,brown] plot[variable=\t,samples=91,domain=-45:45] ({-4.3*sec(\t)+6.1},{3*tan(\t)});
\draw[name path=A, line width= 1 pt,brown] plot[variable=\t,samples=91,domain=-45:45] ({-3.5*sec(\t)+4.95},{3*tan(\t)});
\tikzfillbetween[of=A and B]{gray, opacity=0.3}; 

%\draw[dashed] (0,3)--(0,-3);

\draw[name path=C](0,-3) -- (-3,-3) -- (-3,3) -- (0,3);
\draw[name path=D](0,-3) -- (3,-3) -- (3,3) -- (0,3);

\tikzfillbetween[of=A and C]{blue, opacity=0.1};
\tikzfillbetween[of=B and D]{red, opacity=0.1};

\draw[dashed] (-3,3)--(0,0)--(-3,-3);
\draw[dashed] (3,3)--(1.4,1.4);
\draw[dashed] (3,-3)--(1.4,-1.4);
\draw[dashed] (1.35,-0.6)--(0.75,0)--(1.35,0.6);
%\draw[dotted] (0,0)--(1.15,1.15);
%\draw[dotted] (0,0)--(1.15,-1.15);

%Bousso's wedges
%\draw (1.3,2.3) -- (1.5,2.5) -- (1.7,2.3);
%\draw (-0.2,1.5) -- (0.0,1.7) -- (0.2,1.5);
%\draw (-0.2,-1.5) -- (0,-1.7) -- (0.2,-1.5);
%\draw (-1.7,0.2) -- (-1.5,0) -- (-1.7,-0.2);
%\draw (2.5,0.2) -- (2.3,0) -- (2.5,-0.2);
%\draw (1.3,0.2) -- (1.1,0) -- (1.3,-0.2);
%\draw (0.6,0.2) -- (0.4,0) -- (0.6,-0.2);
%\draw (1.3,-2.3) -- (1.5,-2.5) -- (1.7,-2.3);

\end{scope}
\end{scope}
\end{tikzpicture}
    \caption{\footnotesize Bubble geometry with $(\varepsilon_L,\varepsilon_R)=(-1,-1)$.} \label{fig:Penrose_diag_intro_--}
\end{subfigure}
\caption{\footnotesize The three de Sitter bubble geometries considered in this paper. The blue (red) region is the interior (exterior) of the bubble, which is a part of dS$_{n+1}$ with a smaller (larger) cosmological constant. Each region is bounded by a thin domain wall trajectory (brown lines). The gray-shaded region is not a part of the spacetime. The worldlines of the two antipodal observers correspond to the left and right vertical edges of the diagrams. Each of them has a causal patch bounded by future and past cosmological horizons depicted by the thick dashed lines.}
\label{fig:Penrose_diag_intro}
\end{figure}
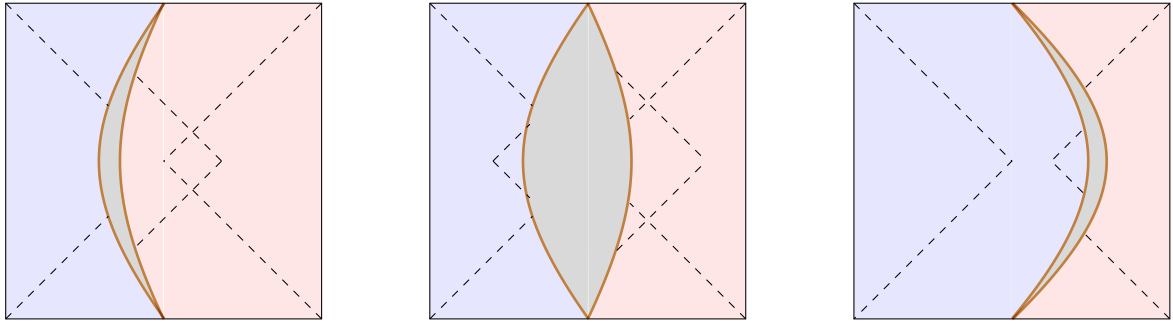

We consider a comoving observer at the center of the bubble, and another observer in the de Sitter ``parent'' region, at the antipodal point of the spatial sphere. The observers' worldlines correspond to the left and right vertical edges of the Penrose diagrams of \Fig{fig:Penrose_diag_intro}, respectively. The bubble region with smaller cosmological constant is shaded blue, while the ``parent'' de Sitter region with higher cosmological constant is shaded red. The causal patches of the two antipodal observers are delimited by the dashed diagonal lines.
Depending on the causal patches associated with the two observers, we classify the bubble geometries into three classes. In the first class, the causal patch of the bubble observer extends in the ``parent'' de Sitter region while the causal patch of the antipodal observer is causally disconnected from the bubble (\Fig{fig:Penrose_diag_intro_++}). In the second class, both causal patches overlap with both the bubble and the ``parent'' de Sitter regions (\Fig{fig:Penrose_diag_intro_+-}). Finally, in the third class, the causal patch of the bubble observer is causally disconnected from the ``parent'' de Sitter region, while the causal patch of the antipodal observer extends in the bubble region (\Fig{fig:Penrose_diag_intro_--}). As we will see in \Sect{sect:bubble_geometry}, the three classes of geometries can be characterized by two signs $\varepsilon_L=\pm 1$, $\varepsilon_R=\pm 1$, whose physical interpretation will be described. The three above cases correspond, respectively, to $(\varepsilon_L,\varepsilon_R)=(+1,+1)$, $(\varepsilon_L,\varepsilon_R)=(+1,-1)$, and $(\varepsilon_L,\varepsilon_R)=(-1,-1)$. In the $(\varepsilon_L,\varepsilon_R)=(+1,+1)$ and $(\varepsilon_L,\varepsilon_R)=(+1,-1)$ cases, the tension of the domain wall is positive and the causal patches of the antipodal observers overlap. In the case $(\varepsilon_L,\varepsilon_R)=(-1,-1)$, the tension of the domain wall is negative, and so special boundary conditions must be imposed, such as those across a non-fluctuating string theory orientifold, in order to avoid instabilities. Moreover, the two patches are causally disconnected. This difference will play a crucial role in the holographic constructions of \Sect{sect:holography}. The dynamical nucleation of bubbles will not be discussed in this paper. See \eg \cite{Brown:1988kg, Brown:1987dd, Maloney:2002rr} for details regarding this case. Only the positive tension cases $(\varepsilon_L,\varepsilon_R)=(+1,\pm 1)$ can describe features of expanding Coleman De Luccia bubbles formed in gravity scalar field theory models via spontaneous nucleation \cite{Coleman:1980aw}. 

Given an SO$(n)$-symmetric bulk Cauchy slice $\Sigma$, we locate two holographic screens $\S_\Le$ and $\S_\Ri$ to describe its parts in the two causal patches, respectively. In particular, the location of $\S_\Le$ ($\S_\Ri$) is determined so as to maximize the extent of the region in the left (right) causal patch to be described holographically, in accordance with the covariant entropy conjecture \cite{Bousso:1999xy,Bousso:2002ju}. The bulk foliation in terms of Cauchy slices yields the screen trajectories, which are nowhere spacelike. The trajectories have segments along the past and future cosmological horizons associated with the observers, but also along the domain wall trajectory separating the bubble from the ``parent'' de Sitter region. The area of the screens is in general not constant as they evolve along their trajectories. Therefore, the number of holographic degrees of freedom on them can change. This suggests a non-unitary evolution of the holographic theory, and a sequence of mappings to Hilbert spaces of varying dimensionalities. See \cite{Cotler:2022weg,Cotler:2023xku,Franken:2023pni} for similar behaviors.

The bilayer proposal can be extended to the $(\varepsilon_L,\varepsilon_R)=(+1,+1)$ and $(\varepsilon_L,\varepsilon_R)=(+1,-1)$ cases, whose Penrose diagrams are depicted in \Figs{fig:Penrose_diag_intro_++} and~\ref{fig:Penrose_diag_intro_+-}. We compute the von Neumann entropy of the two-screen system and the single-screen subsystems to leading geometrical order. We find a vanishing von Neumann entropy for the two-screen system, to leading order. The entanglement wedge comprises complete bulk Cauchy slices. In particular, as the screens evolve in time, the entire spacetime is covered by the entanglement wedge of the two screen system. This implies that the two screens are capable of holographically encoding the whole spacetime. The classical contribution to the von Neumann entropy of the single-screen subsystems arises solely from the exterior causal diamond region of the part of $\Sigma$ between the screens. Since this region has a boundary, extremizing the area functional requires the introduction of Lagrange multipliers and auxiliary fields enforcing the extremal homologous surfaces to lie in the causal diamond, including its boundary \cite{Franken:2023jas,Franken:2023pni}. The entanglement entropy between the single-screen subsystems is constant in the $(\varepsilon_L,\varepsilon_R)=(+1,+1)$ geometries, given by the Gibbons-Hawking entropy of the ``parent'' de Sitter space, but time-dependent in the geometries with $(\varepsilon_L,\varepsilon_R)=(+1,-1)$. 

For de Sitter bubble geometries with $(\varepsilon_L,\varepsilon_R)=(-1,-1)$, whose Penrose diagram is depicted in \Fig{fig:Penrose_diag_intro_--}, the causal patches of two antipodal observers are non-overlapping and causally disconnected. Following the covariant entropy bound \cite{Bousso:1999xy,Bousso:2002ju}, we locate the screens on the cosmological horizons bounding the causal patches of the two observers. As a consequence, the trajectories of the two screens never intersect, in contrast to the bubbles with $(\varepsilon_L,\varepsilon_R)=(+1,+1)$, $(\varepsilon_L,\varepsilon_R)=(+1,-1)$. We argue that the two-screen system is expected to be in a mixed state, and that only the interiors of the causal patches can be encoded holographically on both screens. The degrees of freedom of the two-screen system do not suffice to holographically encode the full bulk spacetime.

 The bilayer proposal can also be extended to flat Minkowski bubbles contained inside a ``parent'' de Sitter spacetime. Such geometries arise from the de Sitter bubble geometries with $(\varepsilon_L,\varepsilon_R)=(+1,+1)$ and $(\varepsilon_L,\varepsilon_R)=(+1,-1)$, by sending the bubble cosmological constant to zero while keeping the cosmological constant of the ``parent'' de Sitter fixed. This procedure yields two classes of Minkowski bubble geometries, depicted in the Penrose diagrams of \Fig{fig:Penrose_diag_Mink}. In both cases, we consider a pair of observers: one sitting at the center of the flat bubble, and the other at the pole of the de Sitter spherical cap. Their causal patches overlap, and the arguments of the holographic construction of \Sect{sect:holography} apply in these two classes of flat bubbles as well. In both cases, the area of the screen associated with the bubble observer is infinite in the far past and the far future. However, the classical, geometrical contribution to the von Neumann entropy of this screen system remains finite throughout the cosmological evolution. In four spacetime dimensions, our construction can incorporate a duality between the Milne patch of the Minkowski bubble and a $2$-dimensional Euclidean CFT on its asymptotic spatial boundary, conjectured in \cite{Freivogel:2006xu}.

The plan of the paper is as follows. In \Sect{sect:bubble_geometry}, we review the construction of de Sitter bubble geometries, and describe in great detail the geometrical features of the $(\varepsilon_L,\varepsilon_R)=(+1,+1)$ and $(\varepsilon_L,\varepsilon_R)=(+1,-1)$ bubbles. In \Sect{sect:holography}, we extend the causal patch holographic proposal of eternal de Sitter space to these two classes of de Sitter bubble geometries. We describe the trajectory of two holographic screens associated with the two antipodal observers, and generalize the bilayer holographic entanglement entropy prescription to these classes of geometries. We compute the leading geometrical contributions to the von Neumann entropy of the two-screen system and the single-screen subsystems. In \Sect{sect:--_geometry}, we describe the $(\varepsilon_L,\varepsilon_R)=(-1,-1)$ cases. We argue that the bilayer proposal is not applicable in these cases, and that more than two holographic screens are needed to holographically encode the entire spacetime. \Sect{sect:flat_bubbles} extends the holographic prescription of \Sect{sect:holography} to flat Minkowski bubbles inside a ``parent'' de Sitter space. Our conclusions are presented in \Sect{sect:conclusions}. In \Appendix{app:global_coord}, we describe the construction of a global conformal coordinate system to cover the whole de Sitter bubble geometry. Quantum corrections are discussed in \Appendix{app:quantum_corrections}. Finally, in \Appendix{app:area_extremization}, we describe the constrained extremization procedure in the exterior causal diamond, used to find the minimal extremal surface homologous to a single-screen system.

\section{de Sitter bubble geometry}
\label{sect:bubble_geometry}
In this section, we briefly review the construction of an $(n+1)$-dimensional de Sitter spacetime containing a de Sitter bubble with a smaller positive cosmological constant. The solutions will be invariant under time reversal and will break the SO$(1,n+1)$ symmetry of the background, ``parent'' de Sitter space to an SO$(1,n)$ symmetry. We will work in the thin-wall approximation, following \cite{Fabinger:2003gp}. In this approximation, the boundary wall of the bubble is taken to be of negligible thickness. As it evolves, it sweeps a codimension-$1$ timelike surface, splitting spacetime into the two regions of different cosmological constants. In the interior of the bubble, the cosmological constant is denoted by $\Lambda_L$, while in the surrounding region, the cosmological constant is denoted by $\Lambda_R$: $\Lambda_R>\Lambda_L$. The two geometries are glued along this timelike surface by suitable matching conditions \cite{Fabinger:2003gp}. Semiclassical bulk computations can be trusted as long as $R_R \gg \ell_p$ (and so, $R_L \gg \ell_p$); that is, the radius of curvature of the ``parent'' de Sitter space, which scales as $1/(\Lambda_R)^{1/2}$, is much greater than the Planck length.

%%%%%%%%%%%%%%%%%%%%%%%%%%%%%%%%%%%%%%%%%%%%%%%%%%%%%%%%%%%%%%%%%

\subsection{Euclidean geometry}
The Lorentzian geometry can be readily obtained by analytic continuation from Euclidean signature. Consider an $(n+1)$-dimensional sphere S$^{n+1}$ of radius $R$ and metric 
\begin{equation}\label{eq:euclid_sphere}
    \d s^2=R^2\left(\d\chi^2+\sin^2\chi \d\Omega^2_{n}\right),
\end{equation}
where $\chi\in[0,\pi]$ is a polar angle and $\d\Omega_{n}^2\!=\!\d\theta^2+\sin^2\theta \d\psi_1^2+\cdots+\sin^2\theta\sin^2\psi_1\cdots\sin^2\psi_{n-2}\d\psi^2_{n-1}$ is the metric on the unit $n$-dimensional sphere S$^{n}$. As usual, we can embed the sphere S$^{n+1}$ in the $(n+2)$-dimensional Euclidean flat space by the mapping
\begin{align} \label{eq:embedding}
    &X_1=R\cos\chi,\nonumber\\
    &X_2=R\sin\chi\cos\theta,\\
    &X_a=R\sin\chi\sin\theta n_a,\quad 3\leq a\leq n+2,\nonumber
\end{align}
where $\sum_an_a^2=1$. The embedding coordinates satisfy 
\begin{equation}
    \sum_{i=1}^{n+2}X_i^2=R^2.
\end{equation}

\begin{figure}[h!]
    %\centering
    \begin{subfigure}[t]{1\linewidth}
\centering
\begin{tikzpicture}[scale=0.9]
\begin{scope}[transparency group]
\begin{scope}[blend mode=multiply]

%right sphere
\draw[name path=A,line width= 1 pt] plot[variable=\t,samples=100,domain=-118:118] 
({3*cos(\t)+12},{3*sin(\t)});
\draw[line width= 1 pt] plot[variable=\t,samples=100,domain=-180:180] 
({3*cos(\t)+12},{3*sin(\t)});
\draw[line width= 1 pt] (12,0)--(13.38,2.65);
\node at (12.35,1.0) [label = right:$R_R$]{};
\node at (12,0) [circle,fill,inner sep=1pt]{};
\draw[name path=B,brown,line width= 1 pt] plot[variable=\t,samples=100,domain=-90:90] 
({0.4*cos(\t)+10.6},{2.63*sin(\t)});
\draw[dashed, brown, line width= 1 pt] plot[variable=\t,samples=100,domain=90:270] 
({0.4*cos(\t)+10.6},{2.63*sin(\t)});
\draw[line width= 1 pt] (10.61,0)--(10.61,2.63);
\node at (10.3,1) [label = right:$R_B$]{};
\node at (10.6,0) [circle,fill,inner sep=1pt]{};
\tikzfillbetween[of=A and B]{red, opacity=0.1};

\draw[->, line width= 1 pt, gray] (12,0) -- (12,3.5); % X_1 arrow 
\node at (12,3.5) [label = right:$X_{1R}$]{};
\draw[->, line width= 1 pt, gray] (12,0) -- (8.5,0); % X_2 arrow 
\node at (8.5,0) [label = above:$X_{2R}$]{};
\draw[->, line width= 1 pt, gray] (12,0) -- (14.2,-2.8); % X_a arrow 
\node at (14.2,-2.8) [label = below:$X_{\{a=3,...,n+2\}R}$]{};

\draw[->, line width= 1 pt, gray] (12,3.6) to[bend right=15] (9.5,2.7);
\node at (9.5,2.7) [label = above:$\chi_R$]{};

\draw[->, line width= 1 pt, gray] (8.6,-0.2) to[bend right=25] (11.5,-1);
\node at (11.5,-1) [label = below:$\theta_R$]{};

%left sphere
\draw[name path=C,line width= 1 pt] plot[variable=\t,samples=100,domain=139:221] ({4*cos(\t)+4},{4*sin(\t)});
\draw[line width= 1 pt] plot[variable=\t,samples=100,domain=-180:180] ({4*cos(\t)+4},{4*sin(\t)});
\draw[line width= 1 pt] (4,0)--(5.5,3.7);
\node at (4.5,1.8) [label = right:$R_L$]{};
\node at (4,0) [circle,fill,inner sep=1pt]{};
\draw[name path=D,brown,line width= 1 pt] plot[variable=\t,samples=100,domain=-90:90] 
({0.4*cos(\t)+1.02},{2.63*sin(\t)});
\draw[dashed, brown, line width= 1 pt] plot[variable=\t,samples=100,domain=90:270] 
({0.4*cos(\t)+1.02},{2.63*sin(\t)});
\draw[line width= 1 pt] (1.02,0)--(1.02,2.63);
\node at (0.7,1) [label = right:$R_B$]{};
\node at (1.02,0) [circle,fill,inner sep=1pt]{};
\tikzfillbetween[of=C and D]{blue, opacity=0.1};

\draw[->, line width= 1 pt, gray] (4,0) -- (4,4.5); % X_1 arrow 
\node at (4,4.5) [label = right:$X_{1L}$]{};
\draw[->, line width= 1 pt, gray] (4,0) -- (-0.5,0); % X_2 arrow 
\node at (-0.5,0) [label = above:$X_{2L}$]{};
\draw[->, line width= 1 pt, gray] (4,0) -- (7,-3.5); % X_a arrow 
\node at (6,-4) [label = right:$X_{\{a=3,...,n+2\}L}$]{};

\draw[->, line width= 1 pt, gray] (4,4.6) to[bend right=15] (1.5,3.8);
\node at (1.5,3.8) [label = above:$\chi_L$]{};

\draw[->, line width= 1 pt, gray] (-0.4,-0.2) to[bend right=30] (3,-1.5);
\node at (3,-1.5) [label = below:$\theta_L$]{};

\end{scope}
\end{scope}
\end{tikzpicture}
    \caption{\footnotesize Euclidean bubble geometries with $(\varepsilon_L,\varepsilon_R)=(+1,+1)$.}
    \label{fig:Euclidean_spheres_++}
\end{subfigure}
\qquad
\begin{subfigure}[t]{0.48\linewidth}
 \centering
\begin{tikzpicture}[scale=0.45]
\begin{scope}[transparency group]
\begin{scope}[blend mode=multiply]

%right sphere
\draw[name path=A,line width= 1 pt] plot[variable=\t,samples=100,domain=-62:62] 
({3*cos(\t)+12},{3*sin(\t)});
\draw[line width= 1 pt] plot[variable=\t,samples=100,domain=-180:180] 
({3*cos(\t)+12},{3*sin(\t)});
\draw[line width= 1 pt] (12,0)--(10.62,2.65);
\node at (9.2,1.5) [label = right:$R_R$]{};
\node at (12,0) [circle,fill,inner sep=1pt]{};
\draw[name path=B,brown,line width= 1 pt] plot[variable=\t,samples=100,domain=-90:90] 
({0.4*cos(\t)+13.3},{2.63*sin(\t)});
\draw[dashed, brown, line width= 1 pt] plot[variable=\t,samples=100,domain=90:270] 
({0.4*cos(\t)+13.3},{2.63*sin(\t)});
\draw[line width= 1 pt] (13.3,0)--(13.3,2.63);
\draw[dashed,line width= 0.5 pt] (12 ,0)--(13.3,0);
\node at (12.9,1) [label = right:$R_B$]{};
\node at (13.3,0) [circle,fill,inner sep=1pt]{};
\tikzfillbetween[of=A and B]{red, opacity=0.1};

\draw[->, line width= 1 pt, gray] (12,0) -- (12,3.5); % X_1 arrow 
\node at (11.7,3.5) [label = right:$X_{1R}$]{};
\draw[->, line width= 1 pt, gray] (12,0) -- (8.5,0); % X_2 arrow 
\node at (10,-1.6) [label = above:$X_{2R}$]{};
\draw[->, line width= 1 pt, gray] (12,0) -- (14.2,-2.8); % X_a arrow 
%\node at (14.2,-2.8) [label = below:$X_{\{a=3,...,n+2\}R}$]{};
\node at (14.2,-2.6) [label = below:$X_{aR}$]{};

%\draw[->, line width= 1 pt, gray] (12,3.6) to[bend right=15] (9.5,2.7);
%\node at (9.5,2.7) [label = above:$\chi_R$]{};

%\draw[->, line width= 1 pt, gray] (8.6,-0.2) to[bend right=25] (11.5,-1);
%\node at (11.5,-1) [label = below:$\theta_R$]{};

%left sphere
\draw[name path=C,line width= 1 pt] plot[variable=\t,samples=100,domain=139:221] ({4*cos(\t)+4},{4*sin(\t)});
\draw[line width= 1 pt] plot[variable=\t,samples=100,domain=-180:180] ({4*cos(\t)+4},{4*sin(\t)});
\draw[line width= 1 pt] (4,0)--(5.5,3.7);
\node at (4.5,2.4) [label = right:$R_L$]{};
\node at (4,0) [circle,fill,inner sep=1pt]{};
\draw[name path=D,brown,line width= 1 pt] plot[variable=\t,samples=100,domain=-90:90] 
({0.4*cos(\t)+1.00},{2.63*sin(\t)});
\draw[dashed, brown, line width= 1 pt] plot[variable=\t,samples=100,domain=90:270] 
({0.4*cos(\t)+1.0},{2.63*sin(\t)});
\draw[line width= 1 pt] (1.0,0)--(1.0,2.63);
\node at (0.8,1) [label = right:$R_B$]{};
\node at (1.0,0) [circle,fill,inner sep=1pt]{};
\tikzfillbetween[of=C and D]{blue, opacity=0.1};

\draw[->, line width= 1 pt, gray] (4,0) -- (4,4.5); % X_1 arrow 
\node at (3.9,4.5) [label = right:$X_{1L}$]{};
\draw[->, line width= 1 pt, gray] (4,0) -- (-0.5,0); % X_2 arrow 
\node at (-0.8,0) [label = above:$X_{2L}$]{};
\draw[->, line width= 1 pt, gray] (4,0) -- (7,-3.5); % X_a arrow 
%\node at (6,-4) [label = right:$X_{\{a=3,...,n+2\}L}$]{};
\node at (6,-4) [label = right:$X_{aL}$]{};

%\draw[->, line width= 1 pt, gray] (4,4.6) to[bend right=15] (1.5,3.8);
%\node at (1.5,3.8) [label = above:$\chi_L$]{};

%\draw[->, line width= 1 pt, gray] (-0.4,-0.2) to[bend right=30] (3,-1.5);
%\node at (3,-1.5) [label = below:$\theta_L$]{};

\end{scope}
\end{scope}
\end{tikzpicture}
    \caption{\footnotesize Euclidean bubble geometries with $(\varepsilon_L,\varepsilon_R)=(+1,-1)$.}
    \label{fig:Euclidean_spheres_+-}
\end{subfigure}
\quad 
 \begin{subfigure}[t]{0.48\linewidth}
    \centering
\raisebox{0.0cm}{
\begin{tikzpicture}[scale=0.45]
\begin{scope}[transparency group]
\begin{scope}[blend mode=multiply]

%right sphere
\draw[name path=A,line width= 1 pt] plot[variable=\t,samples=100,domain=-62:62] 
({3*cos(\t)+12},{3*sin(\t)});
\draw[line width= 1 pt] plot[variable=\t,samples=100,domain=-180:180] 
({3*cos(\t)+12},{3*sin(\t)});
\draw[line width= 1 pt] (12,0)--(10.62,2.65);
\node at (9.2,1.5) [label = right:$R_R$]{};
\node at (12,0) [circle,fill,inner sep=1pt]{};
\draw[name path=B,brown,line width= 1 pt] plot[variable=\t,samples=100,domain=-90:90] 
({0.4*cos(\t)+13.3},{2.63*sin(\t)});
\draw[dashed, brown, line width= 1 pt] plot[variable=\t,samples=100,domain=90:270] 
({0.4*cos(\t)+13.3},{2.63*sin(\t)});
\draw[line width= 1 pt] (13.3,0)--(13.3,2.63);
\draw[dashed,line width= 0.5 pt] (12,0)--(13.3,0);
\node at (12.9,1) [label = right:$R_B$]{};
\node at (13.3,0) [circle,fill,inner sep=1pt]{};
\tikzfillbetween[of=A and B]{red, opacity=0.1};

\draw[->, line width= 1 pt, gray] (12,0) -- (12,3.5); % X_1 arrow 
\node at (11.7,3.5) [label = right:$X_{1R}$]{};
\draw[->, line width= 1 pt, gray] (12,0) -- (8.5,0); % X_2 arrow 
\node at (10,-1.6) [label = above:$X_{2R}$]{};
\draw[->, line width= 1 pt, gray] (12,0) -- (14.2,-2.8); % X_a arrow 
%\node at (14.2,-2.8) [label = below:$X_{\{a=3,...,n+2\}R}$]{};
\node at (14.2,-2.6) [label = below:$X_{aR}$]{};

%\draw[->, line width= 1 pt, gray] (12,3.6) to[bend right=15] (9.5,2.7);
%\node at (9.5,2.7) [label = above:$\chi_R$]{};

%\draw[->, line width= 1 pt, gray] (8.6,-0.2) to[bend right=25] (11.5,-1);
%\node at (11.5,-1) [label = below:$\theta_R$]{};

%left sphere
\draw[name path=C,line width= 1 pt] plot[variable=\t,samples=100,domain=41:329] ({4*cos(\t)+4},{4*sin(\t)});
\draw[line width= 1 pt] plot[variable=\t,samples=100,domain=-180:180] ({4*cos(\t)+4},{4*sin(\t)});
\draw[line width= 1 pt] (4,0)--(5.5,3.7);
\node at (4.5,2.4) [label = right:$R_L$]{};
\node at (4,0) [circle,fill,inner sep=1pt]{};
\draw[name path=D,brown,line width= 1 pt] plot[variable=\t,samples=100,domain=-90:90] 
({0.4*cos(\t)+7.00},{2.63*sin(\t)});
\draw[dashed, brown, line width= 1 pt] plot[variable=\t,samples=100,domain=90:270] 
({0.4*cos(\t)+7.02},{2.63*sin(\t)});
\draw[line width= 1 pt] (7.0,0)--(7.0,2.63);
\draw[dashed, line width= 0.5 pt] (4.0,0)--(7.0,0);
\node at (5.05,1) [label = right:$R_B$]{};
\node at (7.0,0) [circle,fill,inner sep=1pt]{};
\tikzfillbetween[of=C and D]{blue, opacity=0.1};

\draw[->, line width= 1 pt, gray] (4,0) -- (4,4.5); % X_1 arrow 
\node at (3.8,4.5) [label = right:$X_{1L}$]{};
\draw[->, line width= 1 pt, gray] (4,0) -- (-0.5,0); % X_2 arrow 
\node at (-1,0) [label = above:$X_{2L}$]{};
\draw[->, line width= 1 pt, gray] (4,0) -- (7,-3.5); % X_a arrow 
%\node at (6,-4) [label = right:$X_{\{a=3,...,n+2\}L}$]{};
\node at (6,-4) [label = right:$X_{aL}$]{};

%\draw[->, line width= 1 pt, gray] (4,4.6) to[bend right=15] (1.5,3.8);
%\node at (1.5,3.8) [label = above:$\chi_L$]{};

%\draw[->, line width= 1 pt, gray] (-0.4,-0.2) to[bend right=30] (3,-1.5);
%\node at (3,-1.5) [label = below:$\theta_L$]{};

\end{scope}
\end{scope}
\end{tikzpicture}}
    \caption{\footnotesize Euclidean bubble geometries with $(\varepsilon_L,\varepsilon_R)=(-1,-1)$.}
    \label{fig:Euclidean_spheres_--}
    \end{subfigure}
\caption{\footnotesize We cut two different solutions of the Euclidean Einstein's equations with different cosmological constants, $\Lambda_L < \Lambda_R$ ($R_L>R_R$), along a cycle (brown lines) of the same size. In the thin-wall approximation, we glue the parts of the two solutions (shaded regions) together along this cycle. Note that for a specific choice of $R_B$, there are different choices of spherical caps - smaller or larger than a hemisphere - to glue together. Each geometry can be labeled by two signs $\varepsilon_L=\pm 1$, $\varepsilon_R=\pm 1$, as defined in Eq.~\eqref{eq:lorenzian_dom_walls}.
}
\label{fig:Euclidean_spheres}
\end{figure}
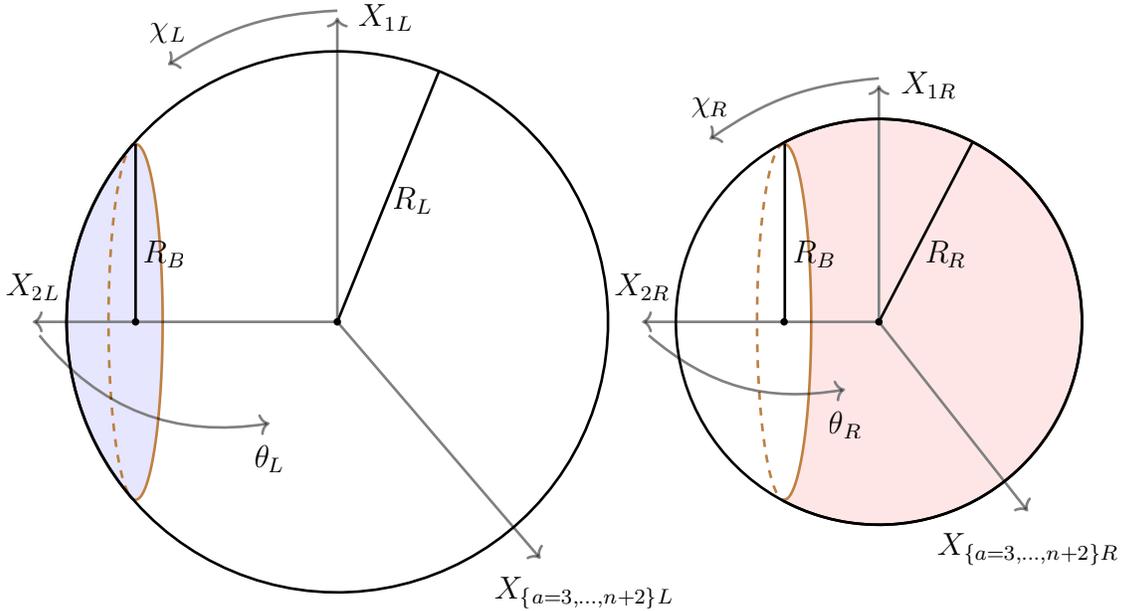

To construct the bubble geometry, S$^{n+1}$ is first cut along an S$^{n}$ sphere of radius $R_B < R$, which we refer to as the domain wall. See \Fig{fig:Euclidean_spheres}.
We may choose this codimension-$1$ domain wall to coincide with the intersection of S$^{n+1}$ with a constant $X_2$ plane, so that
\begin{equation}\label{eq:pos_wall}
    X_2^2+R_B^2=R^2.
\end{equation}
It follows, therefore, from Eqs.~\eqref{eq:embedding} and \eqref{eq:pos_wall} that along the domain wall the following equation holds:
\begin{equation}\label{eq:euclid_domain_wall}
\cos^2\theta=\left(1-\frac{R_B^2}{R^2}\right)\frac{1}{\sin^2\chi}.
\end{equation}

Let us now consider two S$^{n+1}$'s with radii $R_L$ and $R_R$, respectively, such that $R_L>R_R$. For each sphere, we follow the procedure described above, and, therefore, we obtain two spherical caps bounded by an S$^{n}$ of radius $R_B$, along which the coordinates $(\theta_{L},\chi_{ L})$ and $(\theta_{R},\chi_{ R})$ satisfy:
\begin{equation}\label{eq:euclid_dom_wall}
\cos^2\theta_{L}=\left(1-\frac{R_B^2}{R_L^2}\right)\frac{1}{\sin^2\chi_{ L}},\quad \cos^2\theta_{R}=\left(1-\frac{R_B^2}{R_R^2}\right)\frac{1}{\sin^2\chi_{ R}}.
\end{equation} 
The two caps are glued along their S$^{n}$ boundaries, enforcing the Israel junction conditions \cite{Israel:1966rt}. First, the induced metric must be continuous across the domain wall. In particular, the embedding coordinates $X_i$ for each sphere are matched along the domain wall, \eg
\begin{equation}\label{eq:euclid_matching_cond}
    R_L\cos\chi_{L}=R_R\cos\chi_{ R}.
\end{equation} 
Second, there is a jump discontinuity in the extrinsic curvature, which must be supported by a non-trivial energy-momentum tensor localized along the domain wall. Essentially, the domain wall behaves as an $(n-1)$-dimensional brane with tension given by the jump discontinuity in the trace of the extrinsic curvature \cite{Brown:1988kg,Brown:1987dd}.\footnote{Following \cite{Brown:1988kg,Brown:1987dd}, we take the unit normal to the domain wall to point from the region of smaller cosmological constant ($\Lambda_L$) to the region of bigger cosmological constant ($\Lambda_R$).} There are three possible cases that lead to time-symmetric de Sitter bubble geometries upon analytic continuation to Lorentzian signature, shown in \Fig{fig:Euclidean_spheres}. In cases (a) and (b), the left spherical cap with a smaller cosmological constant ($\Lambda_L < \Lambda_R$) is smaller than a hemisphere. In case (a), the right spherical cap with a bigger cosmological constant is larger (or equal) to a hemisphere, while in case (b) it is smaller (or equal) to a hemisphere. In case (c), the left spherical cap is larger (or equal) to a hemisphere, while the right spherical cap is smaller than a hemisphere. A fourth case, where both the left and right spherical caps are larger than a hemisphere, does not lead to a well-defined Lorentzian bubble geometry. See the following subsection.  

The tension of the domain will be discussed in more detail in the following subsection. For cases (a) and (b), the tension is positive. Geometries in both of these two classes can be obtained from instanton solutions of gravity scalar field theory models \cite{Coleman:1980aw}, in the thin-wall approximation. For case (c), on the other hand, the tension is negative, and so the discontinuity in the extrinsic curvature must be supported by a non-fluctuating negative tension brane such as a string theory orientifold. Finally, let us note that the gluing procedure breaks the initial SO$(n+2)$ symmetry of the S$^{n+1}$'s to SO$(n+1)$, the symmetry of the domain wall (which is an S$^{n}$ sphere).
%%%%%%%%%%%%%%%%%%%%%%%%%%%%%%%%%%%%%%%%%%%%%%%%%%%%%%%%%%%%%%%%

\subsection{Lorentzian geometry}
\label{sect:lorentzian_geometry}
Upon the analytic continuation 
\begin{equation}\label{eq:analytic_conti}
\chi\rightarrow i\tau+\frac{\pi}{2}
\end{equation}
the S$^{n+1}$ sphere \eqref{eq:euclid_sphere} yields the $(n+1)$-dimensional de Sitter spacetime 
\begin{equation}
    \d s^2=R^2\left(-\d\tau^2+\cosh^2\tau \d\Omega^2_{n}\right),
\end{equation}
where $\tau\in(-\infty,+\infty)$ is the global time and $R$ is the radius of curvature. The time reparametrization
\begin{equation}\label{eq:time_reparam}
    \frac{1}{\cos \sigma}=\cosh\tau
\end{equation}
brings the metric into the conformal form
\begin{equation}\label{eq:dS_metric}
    \d s^2=\frac{R^2}{\cos^2 \sigma}\left(-\d \sigma^2+\d\theta^2+\sin^2\theta\d\Omega^2_{n-1}\right),
\end{equation}
where $\sigma\in(-\pi/2,\pi/2)$ is the conformal time and $\theta\in[0,\pi]$ is a polar angle. In this form, the metric has a manifest SO$(n)$ symmetry, which is the rotational symmetry of the constant $\sigma$ and $\theta$ $(n-1)$-dimensional spherical surfaces. 

The analytic continuation \eqref{eq:analytic_conti} can also be applied to the two spherical caps of different radii described in the previous section, yielding two parts of $(n+1)$-dimensional de Sitter spacetimes with cosmological constants $\Lambda_L$ and $\Lambda_R$, respectively, such that $\Lambda_L< \Lambda_R$. (Recall that in terms of the radius of curvature, the cosmological constant is given by $\Lambda=n(n-1)/(2R^2)$). Each de Sitter part is bounded by an $n$-dimensional timelike surface, corresponding to the worldvolume of the domain wall. Along the worldvolumes of the walls, the following relations hold:
\begin{equation}\label{eq:lorenzian_dom_walls}
    \cos\theta_{L}=\varepsilon_L\sqrt{1-\frac{R_B^2}{R^2_L}}\cos \sigma_L,\quad \cos\theta_{R}=\varepsilon_R\sqrt{1-\frac{R_B^2}{R^2_R}}\cos \sigma_R,
\end{equation}
where $\theta_L$ $(\theta_R)$ and $\sigma_L$ $(\sigma_R)$ are the polar angle and conformal time coordinates for the left (right) de Sitter part, and $\varepsilon_L=\pm 1$, $\varepsilon_R=\pm 1$. These relations follow from Eqs.~\eqref{eq:analytic_conti}, \eqref{eq:time_reparam} and \eqref{eq:euclid_dom_wall}. For $i=\{L,R\}$, the choice $\varepsilon_i=+1$ ($\varepsilon_i=-1$) describes a domain wall along which the polar coordinate $\theta_i<\pi/2$ $(\theta_i>\pi/2)$. In particular for $\varepsilon_i=+1$ ($\varepsilon_i=-1$), as $\sigma_i$ increases from $-\pi/2$ to $0$, the domain wall trajectory is such that $\theta_i$ decreases (increases) from $\pi/2$ to a minimum (maximum) value, and then it increases (decreases) back to $\pi/2$ as $\sigma_i \to \pi/2$. That is, the domain wall trajectory is concave with respect to the right (left) region with bigger (smaller) cosmological constant. See \Fig{fig:Penrose_diag_intro}. All combinations are possible except for the case $(\varepsilon_L, \varepsilon_R)=(-1,+1)$, since then $\theta_L$ would be greater than $\theta_R$.\footnote{For $(\varepsilon_L, \varepsilon_R)=(-1,+1)$, the left trajectory should be concave with respect to the left region and the right trajectory with respect to the right region. However, the left and right wall trajectories must also have the same endpoints at $(\sigma_i, \theta_i)=(\pm \pi/2, \theta_i=\pi/2)$, which cannot be simultaneously realized.} The Euclidean counterparts of the three allowable cases $(\varepsilon_L,\varepsilon_R)=(+1,+1)$, $(\varepsilon_L,\varepsilon_R)=(+1,-1)$ and $(\varepsilon_L,\varepsilon_R)=(-1,-1)$ correspond to cases (a), (b) and (c), respectively, in \Fig{fig:Euclidean_spheres}.

The tension $T$ of the domain wall can be obtained from the jump discontinuity in the trace of the extrinsic curvature \cite{Brown:1987dd, Brown:1988kg}. It is given by 
\begin{equation}
    8\pi GT/(n-1)=-\varepsilon_R\sqrt{R_B^{-2}-2\Lambda_R/(n(n-1))}+\varepsilon_L\sqrt{R_B^{-2}-2\Lambda_L/(n(n-1))}.
\end{equation}
The contribution of the left de Sitter region is positive (negative) if the corresponding Euclidean spherical cap is smaller (larger) than a hemisphere, and vice versa for the right de Sitter region \cite{Brown:1987dd,Brown:1988kg}. Since $R_B \leq R_R$, there is a critical value given by
\begin{equation}
    T_C=(8\pi G)^{-1}\sqrt{2(\Lambda_R-\Lambda_L)(n-1)/n}.
\end{equation}
Therefore, for both the $(\varepsilon_L,\varepsilon_R)=(+1,+1)$ and $(\varepsilon_L,\varepsilon_R)=(+1,-1)$ cases, the tension of the domain wall is positive. Since $\Lambda_L < \Lambda_R$, the contribution of the left de Sitter region (which is positive) is always larger than the absolute value of the contribution of the right de Sitter region. The $(\varepsilon_L,\varepsilon_R)=(+1,+1)$ geometries are subcritical. Indeed, as $R_B$ increases from zero to its maximum value $R_R$, the tension increases from zero to $T_C$. On the other hand, the $(\varepsilon_L,\varepsilon_R)=(+1,-1)$ geometries are supercritical. As $R_B$ decreases from $R_R$ to zero, the tension increases from $T_C$ to infinity. For these geometries, we must keep $R_B$ close to its maximum value $R_R$, keeping the tension close to $T_C$, in order to ensure the validity of the semiclassical approximation.\footnote{It was argued in \cite{Maloney:2002rr} that the spontaneous nucleation of Coleman-De Luccia bubbles with supercritical tension cannot be analyzed in the semiclassical regime. Indeed, as can be seen in \Fig{fig:Penrose_diag_intro}, since such bubbles are large, their nucleation amounts to a large perturbation of the parent de Sitter geometry. In our work however, we study time-symmetric solutions, where the bubble is present for the entire cosmological evolution. It would be interesting to study the stability of such supercritical bubble geometries against small perturbations, to see if the asymptotic structure at null future infinity would remain the same.}

For the $(\varepsilon_L,\varepsilon_R)=(-1,-1)$ case, the tension is negative and could be realized by using a non-fluctuating negative tension brane such as a string theory orientifold. Indeed, the consistency of such bubble geometries would require special boundary conditions across the domain wall in order to avoid instabilities -- see \eg \cite{Marolf:2002np} for discussions.\footnote{The fact that the domain wall trajectory is hyperbolic retaining an SO$(1,n)$ symmetry may facilitate the construction.} Even though we do not have a concrete, explicit construction, we analyze this case as well to highlight that the resulting causal structure renders the holographic description in terms of two screens not possible.   

The Penrose diagrams for the bubble geometries with $(\varepsilon_L,\varepsilon_R)=(+1,+1)$ and $(\varepsilon_L,\varepsilon_R)=(+1,-1)$ are shown in \Fig{fig:Penrose_diag_++} and \Fig{fig:Penrose_diag_+-}, respectively. We will describe their geometrical features in great detail in \Sect{sect:++_+-_geometry}. For both cases, a part of the right de Sitter region (as well as a segment of the domain wall trajectory) lies in the causal patch of a comoving observer at the center of the bubble. In other words, an observer at the center of the bubble has access to the ``parent'' de Sitter space, which has a larger cosmological constant. The holographic proposal of \Sect{sect:holography} in terms of a two-screen system and, in particular, the bilayer proposal for holographic entanglement entropy computations apply only for these bubble geometries. The Penrose diagram for the $(\varepsilon_L, \varepsilon_R)=(-1,-1)$ case is shown in \Fig{fig:Penrose_diag_--}. We will describe features of it in \Sect{sect:--_geometry}. This is the case of a large bubble; the causal patch of a comoving observer at the center lies entirely in the left de Sitter bubble, having no overlap with the right de Sitter region. As we will argue, more than two holographic screens are needed to describe all regions of this spacetime.    

In all three cases, the trajectories of the domain walls are depicted in the Penrose diagrams by the thick brown curves. Points on the left and right walls are identified using the matching condition 
\begin{equation}\label{eq:matching_cond}
R_L\tan\sigma_L=R_R\tan\sigma_R,
\end{equation}
where $\sigma_L$ and $\sigma_R$ are the conformal times on the left and right walls. This follows from the continuation of Eq.~(\ref{eq:euclid_matching_cond}) to Lorentzian signature.

One may also describe the full spacetime using a global coordinate system $(\theta_G,\sigma_G)$, instead of two different ones for the left and right de Sitter patches. For instance, we can keep the right coordinates and perform a conformal transformation on the left coordinate system. The resulting Penrose diagrams are shown in \Figs{fig:Penrose_diag_++_connected}, \ref{fig:Penrose_diag_+-_connected} and ~\ref{fig:Penrose_diag_--_connected}, for the bubble geometries with $(\varepsilon_L,\varepsilon_R)=(+1,+1)$, $(\varepsilon_L,\varepsilon_R)=(+1,-1)$ and $(\varepsilon_L,\varepsilon_R)=(-1,-1)$, respectively. More details about the global coordinate system are given in \Appendix{app:global_coord}.

All three types of solutions are invariant under time-reversal and retain an SO$(1,n)$ symmetry, the symmetry of the codimension-$1$ timelike surface swept by the domain wall as it evolves. The bubble starts from infinite size in the far past and contracts to a finite size. It then bounces and expands to infinite size in the far future. This is independently of the choice $(\varepsilon_L, \varepsilon_R)$. However, only the positive tension cases $(\varepsilon_L, \varepsilon_R)=(+1, \pm 1)$ may describe features of expanding Coleman De Luccia bubbles formed by spontaneous nucleation. It would be interesting to check for the stability of these time-symmetric geometries, and in particular the supercritical ones, against small perturbations. Coleman de Luccia bubbles that collapse in a finite amount of time will not be analyzed in this work, since for these cases the asymptotic structure at null future infinity is that of the parent de Sitter space, whose holographic description was analyzed in \cite{Franken:2023pni}.  

%%%%%%%%%%%%%%%%%%%%%%%%%%%%%%%%%%%%%%%%%%%%%%%%%%%%%%%%%%%%%%%

\subsection{Geometrical features of the $(\varepsilon_L,\varepsilon_R)=(+1,\pm 1)$ bubbles}
\label{sect:++_+-_geometry}
In this section we describe a number of geometrical features of the positive tension $(\varepsilon_L,\varepsilon_R)=(+1,+1)$ and $(\varepsilon_L,\varepsilon_R)=(+1,-1)$ bubble geometries, which will be useful in motivating the holographic construction of \Sect{sect:holography}. The spacetime can be foliated with SO$(n)$-symmetric Cauchy slices, which have spherical topology. The domain wall trajectory divides each of them into two spherical caps, one lying inside the bubble and the other in the region exterior to the bubble. Following \cite{Susskind:2021omt}, we refer to the pole of the left spherical cap inside the bubble as the ``pode'', and to the pole of the right spherical cap as the ``antipode''. The pode and antipode trajectories are represented by the left and right vertical edges of the Penrose diagram, which we denote by $AA'$ and $BB'$, respectively. See \Figs{fig:Penrose_diag_++} and \ref{fig:Penrose_diag_+-}. We consider a pair of comoving observers, one sitting at the pode and the other at the antipode. 

Points in the Penrose diagram correspond to codimension-$2$ surfaces of constant $(\theta,\sigma)$, which are SO$(n)$-symmetric spheres. Their areas in the left and right regions are given, respectively, by:
\begin{equation}\label{eq:area}
    {\rm Area}(\theta_L,\sigma_L)=\omega_{n-1}\left(\frac{R_L\sin\theta_L}{\cos\sigma_L}\right)^{n-1},\quad 
    {\rm Area}(\theta_R,\sigma_R)=\omega_{n-1}\left(\frac{R_R\sin\theta_R}{\cos\sigma_R}\right)^{n-1},
\end{equation}
where $\omega_{n-1}$ is the area of the unit $(n-1)$-dimensional sphere. The matching condition \eqref{eq:matching_cond} ensures that the area function is continuous across the domain wall. 

Along the wall trajectory, the area is given by
\begin{equation}\label{eq:area_domain_wall}
\begin{aligned}
    &{\rm Area}(\theta_L,\sigma_L)=\omega_{n-1}\left(R_L^2\tan^2\sigma_L+R_B^2\right)^{\frac{n-1}{2}},\\
    &{\rm Area}(\theta_R,\sigma_R)=\omega_{n-1}\left(R_R^2\tan^2\sigma_R+R_B^2\right)^{\frac{n-1}{2}},
\end{aligned}
\end{equation}
independently of the choice of $(\varepsilon_L,\varepsilon_R)$. The matching condition \eqref{eq:matching_cond} ensures that the expressions are equal along the wall, once the points are identified.
The area of the domain wall decreases from $+\infty$ at $\sigma_{L(R)}=-\pi/2$ to its minimum value $\omega_{n-1}R_B^{n-1}$ at $\sigma_{L(R)}=0$, and then increases to $+\infty$ at $\sigma_{L(R)}=\pi/2$.

\subsubsection{The $(\varepsilon_L,\varepsilon_R)=(+1,+1)$ bubbles}

\noindent $\bullet$ {\bf Cosmological horizons}

\begin{figure}[h!]
 \begin{subfigure}[t]{0.48\linewidth}
    \centering
\raisebox{0.9cm}{
\begin{tikzpicture}
\begin{scope}[transparency group]
\begin{scope}[blend mode=multiply]

%wall
\draw[name path=A, line width= 1 pt,brown] plot[variable=\t,samples=91,domain=-45:45] ({3*sec(\t)-4.23},{3*tan(\t)});
\draw[name path=B, line width= 1 pt,brown] plot[variable=\t,samples=91,domain=-45:45] ({2*sec(\t)-2.83},{3*tan(\t)});
\tikzfillbetween[of=A and B]{gray, opacity=0.3}; 

\draw[name path=C](0,-3) -- (-3,-3) -- (-3,3) -- (0,3);
\draw[name path=D](0,-3) -- (3,-3) -- (3,3) -- (0,3);

\tikzfillbetween[of=A and C]{blue, opacity=0.1};
\tikzfillbetween[of=B and D]{red, opacity=0.1};

\draw[dashed] (-3,3)--(-1.05,1.05);
\draw[dashed] (-3,-3)--(-1.05,-1.05);
\draw[dashed] (-0.5,1.6)--(1.1,0);
\draw[dashed] (-0.5,-1.6)--(1.1,0);
\draw[dashed] (3,3)--(0,0);
\draw[dashed] (3,-3)--(0,0);
\draw[dotted] (0,0)--(-0.7,0.7);
\draw[dotted] (0,0)--(-0.7,-0.7);

%Bousso's wedges
\draw (0.55,1.8) -- (0.75,2) -- (0.95,1.8);
\draw (-0.2,0.5) -- (-0.0,0.7) -- (0.2,0.5);
\draw (-2.2,0.2) -- (-2,0) -- (-2.2,-0.2);
\draw (-0.6,0.2) -- (-0.4,0) -- (-0.6,-0.2);
\draw (-1.5,2.2) -- (-1.3,2.4) -- (-1.1,2.2);
\draw (-1.5,-2.2) -- (-1.3,-2.4) -- (-1.1,-2.2);
\draw (2.2,0.2) -- (2,0) -- (2.2,-0.2);
\draw (0.6,0.2) -- (0.4,0) -- (0.6,-0.2);
\draw (0.55,-1.8) -- (0.75,-2) -- (0.95,-1.8);
\draw (-0.2,-0.5) -- (0,-0.7) -- (0.2,-0.5);

\node at (-3,-3) [label=left:$A$]{};
\node at (-3,3) [label=left:$A'$]{};
\node at (3,-3) [label=right:$B$]{};
\node at (3,3) [label=right:$B'$]{};

\node at (-1.05,-1.05) [circle,fill,inner sep=1.5pt,label=left:$P$]{};
\node at (-0.75,-0.75) [circle,fill,inner sep=1.5pt]{};
\node at (-0.9,-0.75) [label=right:$P'$]{};
\node at (-1.05,1.05) [circle,fill,inner sep=1.5pt,label=left:$Q$]{};
\node at (-0.75,0.75) [circle,fill,inner sep=1.5pt]{};
\node at (-0.9,0.75) [label=right:$Q'$]{};

\node at (-0.55,-1.65) [circle,fill,inner sep=1.5pt,label=right:$P$]{};
\node at (-0.55,1.65) [circle,fill,inner sep=1.5pt,label=right:$Q$]{};

\node at (1.1,0) [circle,fill,inner sep=1.5pt]{};
\node at (1.2,-0.1) [label = above:$O$]{};
\node at (0,0) [circle,fill,inner sep=1.5pt]{};
\node at (0.1,-0.1) [label = above:$O'$]{};

\node at (0.57,-0.57) [circle,fill,inner sep=1.5pt, label = below:$T$]{};

%axes
\node at (-3, 0) [label={[rotate=90]$\theta_L = 0$}] {};
\node at (3.6, 0) [label={[rotate=90]$\theta_R = \pi$}] {};
\node at (0,-3.75) [label={$\sigma_L=\sigma_R=-\pi/2$}] {};
\node at (0,2.75) [label={$\sigma_L=\sigma_R=\pi/2$}] {};
\node at (1.7, 1.5) [label={[rotate=45] \scriptsize 
 $\theta_R-\sigma_R = \pi/2$}] {};
\node at (-2,1.8) [label={[rotate=-45]\scriptsize $\theta_L+\sigma_L = \pi/2$}] {};
\node at (-1.6, -2.4) [label={[rotate=45]\scriptsize $ \theta_L-\sigma_L = \pi/2$}] {};
\node at (1.6, -1.8) [label={[rotate=-45]\scriptsize $ \theta_R+\sigma_R = \pi/2$}] {};

\end{scope}
\end{scope}
\end{tikzpicture}}
    \caption{\footnotesize Two different conformal coordinate systems $(\theta_L,\sigma_L)$ and $(\theta_R,\sigma_R)$, covering respectively the blue and red de Sitter regions. The blue region represents the bubble with the smaller cosmological constant. Each region is bounded by a thin domain wall trajectory (brown lines). The gray-shaded region is not a part of the spacetime. Points on the left and right domain wall trajectories are identified via the matching condition $R_L\tan\sigma_L=R_R\tan\sigma_R$.}
    \label{fig:Penrose_diag_++_disconnected}
    \end{subfigure}
\quad \
\begin{subfigure}[t]{0.48\linewidth}
 \centering
\begin{tikzpicture}
\begin{scope}[transparency group]
\begin{scope}[blend mode=multiply]
\path
       +(-3,4.5) coordinate (IItophat)
       +(0,3) coordinate (IItophatright)
       +(3,3)  coordinate (IIsquaretopright)
       +(3,-3) coordinate (IIsquarebotright)
       +(0,-3) coordinate (IIbothatright)
       +(-3,-4.5) coordinate (IIbothat)
       +(-3,3) coordinate (IIsquaretopleft)
       +(-3,-3) coordinate(IIsquarebotleft)
      
       ;

\draw[name path=A,line width= 1 pt,brown] plot[variable=\t,samples=90,domain=-44.8:44.8] ({3*sec(\t)-4.23},{3.02*tan(\t)});

\draw[name path=C](0,-3) to [bend left=20] (-3,-4.5) -- (-3,4.5) to [bend left=20] (0,3);
\draw[name path=D](0,-3) -- (3,-3) -- (3,3) -- (0,3);

\tikzfillbetween[of=A and C]{blue, opacity=0.1};
\tikzfillbetween[of=A and D]{red, opacity=0.1};

\draw[dashed] (IIsquaretopright) -- (0,0) -- (IIsquarebotright);
\draw[dashed] (IItophat) -- (1.5,0) -- (IIbothat);
\draw[dotted] (-1,-1) -- (0,0) -- (-1,1);

%\draw (0,-3) to[bend left=44] (0,3);

\draw (0.55,1.8) -- (0.75,2) -- (0.95,1.8);
\draw (-0.2,0.8) -- (0,1) -- (0.2,0.8);
\draw (-2.2,0.2) -- (-2,0) -- (-2.2,-0.2);
\draw (-0.8,0.2) -- (-0.6,0) -- (-0.8,-0.2);
\draw (-1.7,3.6) -- (-1.5,3.8) -- (-1.3,3.6);
\draw (-1.7,-3.6) -- (-1.5,-3.8) -- (-1.3,-3.6);
\draw (2.2,0.2) -- (2,0) -- (2.2,-0.2);
\draw (0.85,0.2) -- (0.65,0) -- (0.85,-0.2);
\draw (0.55,-1.8) -- (0.75,-2) -- (0.95,-1.8);
\draw (-0.2,-0.8) -- (0,-1) -- (0.2,-0.8);

\node at (-3,-4.5) [label=left:$A$]{};
\node at (-3,4.5) [label=left:$A'$]{};
\node at (3,-3) [label=right:$B$]{};
\node at (3,3) [label=right:$B'$]{};

%\node at (-0.5,-2.08) [label=$P$]{};
\node at (-0.58,-2.08) [circle,fill,inner sep=1.5pt, label=left:$P$]{};
%\node at (-0.60,2) [label=$Q$]{};
\node at (-0.58,2.08) [circle,fill,inner sep=1.5pt, label=left:$Q$]{};

\node at (-1.05,-1.05) [circle,fill,inner sep=1.5pt,label=left:$P'$]{};
\node at (-1.05,1.05) [circle,fill,inner sep=1.5pt,label=left:$Q'$]{};

\node at (1.5,0) [circle,fill,inner sep=1.5pt]{};
\node at (1.6,0) [label = above:$O$]{};
\node at (0,0) [circle,fill,inner sep=1.5pt]{};
\node at (0.1,0) [label = above:$O'$]{};

\node at (0.75,-0.75) [circle,fill,inner sep=1.5pt, label = below:$T$]{};

\end{scope}
\end{scope}
\end{tikzpicture}
    \caption{\footnotesize A single global conformal coordinate system $(\theta_G,\sigma_G)$ covering the entire spacetime. The brown curve is a thin wall trajectory that separates the two de Sitter regions with different cosmological constants.}
    \label{fig:Penrose_diag_++_connected}
\end{subfigure}
\caption{\footnotesize The Penrose diagrams for the $(\varepsilon_L,\varepsilon_R)=(+1,+1)$ bubble geometry. The blue (red) region is the interior (exterior) of the bubble, which is a part of dS$_{n+1}$ with a smaller (larger) cosmological constant. Appropriate boundary conditions enforce the identification of the points on the left and right domain wall trajectories bounding the two de Sitter regions. The left and right vertical edges correspond to the worldlines of two antipodal observers. Each of them has a causal patch bounded by future and past cosmological horizons depicted by the thick dashed lines. The apparent horizons associated with the left observer at the center of the bubble are depicted by the thin dotted lines. Bousso wedges are drawn in each region of the diagrams.
}
\label{fig:Penrose_diag_++}
\end{figure}
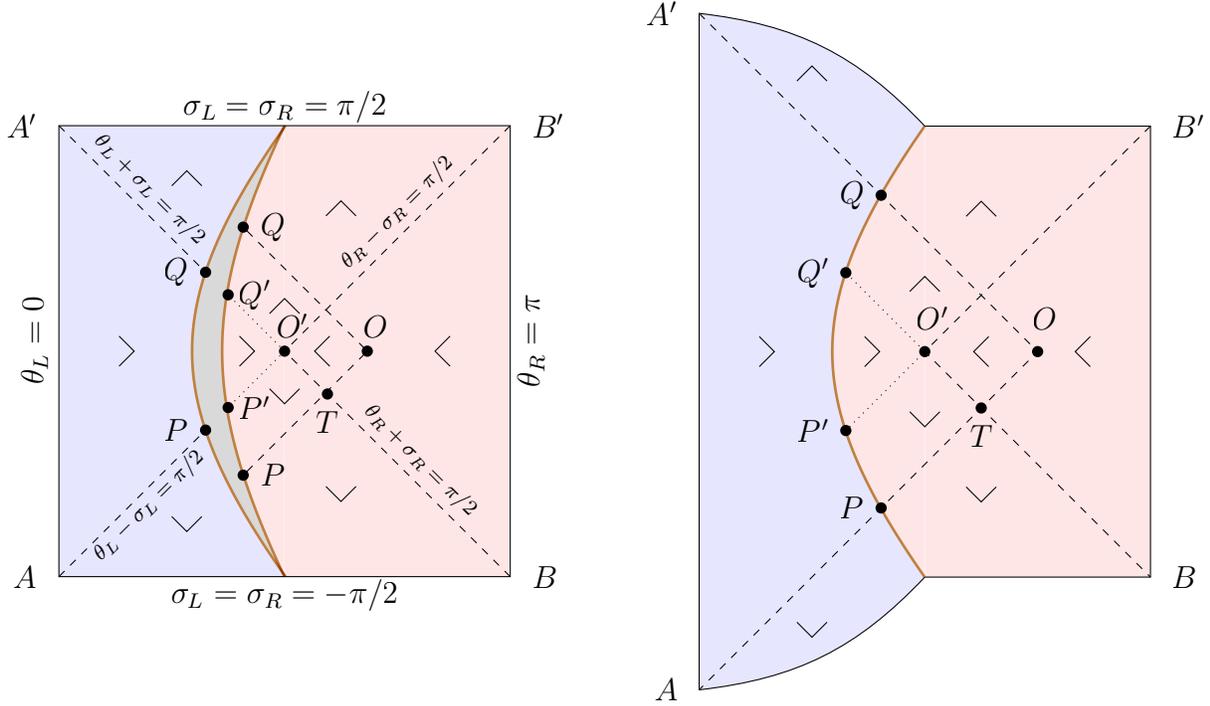

The Penrose diagram for the bubble geometry with $(\varepsilon_L,\varepsilon_R)=(+1,+1)$ is depicted in \Fig{fig:Penrose_diag_++}.
Along the right domain wall trajectory $\theta_R < \pi/2$, and therefore the causal patch of the antipode observer is outside the bubble (and inside the right de Sitter region). The antipode causal patch is delimited by future and past cosmological horizons, the null line segments $O'B'$ and $BO'$ of \Fig{fig:Penrose_diag_++}, described, respectively, by the equations
\begin{equation}
 \left\{\!\begin{array}{ll}
\dis \sigma_R-\theta_R=-\frac{\pi}{2}\, , &\mbox{if } \dis 0\le \sigma_R <\frac{\pi}{2}\, , \\
\dis \sigma_R+\theta_R=\frac{\pi}{2}\, , &\mbox{if } \dis -\frac{\pi}{2}<\sigma_R \le 0\, .\esp
\end{array}\right.
\label{eq:right_horizons}
\end{equation}
We denote by $O'$ the bifurcation point of these horizons, which lies in the region exterior to the bubble and has conformal coordinates $(\theta_{RO'}, \sigma_{RO'})=(\pi/2, 0)$. We will refer to the union of these future and past horizons associated with the antipode observer as the ``antipode horizon''. As we evolve along the antipode horizon, the area of the spherical sections remains constant, equal to $\omega_{n-1} R_R^{n-1}$, as can be verified by Eq.~(\ref{eq:area}). 

It will be useful to denote by $P'$ and $Q'$ the intersections between the future and past cosmological horizons of the observer at the antipode with the trajectory of the domain wall, \Eqs{eq:lorenzian_dom_walls}. The right conformal coordinates of $P'$ and $Q'$ are given by
\begin{equation}\label{eq:rp'_rq'_coordinates_++}
\theta_{RP'}=\theta_{RQ'}=\arctan\frac{R_R}{\sqrt{R_R^2-R_B^2}}, \quad \sigma_{RP'}=-\sigma_{RQ'}=-\arctan\frac{\sqrt{R_R^2-R_B^2}}{R_R}.
\end{equation}
Notice that both $\theta_{RP'}$ and $\theta_{RQ'}$ are smaller than $\pi/2$ as expected.
Using \eqref{eq:lorenzian_dom_walls}, the matching condition \eqref{eq:matching_cond} and \eqref{eq:rp'_rq'_coordinates_++}, we can obtain the corresponding left conformal coordinates:
\begin{equation}
\theta_{LP'}=\theta_{LQ'}=\arctan\frac{R_R}{\sqrt{R_L^2-R_B^2}}, \quad \sigma_{LP'}=-\sigma_{LQ'}=-\arctan\frac{\sqrt{R_R^2-R_B^2}}{R_L}.
\label{eq:lp'_lq'_coordinates_++}
\end{equation}

As shown in \Fig{fig:Penrose_diag_++}, the future and past cosmological horizons delimiting the causal patch of the pode observer, depicted by the thick dashed lines, have segments inside and outside the bubble. This is because $\theta_{L}<\pi/2$ along the left domain wall trajectory. As a result, a part of the right de Sitter region and a segment of the domain wall trajectory lie in the causal patch of the pode observer. 

Denote by $Q$ and $P$ the intersections between the horizons and the trajectory of the domain wall. See \Fig{fig:Penrose_diag_++}.
Inside the bubble, the future and past pode cosmological horizons are represented, respectively, by the null line segments $QA'$ and $AP$, with equations:
\begin{equation}
 \left\{\!\begin{array}{ll}
\dis \sigma_L+\theta_L=\frac{\pi}{2}\, , &\mbox{if } \dis \sigma_{LQ}\le \sigma_L <\frac{\pi}{2}\, , \\
\dis \sigma_L-\theta_L=-\frac{\pi}{2}\, , &\mbox{if } \dis -\frac{\pi}{2}<\sigma_L \le \sigma_{LP}\, ,\esp
\end{array}\right.
\label{eq:left_horizon_left}
\end{equation}
where $\sigma_{LQ}$ and $\sigma_{LP}$ are the left conformal time coordinates of $Q$ and $P$. Using the horizon equations above and Eq.~\eqref{eq:lorenzian_dom_walls} for the left domain wall trajectory, the left conformal coordinates of $Q$ and $P$ are given explicitly by
\begin{equation}\label{eq:lp_lq_coordinates_++}
\theta_{LQ}=\theta_{LP}=\arctan\frac{R_L}{\sqrt{R_L^2-R_B^2}}, \quad \sigma_{LQ}=-\sigma_{LP}=\arctan\frac{\sqrt{R_L^2-R_B^2}}{R_L}.
\end{equation}
Furthermore, using \eqref{eq:lorenzian_dom_walls} for the right domain wall trajectory, the matching condition \eqref{eq:matching_cond} and \eqref{eq:lp_lq_coordinates_++}, the right conformal coordinates of $Q$ and $P$ are given by 
\begin{equation}
\theta_{RQ}=\theta_{RP}=\arctan\frac{R_L}{\sqrt{R_R^2-R_B^2}}, \quad \sigma_{RQ}=-\sigma_{RP}=\arctan\frac{\sqrt{R_L^2-R_B^2}}{R_R}.
\label{eq:rp_rq_coordinates_++}
\end{equation}
 
Outside the bubble, the future and past horizons are represented by the null line segments $OQ$ and $PO$, with $O$ being their bifurcation point. Utilizing the right conformal coordinate system, these segments of the horizons are described, respectively, by the equations:
\begin{equation}
 \left\{\!\begin{array}{ll}
\dis \sigma_R+\theta_R=\sigma_{RQ}+\theta_{RQ}\, , &\mbox{if } \dis \sigma_{RO} \le \sigma_R \leq \sigma_{RQ}\, , \\
\dis \sigma_R-\theta_R=\sigma_{RP}-\theta_{RP}\, , &\mbox{if } \dis \sigma_{RP}\leq\sigma_R \le \sigma_{RO}\, ,\esp
\end{array}\right.
\label{eq:left_horizon_right}
\end{equation}
where $\sigma_{RQ}$ and $\sigma_{RP}$ are given in \Eq{eq:rp_rq_coordinates_++}. We will refer to the union of the above lightlike line segments as the ``pode horizon''. Using the horizon equations above and \Eq{eq:rp_rq_coordinates_++}, it is easy to derive the conformal coordinates of the bifurcation point $O$: 
\begin{equation}
\theta_{RO}=\pi-\arctan\frac{R_L R_R+\sqrt{R_R^2-R_B^2}\sqrt{R_L^2-R_B^2}}{R_L\sqrt{R_L^2-R_B^2}-R_R\sqrt{R_R^2-R_B^2}}, \,\,\,\,\, \sigma_{RO}=0. 
\end{equation}
Since $R_L>R_R$, $\pi/2<\theta_{RO} <\pi$ and $\sigma_{RP}<\sigma_{RP'}$, as can be seen from \Eqs{eq:rp_rq_coordinates_++} and~\eqref{eq:rp'_rq'_coordinates_++}. These imply that the pode and antipode causal patches have a non-trivial overlap in the region exterior to the bubble. See \Fig{fig:Penrose_diag_++}.

The area of the spherical section of the pode horizon remains constant along the segments $AP$ and $QA'$ inside the bubble, given by $\omega_{n-1}R_L^{n-1}$. Outside the bubble on the other hand, the area varies with time according to:
\begin{equation} 
{\rm Area}(\theta_R,\sigma_R)=\omega_{n-1}\times
 \left\{\!\begin{array}{ll}
\dis \left(\frac{R_R\sin(-\sigma_R+\sigma_{RQ}+\theta_{RQ})}{\cos\sigma_R}\right)^{n-1}\, , &\mbox{if } \dis 0 \le \sigma_R \leq \sigma_{RQ}\, , \\
\dis \left(\frac{R_R\sin(\sigma_R-\sigma_{RP}+\theta_{RP})}{\cos\sigma_R}\right)^{n-1}\, , &\mbox{if } \dis \sigma_{RP}\leq\sigma_R \le 0\, .\esp
\end{array}\right.
\label{areapode}
\end{equation}
The area is decreasing for $\sigma_{RP}\leq \sigma_R \leq 0$ along $PO$ and increasing for $0\leq \sigma_{R}\leq \sigma_{RQ}$ along $OQ$\footnote{Note that the argument of the sin in these expressions is between $[0,\pi]$.}. 

Using the expression of the polar angle $\theta_{RO}$ of the bifurcation point and $\sigma_{RO}=0$, we can obtain the area of the horizon sphere at $O$:
\begin{equation}
{\rm Area}(O)=\omega_{n-1}R_R^{n-1}\left(\frac{R_L R_R+\sqrt{R_R^2-R_B^2}\sqrt{R_L^2-R_B^2}}{R_R^2+R_L^2-R_B^2}\right)^{n-1}.
\end{equation}
For $R_B/R_R \ll 1$ (and thus $R_B/R_L \ll 1)$, the area of the bubble wall at $\sigma_L=0$, which is given by $\omega_{n-1}R_B^{n-1}$, becomes smaller than the area of the bifurcation sphere at $O$. For larger values of $R_B$, close to $R_R$, the area of the bubble wall at $\sigma_L=0$ becomes larger than the area of $O$. Therefore, the class of geometries defined by $(\varepsilon_L,\varepsilon_R)=(+1,+1)$ includes  \emph{subhorizon} (for $R_B/R_R \ll 1$) and \emph{superhorizon} (for $R_B\sim R_R$) bubbles from the point of view of the pode observer at the center of the bubble. 

\noindent $\bullet$ {\bf Bousso wedges}

The holographic proposal of \Sect{sect:holography} relies on the Bousso covariant entropy conjecture\cite{Bousso:1999xy,Bousso:2002ju} and the Bousso entropy bound, which require us to determine the codimension-$1$ lightlike surfaces of non-positive expansion emanating from a given codimension-$2$ spacelike surface $A$. As we already mentioned, SO$(n)$-symmetric surfaces are represented by points in the Penrose diagram. The four codimension-$1$ null hypersurfaces emanating from $A$ are represented by $45$ degree lines. Among the four null lines, the \emph{Bousso wedges} indicate the two directions along which the area decreases. The computation is easily carried out in light-cone coordinates:
\begin{equation}
    x^\pm=\sigma\pm\theta.
\end{equation}
The derivatives of the area \eqref{eq:area} along $x^\pm$ are given by
\begin{equation}\label{eq:derivative_of_area}
\partial_\pm{\rm Area}(\theta,\sigma)=\pm\omega_{n-1}R^{n-1}\frac{n-1}{2}\cos(x^\mp)\frac{(\sin\theta)^{n-2}}{(\cos\sigma)^n},
\end{equation}
which hold for both the $(\theta_L,\sigma_L)$ and $(\theta_R,\sigma_R)$ coordinate systems. Since $\theta_{L,R}\in[0,\pi]$ and $\sigma_{L,R}\in(-\pi/2,\pi/2)$, the sign of $\partial_\pm{\rm Area}(\theta,\sigma)$ is determined by the sign of $\pm\cos(x^\mp)$. In each de Sitter region, the diagonal lines of the Penrose diagram are given by
\begin{equation}
   \theta=\sigma+\frac{\pi}{2} ~\Leftrightarrow~ x^-=-\frac{\pi}{2},\qquad \theta=-\sigma+\frac{\pi}{2} ~\Leftrightarrow~ x^+=\frac{\pi}{2}.
\end{equation}
They split the spacetime into four regions. In each region, the signs of $\pm\cos(x^\mp)$ determine the orientation of the Bousso wedges, as shown in \Figs{fig:Penrose_diag_++} and \ref{fig:Penrose_diag_+-}. 

\noindent $\bullet$ {\bf Apparent horizons}

The regions of trapped and anti-trapped surfaces are delimited by apparent horizons. In the case of eternal de Sitter space, the apparent horizons coincide with the cosmological horizons. When de Sitter space is deformed, this property no longer holds. It follows from the orientation of the Bousso wedges that the region of trapped surfaces is delimited by the union of segments $AP\cup PP' \cup P'O'$ and $BO'$. The region of anti-trapped surfaces is delimited by the union of segments $O'Q' \cup Q'Q \cup QA$ and $O'B'$. 

The apparent horizon of the pode consists of the union of $AP\cup PP' \cup P'O'$ and $O'Q' \cup Q'Q \cup QA$, lying in the causal patch of the pode. Outside the bubble, it does not coincide with the pode cosmological horizon. Their parts inside the bubble coincide. The pode apparent horizon is timelike along the segments $PP'$ and $Q'Q$, which lie on the domain wall trajectory. 

The apparent horizon of the antipode is the union of the line segments $BO'$ and $O'B'$ and coincides with the cosmological horizon of the antipode.

\subsubsection{The $(\varepsilon_L,\varepsilon_R)=(+1,-1)$ bubbles}

The preceding analysis can be extended to describe geometrical features and, in particular, the causal structure of the $(\varepsilon_L,\varepsilon_R)=(+1,-1)$ bubble geometries. Since for these cases $\theta_R > \pi/2$ along the right domain wall trajectory, the causal patch of the antipode observer also contains a segment of the domain wall trajectory and a part of the left de Sitter bubble. Therefore, in this case, both antipodal observers are aware of a ``parent'' and ``daughter'' de Sitter region separated by a domain wall. The Penrose diagrams are depicted in \Fig{fig:Penrose_diag_+-}.
\begin{figure}[h!]
 \begin{subfigure}[t]{0.48\linewidth}
    \centering
    \raisebox{1.6cm}{
\begin{tikzpicture}
\begin{scope}[transparency group]
\begin{scope}[blend mode=multiply]

%wall
\draw[name path=A, line width= 1 pt,brown] plot[variable=\t,samples=91,domain=-45:45] ({3*sec(\t)-4.23},{3*tan(\t)});
\draw[name path=B, line width= 1 pt,brown] plot[variable=\t,samples=91,domain=-45:45] ({-2*sec(\t)+2.83},{3*tan(\t)});
\tikzfillbetween[of=A and B]{gray, opacity=0.3}; 

\draw[name path=C](0,-3) -- (-3,-3) -- (-3,3) -- (0,3);
\draw[name path=D](0,-3) -- (3,-3) -- (3,3) -- (0,3);

\tikzfillbetween[of=A and C]{blue, opacity=0.1};
\tikzfillbetween[of=B and D]{red, opacity=0.1};

\draw[dashed] (-3,3)--(-1.05,1.05);
\draw[dashed] (-3,-3)--(-1.05,-1.05);
\draw[dashed] (0.55,1.7)--(2.25,0);
\draw[dashed] (0.55,-1.7)--(2.25,0);
\draw[dashed] (3,3)--(0.8,0.8);
\draw[dashed] (3,-3)--(0.8,-0.8);

\draw[dashed] (-1.8,0)--(-1.1,0.7);
\draw[dashed] (-1.8,0)--(-1.1,-0.7);

%Bousso wedges
\draw (1.5,2.2) -- (1.3,2.4) -- (1.1,2.2);
%\draw (-0.2,0.5) -- (-0.0,0.7) -- (0.2,0.5);
\draw (-2.7,0.2) -- (-2.5,0) -- (-2.7,-0.2);
\draw (-1.5,0.2) -- (-1.3,0) -- (-1.5,-0.2);
\draw (-1.5,2.2) -- (-1.3,2.4) -- (-1.1,2.2);
\draw (-1.5,-2.2) -- (-1.3,-2.4) -- (-1.1,-2.2);
\draw (2.7,0.2) -- (2.5,0) -- (2.7,-0.2);
\draw (1.35,0.2) -- (1.15,0) -- (1.35,-0.2);
\draw (1.5,-2.2) -- (1.3,-2.4) -- (1.1,-2.2);
\draw (0.75,-1.1) -- (0.85,-1.2) -- (0.95,-1.1);
\draw (0.75,1.1) -- (0.85,1.2) -- (0.95,1.1);
%\draw (-0.2,-0.5) -- (0,-0.7) -- (0.2,-0.5);

\node at (-3,-3) [label=left:$A$]{};
\node at (-3,3) [label=left:$A'$]{};
\node at (3,-3) [label=right:$B$]{};
\node at (3,3) [label=right:$B'$]{};

\node at (-1.05,-1.05) [circle,fill,inner sep=1.5pt,label=right:$P$]{};
\node at (-1.05,1.05) [circle,fill,inner sep=1.5pt,label=right:$Q$]{};

\node at (0.75,-0.76) [circle,fill,inner sep=1.5pt,label=left:$Q'$]{};
\node at (0.75,0.76) [circle,fill,inner sep=1.5pt,label=left:$P'$]{};

\node at (-1.15,-0.7) [circle,fill,inner sep=1.5pt]{};
\node at (-1.25,-0.5) [label=right:$Q'$]{};
\node at (-1.15,0.7) [circle,fill,inner sep=1.5pt]{};
\node at (-1.25,0.5) [label=right:$P'$]{};

\node at (0.52,-1.73) [circle,fill,inner sep=1.5pt,label=left:$P$]{};
\node at (0.52,1.73) [circle,fill,inner sep=1.5pt,label=left:$Q$]{};

\node at (2.25,0) [circle,fill,inner sep=1.5pt, label = above:$O$]{};
\node at (-1.8,0) [circle,fill,inner sep=1.5pt, label = above:$O'$]{};

\node at (1.13,-1.13) [circle,fill,inner sep=1.5pt, label = below:$T$]{};

%axes
\node at (-3, 0) [label={[rotate=90]$\theta_L = 0$}] {};
\node at (3.6, 0) [label={[rotate=90]$\theta_R = \pi$}] {};
\node at (0,-3.75) [label={$\sigma_L=\sigma_R=-\pi/2$}] {};
\node at (0,2.75) [label={$\sigma_L=\sigma_R=\pi/2$}] {};
\node at (2, 1.8) [label={[rotate=45] \scriptsize  $\theta_R-\sigma_R = \pi/2$}] {};
\node at (-2,1.8) [label={[rotate=-45]\scriptsize $\theta_L+\sigma_L = \pi/2$}] {};
\node at (-1.6, -2.4) [label={[rotate=45]\scriptsize $ \theta_L-\sigma_L = \pi/2$}] {};
\node at (1.9, -2.1) [label={[rotate=-45]\scriptsize $ \theta_R+\sigma_R = \pi/2$}] {};

\end{scope}
\end{scope}
\end{tikzpicture}}
    \caption{\footnotesize Two different conformal coordinate systems $(\theta_L,\sigma_L)$ and $(\theta_R,\sigma_R)$, covering, respectively, the blue and red de Sitter regions (with different cosmological constants). Each region is bounded by a domain wall trajectory (brown lines). The gray shaded region is not a part of the spacetime. Points on the left and right walls are identified appropriately.}
    \label{fig:Penrose_diag_+-_disconnected}
    \end{subfigure}
\quad \
\begin{subfigure}[t]{0.48\linewidth}
 \centering
\begin{tikzpicture}
\begin{scope}[transparency group]
\begin{scope}[blend mode=multiply]
\path
       +(-3,5.25) coordinate (IItophat)
       +(0,3) coordinate (IItophatright)
       +(3,3)  coordinate (IIsquaretopright)
       +(3,-3) coordinate (IIsquarebotright)
       +(0,-3) coordinate (IIbothatright)
       +(-3,-5.25) coordinate (IIbothat)
       +(-3,3) coordinate (IIsquaretopleft)
       +(-3,-3) coordinate(IIsquarebotleft)
      
       ;
%\draw (IItophat) to
%          [bend left=20] (IItophatright) --
%          (IIsquaretopright) --       (IIsquarebotright) --
%          (IIbothatright) to 
%          [bend left=20] (IIbothat)--
%          (IItophat)--cycle;

%\draw[name path=A,line width= 1 pt,brown] plot[variable=\t,samples=90,domain=-44.8:44.8] ({-2*sec(\t)+2.83},{3*tan(\t)});

\draw[name path=A, line width= 1 pt,brown] plot[variable=\t,samples=91,domain=-45:45] ({-2*sec(\t)+2.83},{3*tan(\t)});

\draw[name path=C](IIbothatright) to [bend left=20] (IIbothat) -- (IItophat) to [bend left=20] (IItophatright);
\draw[name path=D](0,-3) -- (3,-3) -- (3,3) -- (0,3);

\tikzfillbetween[of=A and C]{blue, opacity=0.1};
\tikzfillbetween[of=A and D]{red, opacity=0.1};

%\draw[dotted] (IItophatright) -- (3,0);
%\draw[dotted] (IIbothatright) -- (3,0);
%\draw[dotted] (IItophatright) -- (-3,0);
%\draw[dotted] (IIbothatright) -- (-3,0);

\draw[dashed] (IIsquaretopright) -- (0,0) -- (IIsquarebotright);
\draw[dashed] (IItophat) -- (2.25,0) -- (IIbothat);

%Bousso wedges
\draw (1.05,2.3) -- (1.25,2.5) -- (1.45,2.3);
%\draw (-0.2,0.8) -- (0,1) -- (0.2,0.8);
\draw (-2.2,0.2) -- (-2,0) -- (-2.2,-0.2);
\draw (0.4,0.2) -- (0.6,0) -- (0.4,-0.2);
\draw (-1.2,3.6) -- (-1,3.8) -- (-0.8,3.6);
\draw (-1.2,-3.6) -- (-1,-3.8) -- (-0.8,-3.6);
\draw (2.7,0.2) -- (2.5,0) -- (2.7,-0.2);
\draw (1.35,0.2) -- (1.15,0) -- (1.35,-0.2);
\draw (1.05,-2.3) -- (1.25,-2.5) -- (1.45,-2.3);
\draw (0.75,-1.1) -- (0.85,-1.2) -- (0.95,-1.1);
\draw (0.75,1.1) -- (0.85,1.2) -- (0.95,1.1);
%\draw (-0.2,-0.8) -- (0,-1) -- (0.2,-0.8);

\node at (1.13,-1.13) [circle,fill,inner sep=1.5pt, label = below:$T$]{};

\node at (-3,-5.25) [label=left:$A$]{};
\node at (-3,5.25) [label=left:$A'$]{};
\node at (3,-3) [label=right:$B$]{};
\node at (3,3) [label=right:$B'$]{};

\node at (0.52,-1.73) [circle,fill,inner sep=1.5pt,label=left:$P$]{};
\node at (0.52,1.73) [circle,fill,inner sep=1.5pt,label=left:$Q$]{};

\node at (0.76,-0.76) [circle,fill,inner sep=1.5pt,label=left:$Q'$]{};
\node at (0.75,0.76) [circle,fill,inner sep=1.5pt,label=left:$P'$]{};

\node at (2.25,0) [circle,fill,inner sep=1.5pt, label = above:$O$]{};
\node at (0,0) [circle,fill,inner sep=1.5pt]{};
\node at (0.1,0) [label = left:$O'$]{};

\end{scope}
\end{scope}
\end{tikzpicture}
    \caption{\footnotesize A single global conformal coordinate system $(\theta_G,\sigma_G)$ covering the whole spacetime. The brown curve depicts the domain wall trajectory.}
    \label{fig:Penrose_diag_+-_connected}
\end{subfigure}
\caption{\footnotesize 
The Penrose diagrams for the $(\varepsilon_L,\varepsilon_R)=(+1,-1)$ bubble geometry. The blue (red) region is the interior (exterior) of the bubble, which is a part of dS$_{n+1}$ with a smaller (larger) cosmological constant. Appropriate boundary conditions enforce the identification of the points on the left and right domain wall trajectories bounding the two de Sitter regions. The  causal patches of the pode and antipode observers are bounded by future and past cosmological horizons, depicted by the thick dashed lines. Bousso wedges are drawn in each region of the diagrams.
}
\label{fig:Penrose_diag_+-}
\end{figure}

\noindent $\bullet$ {\bf Cosmological horizons}

In the right de Sitter region, the future and past cosmological horizons delimitating the causal patch of the antipode observer are represented, respectively, by the null line segments $P'B'$ and $BQ'$. $P'$ and $Q'$ are the end points of these segments on the domain wall trajectory. The equations describing $P'B'$ and $BQ'$ are given by
\begin{equation}
 \left\{\!\begin{array}{ll}
\dis \sigma_R-\theta_R=-\frac{\pi}{2}\, , &\mbox{if } \dis \sigma_{RP'}\le \sigma_R <\frac{\pi}{2}\, , \\
\dis \sigma_R+\theta_R=\frac{\pi}{2}\, , &\mbox{if } \dis -\frac{\pi}{2}<\sigma_R \le \sigma_{RQ'}\, ,\esp
\end{array}\right.
\label{eq:right_horizons_right}
\end{equation}
where $\sigma_{RP'}$ and $\sigma_{RQ'}$ are the right conformal time coordinates of $P'$ and $Q'$. The full set of right conformal coordinates of these points are given by
\begin{equation}\label{eq:rp'_rq'_coordinates_+-}
\theta_{RP'}=\theta_{RQ'}=\pi-\arctan\frac{R_R}{\sqrt{R_R^2-R_B^2}}, \quad \sigma_{RP'}=-\sigma_{RQ'}=\arctan\frac{\sqrt{R_R^2-R_B^2}}{R_R}.
\end{equation}
As expected, $\theta_{RP'},\,\theta_{RQ'}$ are greater than $\pi/2$. Using the matching condition Eq.~\eqref{eq:matching_cond}, we can obtain the corresponding left conformal coordinates
\begin{equation}
\theta_{LP'}=\theta_{LQ'}=\arctan\frac{R_R}{\sqrt{R_L^2-R_B^2}}, \quad \sigma_{LP'}=-\sigma_{LQ'}=\arctan\frac{\sqrt{R_R^2-R_B^2}}{R_L}.
\label{eq:lp'_lq'_coordinates_+-}
\end{equation}

Inside the left de Sitter region, the future and past cosmological horizons of the antipodal observer are represented by the null line segments $O'P'$ and $Q'O'$, where $O'$ is the bifurcation point. These are described, respectively, by the equations:
\begin{equation}\label{right_horizon_left}
 \left\{\!\begin{array}{ll}
\dis \sigma_L-\theta_L=\sigma_{LP'}-\theta_{LP'}\, , &\mbox{if } \dis \sigma_{LO'} \le \sigma_L \leq \sigma_{LP'}\, , \\
\dis \sigma_L+\theta_L=\sigma_{LQ'}+\theta_{LQ'}\, , &\mbox{if } \dis \sigma_{LQ'}\leq\sigma_L \le \sigma_{LO'}\, ,\esp
\end{array}\right.
\end{equation}
where $\sigma_{LP'}$ and $\sigma_{LQ'}$ are given in \Eq{eq:lp'_lq'_coordinates_+-}, and the coordinates of the bifurcation point $O'$ are given by 
\begin{equation}\label{O'}
\theta_{LO'}=\arctan\frac{R_L R_R-\sqrt{R_R^2-R_B^2}\sqrt{R_L^2-R_B^2}}{R_L\sqrt{R_L^2-R_B^2}+R_R\sqrt{R_R^2-R_B^2}}, \,\,\,\,\, \sigma_{LO'}=0. 
\end{equation}
Using these expressions, we can obtain the area of the horizon sphere at $O'$:
\begin{equation}
{\rm Area}(O')=\omega_{n-1}R_L^{n-1}\left(\frac{R_L R_R-\sqrt{R_R^2-R_B^2}\sqrt{R_L^2-R_B^2}}{R_R^2+R_L^2-R_B^2}\right)^{n-1}.
\end{equation}

The causal patch of the pode observer is delimited by the null line segments $QA'$ and $AP$ within the bubble, with equations given by \eqref{eq:left_horizon_left}. The left conformal coordinates of the end points $Q$ and $P$ on the domain wall trajectory are given by Eq.~\eqref{eq:lp_lq_coordinates_++}. The corresponding right conformal coordinates are 
\begin{equation}
\theta_{RQ}=\theta_{RP}=\pi-\arctan\frac{R_L}{\sqrt{R_R^2-R_B^2}}, \quad \sigma_{RQ}=-\sigma_{RP}=\arctan\frac{\sqrt{R_L^2-R_B^2}}{R_R}.
\label{eq:rp_rq_coordinates_+-}
\end{equation}
Notice that $\theta_{RQ},\, \theta_{RP}>\pi/2$ for this case. In the right de Sitter region, the future and past pode cosmological horizons are represented by the null line segments $OQ$ and $PO$, with equations given by \eqref{eq:left_horizon_right}. The coordinates of the bifurcation point $O$ are given by 
\begin{equation}
\theta_{RO}=\pi-\arctan\frac{R_L R_R-\sqrt{R_R^2-R_B^2}\sqrt{R_L^2-R_B^2}}{R_L\sqrt{R_L^2-R_B^2}+R_R\sqrt{R_R^2-R_B^2}}, \,\,\,\,\, \sigma_{RO}=0, 
\end{equation}
and the area of the corresponding sphere by the expression
\begin{equation}
{\rm Area}(O)=\omega_{n-1}R_R^{n-1}\left(\frac{R_L R_R-\sqrt{R_R^2-R_B^2}\sqrt{R_L^2-R_B^2}}{R_R^2+R_L^2-R_B^2}\right)^{n-1}.
\end{equation}

The pode and antipode causal patches overlap in both de Sitter regions. The area of the bubble wall at $\sigma_L=\sigma_R=0$, given by $\omega_{n-1}R_B^{n-1}$, is bigger than the area of both bifurcate horizons at $O$ and $O'$. Therefore, this is an example of a superhorizon bubble. As we mentioned before, the tension of these bubbles is supercritical. Therefore, we must keep the value of $R_B$ close to its maximum value ($R_R$) in order for the tension to remain close to the critical value and ensure the validity of the semiclassical approximation. 

\noindent $\bullet$ {\bf Bousso wedges and apparent horizons}

In the Penrose diagrams of \Fig{fig:Penrose_diag_+-}, we draw the Bousso wedges in all causal regions. The region of trapped surfaces is bounded by the null line segment $AP$, the segment $PQ'$ of the domain wall trajectory and the null line segment $BQ'$. The region of anti-trapped surfaces is bounded by the segment $P'Q$ along the domain wall trajectory, and the null segments $QA'$ and $P'B'$. Based on these, the apparent horizon of the pode is given by the union $AP\cup PQ \cup QA'$. The apparent horizon of the antipode is given by $BQ'\cup Q'P' \cup P'B'$. Along the domain wall trajectory, the cross sectional area of the apparent horizons changes.

\section{Holographic proposal for the $(\varepsilon_L,\varepsilon_R)=(+1,\pm 1)$ bubbles}
\label{sect:holography}
In this section we generalize the static patch holographic proposal for de Sitter space \cite{Susskind:2021omt}, as well as more recent proposals \cite{Franken:2023pni,Franken:2023jas} concerning closed FRW cosmologies, to the time-symmetric de Sitter bubble geometries with $(\varepsilon_L,\varepsilon_R)=(+1,+1)$ and $(\varepsilon_L,\varepsilon_R)=(+1,-1)$. What is special for these cosmologies is that the union of the pode and antipode causal patches contains complete Cauchy slices. The proposal entails the use of two holographic screen systems following timelike or lightlike trajectories in the bulk\footnote{The screen trajectories must be inside the bulk, because the spacetime has no boundary.}. The left screen, $\S_\Le$, lies in the causal patch of the pode observer at $\theta_L=0$, while the right screen, $\S_\Ri$, lies in the causal patch of the antipode observer at $\theta_R=\pi$. 

\subsection{The screen trajectories}
\label{sect:screen_trajectories}

We consider an arbitrary foliation $\F$ of spacetime in terms of SO$(n)$-symmetric Cauchy slices, which we denote by $\Sigma$. These slices have a spherical topology and, moreover, they have the pode and antipode as poles. Throughout this paper, we assume that the state on $\Sigma$ is pure and thus of vanishing fine-grained entropy. Our considerations can be generalized to the cases for which the state on $\Sigma$ is thermal or mixed. In the latter cases, the calculation of the next to leading corrections to the entropy ($\mathcal{O}(G\hbar)^0$) is more difficult. We will focus on the $(\varepsilon_L,\varepsilon_R)=(+1,+1)$ case to motivate our proposal and summarize the results concerning the screen trajectories in the $(\varepsilon_L,\varepsilon_R)=(+1,-1)$ bubble cases in the end.

\noindent {\it 1) The $(\varepsilon_L,\varepsilon_R)=(+1,+1)$ case}

Pick an arbitrary Cauchy slice $\Sigma$ and consider an SO$(n)$-symmetric spherical cap $\Sigma_1\subset \Sigma$, which lies entirely in the causal patch of the pode. The boundary of $\Sigma_1$ is a sphere of dimension $n-1$ (corresponding to a point in the Penrose diagram), denoted by $\S_1$. The pole of this spherical cap $\Sigma_1$ is the pode. See \Fig{fig:entropy_bound}.
\begin{figure}[h!]
    \centering
\begin{tikzpicture}
\begin{scope}[transparency group]
\begin{scope}[blend mode=multiply]
\path
       +(-3,4.5) coordinate (IItophat)
       +(0,3) coordinate (IItophatright)
       +(3,3)  coordinate (IIsquaretopright)
       +(3,-3) coordinate (IIsquarebotright)
       +(0,-3) coordinate (IIbothatright)
       +(-3,-4.5) coordinate (IIbothat)
       +(-3,3) coordinate (IIsquaretopleft)
       +(-3,-3) coordinate(IIsquarebotleft)
      
       ;
\draw (IItophat) to
          [bend left=20] (IItophatright) --
          (IIsquaretopright) --       (IIsquarebotright) --
          (IIbothatright) to 
          [bend left=20] (IIbothat)--
          (IItophat)--cycle;

%\draw[dotted] (IItophatright) -- (3,0);
%\draw[dotted] (IIbothatright) -- (3,0);
%\draw[dotted] (IItophatright) -- (-3,0);
%\draw[dotted] (IIbothatright) -- (-3,0);

\draw[dashed] (IIsquaretopright) -- (0,0) -- (IIsquarebotright);
\draw[dashed] (IItophat) -- (1.5,0) -- (IIbothat);
\draw[dotted] (-1,-1) -- (0,0) -- (-1,1);

\draw[line width= 1 pt,brown] plot[variable=\t,samples=91,domain=-45:45] ({3*sec(\t)-4.2},{3*tan(\t)});

\draw (0.55,1.8) -- (0.75,2) -- (0.95,1.8);
\draw (-0.2,0.8) -- (0,1) -- (0.2,0.8);
\draw (-2.2,0.2) -- (-2,0) -- (-2.2,-0.2);
\draw (-0.8,0.2) -- (-0.6,0) -- (-0.8,-0.2);
\draw (-1.7,3.6) -- (-1.5,3.8) -- (-1.3,3.6);
\draw (-1.7,-3.6) -- (-1.5,-3.8) -- (-1.3,-3.6);
\draw (2.2,0.2) -- (2,0) -- (2.2,-0.2);
\draw (0.85,0.2) -- (0.65,0) -- (0.85,-0.2);
\draw (0.55,-2.1) -- (0.75,-2.3) -- (0.95,-2.1);
\draw (-0.2,-0.8) -- (0,-1) -- (0.2,-0.8);

%draw cauchy slice and label on the slice
\draw[line width=0.8mm,red,smooth] plot[variable=\t,samples=100,domain=3:4.6] ({\t-6},{0.1*\t*cos(40*\t)-1.5});
\draw[red, smooth] plot[variable=\t,samples=100,domain=4.6:9] ({\t-6},{0.1*\t*cos(40*\t)-1.5});
\node at (-1.4,-1.95) [circle,fill,inner sep=1.5pt, label=below:$\S_1$]{};
\node at (-2.3,-2.7) [label = $\color{red} \Sigma_1$]{};
\node at (2.2,-1.65) [label = $\color{red} \Sigma$]{};

%future lightsheet
\draw[blue] (-1.4,-1.95)--(-3,-0.35);
\node at (-1.9,-1.4) [label = $\color{blue} \Omega$]{};

%x+, x- axes
\draw[->] (-3,0)--(-2.5,0.5);
\draw[->] (-3,0)--(-3.5,0.5);
\node at (-2.6,0.5) [label = right:$x^+$]{};
\node at (-3.4,0.5) [label = left:$x^-$]{};

\end{scope}
\end{scope}
\end{tikzpicture}
    \caption{\footnotesize An SO$(n)$-symmetric Cauchy slice $\Sigma$. A screen $\S_1$ is located in the part of $\Sigma$ lying in the causal patch of the pode. $\Sigma_1$ is the part of $\Sigma$ to the left of $\S_1$. As shown by the Bousso wedges, $\S_1$ has a future-directed lightsheet $\Omega$ depicted by the blue segment.}
    \label{fig:entropy_bound}
\end{figure}
Since $\Sigma_1$ lies in the causal patch of the pode, the information on it can be accessed by the cosmological observer at the pode. Now suppose that $\S_1$ coincides with the tip of a Bousso wedge containing a {\it future directed} lightsheet of non-positive expansion, denoted by $\Omega$, which is parallel to the positive $x^{-}$ axis and terminates on the worldline of the pode. $\Omega$ is depicted by the blue line segment in \Fig{fig:entropy_bound}. Then, according to the holographic conjecture \cite{tHooft:1993dmi,Susskind:1994vu} and the Bousso covariant entropy bound \cite{Bousso:1999xy,Bousso:2002ju}, the gravitational system associated with $\Sigma_1$ can be described holographically in terms of a dual quantum mechanical system ``living'' on (an $(n-1)$-dimensional manifold isomorphic to) the boundary sphere $\S_1$. We will refer to the spatial manifold where the dual holographic system resides as the ``holographic screen''. In the case at hand, the screen is located at (or identified with) the boundary sphere $\S_1$. The number of holographic degrees of freedom on the spherical screen is given by the area of $\S_1$ divided by $4G\hbar$ \cite{tHooft:1993dmi,Susskind:1994vu}.  

Indeed, if $\S_1$ has such a lightsheet $\Omega$, the Bousso covariant entropy bound \cite{Bousso:1999xy,Bousso:2002ju} ensures that the holographic screen at $\S_1$ has enough degrees of freedom to encode the state on $\Sigma_1$. This is because the area of $\S_1$ divided by $4G\hbar$ (which is a measure of the number of holographic degrees of freedom) bounds from above the thermodynamic, coarse-grained entropy passing through the lightsheet $\Omega$ (since the latter is a lightsheet of non-positive expansion):
\begin{equation}\label{eq:BCEB}
S_{\rm coarse\,grained}(\Omega)\leq \frac{{\rm Area}(\S_1)}{4G\hbar}.
\end{equation}
But since $\Omega$ is a future directed lightsheet that emanates from $\S_1$ and bounds the future causal diamond of $\Sigma_1$, the thermodynamic, coarse-grained entropy that passes through it bounds from above the coarse-grained entropy of $\Sigma_1$:
\begin{equation}\label{eq:BCEB2}
S_{\rm coarse\,grained}(\Sigma_1)\leq S_{\rm coarse\,grained}(\Omega).
\end{equation}
The latter inequality follows from the second law of thermodynamics. It follows from Eq.~(\ref{eq:BCEB}) and Eq.~(\ref{eq:BCEB2}) that the area of $\S_1$ divided by $4G\hbar$ bounds from above the coarse-grained entropy of $\Sigma_1$:
\begin{equation}
    S_{\rm coarse\,grained}(\Sigma_1)\leq \frac{{\rm Area}(\S_1)}{4G\hbar}.
\end{equation}
Now the maximal value of the coarse-grained entropy of $\Sigma_1$ is a measure of the number of degrees of freedom needed to describe this gravitational system, including the degrees of freedom associated with the observer at the pode, and, therefore, the holographic system on the screen at $\S_1$ has adequate (or even more) number of degrees of freedom to effectively describe it. 

Based on the considerations above, we would like to determine the maximal possible spherical cap $\Sigma_1\subset \Sigma$ that can be described via an effective holographic theory on its boundary. Presumably then, the holographic description pertaining to smaller spherical caps inside $\Sigma_1$ can be obtained by integrating out a number of degrees of freedom from this ``parent'' theory. Recapitulating, two criteria must be met:   
\begin{itemize}
\item $\Sigma_1$ (and the boundary sphere $\S_1$) must lie in the causal patch of the pode, in order for the information associated with it to be accessible by the pode observer. 
\item $\S_1$ must be the starting point of a future directed lightsheet of non-positive expansion that terminates on the worldline of the pode, in order for the coarse-grained entropy on $\Sigma_1$ to be bounded from above by the area of $\S_1$ divided by $4G\hbar$.
\end{itemize}

\begin{figure}[h!]
%\begin{subfigure}[t]{0.48\linewidth}
    \centering
\begin{tikzpicture}
\begin{scope}[transparency group]
\begin{scope}[blend mode=multiply]
\path
       +(-3,4.5) coordinate (IItophat)
       +(0,3) coordinate (IItophatright)
       +(3,3)  coordinate (IIsquaretopright)
       +(3,-3) coordinate (IIsquarebotright)
       +(0,-3) coordinate (IIbothatright)
       +(-3,-4.5) coordinate (IIbothat)
       +(-3,3) coordinate (IIsquaretopleft)
       +(-3,-3) coordinate(IIsquarebotleft)
      
       ;
\draw (IItophat) to
          [bend left=20] (IItophatright) --
          (IIsquaretopright) --       (IIsquarebotright) --
          (IIbothatright) to 
          [bend left=20] (IIbothat)--
          (IItophat)--cycle;

%\draw[dotted] (IItophatright) -- (3,0);
%\draw[dotted] (IIbothatright) -- (3,0);
%\draw[dotted] (IItophatright) -- (-3,0);
%\draw[dotted] (IIbothatright) -- (-3,0);

%\draw[dashed] (IIsquaretopright) -- (0,0) -- (IIsquarebotright);
\draw[dashed] (-0.55,2.05) -- (1.5,0) -- (0.75,-0.75);
\draw[dotted] (-1,-1) -- (0,0);

%wall
\draw[line width= 1 pt,brown] plot[variable=\t,samples=90,domain=-45:45] ({3*sec(\t)-4.2},{3*tan(\t)});

%segments of the screen trajectory
\node at (IIbothat) [circle,fill,inner sep=1.5pt, label = left:$A$]{};
\node at (IItophat) [circle,fill,inner sep=1.5pt, label = left:$A'$]{};
\node at (-0.55,-2.05) [circle,fill,inner sep=1.5pt, label = left:$P$]{};
\node at (0.75,-0.75) [circle,fill,inner sep=1.5pt, label = right:$T$]{};
\node at (0,0) [circle,fill,inner sep=1.5pt]{};
\node at (0.1,0) [label = above:$O'$]{};
\node at (-1.03,1.03) [circle,fill,inner sep=1.5pt, label = left:$Q'$]{};
\node at (-0.55,2.05) [circle,fill,inner sep=1.5pt, label = right:$Q$]{};

\draw (0.55,1.8) -- (0.75,2) -- (0.95,1.8);
\draw (-0.2,0.8) -- (0,1) -- (0.2,0.8);
\draw (-2.2,0.2) -- (-2,0) -- (-2.2,-0.2);
\draw (-0.8,0.2) -- (-0.6,0) -- (-0.8,-0.2);
\draw (-1.7,3.6) -- (-1.5,3.8) -- (-1.3,3.6);
\draw (-1.7,-3.6) -- (-1.5,-3.8) -- (-1.3,-3.6);
\draw (2.2,0.2) -- (2,0) -- (2.2,-0.2);
\draw (0.85,0.2) -- (0.65,0) -- (0.85,-0.2);
\draw (0.55,-1.8) -- (0.75,-2) -- (0.95,-1.8);
\draw (-0.2,-0.8) -- (0,-1) -- (0.2,-0.8);

%trajectory of left screen
\draw[violet,line width=0.8mm] (IIbothat) -- (0.75,-0.75) --  (-1.05,1.05);
\draw[violet,line width=0.8mm] plot[variable=\t,samples=18,domain=18.7:34.5] ({3*sec(\t)-4.2},{3*tan(\t)});
\draw[violet,line width=0.8mm] (-0.55,2.05) -- (IItophat); 

\fill[fill=violet!20] (IIbothat) -- (0.75,-0.75) -- (-1.05,1.05)  to [bend left=8] (-0.55,2.05) -- (IItophat) -- cycle;

%\draw[name path=A,violet,line width=0.8mm] (-3,-4.5)--(0.75,-0.75)--(-1.05,1.05);
%\draw[name path=B, violet,line width=0.8mm] plot[variable=\t,samples=18,domain=18.7:34.5] ({3*sec(\t)-4.2},{3*tan(\t)});
%\draw[name path=C, violet,line width=0.8mm] (-3,4.5)--(-0.55,2.05);
%\draw[name path =D] (-3,4.5)--(-3,-4.5);
%\tikzfillbetween[of=A and D]{violet, opacity=0.1};
%\tikzfillbetween[of=B and C]{violet, opacity=0.1};
%\tikzfillbetween[of=C and D]{violet, opacity=0.1};

%draw right horizon
\draw[orange,line width=0.8mm] (IIsquaretopright) -- (0,0) -- (IIsquarebotright);
\fill[fill=orange!20] (IIsquaretopright) -- (0,0) -- (IIsquarebotright) -- cycle;

%\draw[name path=E, orange, line width=0.8mm](3,3)--(0,0)--(3,-3);
%\draw[name path=F] (3,3)--(3,-3);
%\tikzfillbetween[of=E and F]{orange, opacity=0.1};

%draw cauchy slice and label on the slice
\draw[line width=0.8mm,red,smooth] plot[variable=\t,samples=100,domain=0:3] ({\t-3},{0.1*\t*cos(45*\t)-1.2});
\draw[line width=0.8mm,red, smooth] plot[variable=\t,samples=100,domain=3:4.52] ({\t-3},{0.1*\t*cos(45*\t)-1.2});
\draw[line width=0.8mm,red, smooth] plot[variable=\t,samples=100,domain=4.52:6] ({\t-3},{0.1*\t*cos(45*\t)-1.2});
\node at (0.06,-1.44) [circle,fill,inner sep=2pt, label=below:$\S_\Le$]{};
\node at (1.62,-1.62) [circle,fill,inner sep=2pt]{};
\node at (1.6,-1.6) [label=below:$\S_\Ri$]{};
\node at (-1.5,-1.3) [label = $\color{red} \Sigma_\Le$]{};
\node at (2.3,-1.6) [label = $\color{red} \Sigma_\Ri$]{};
\node at (0.8,-1.8) [label = $\color{red} \Sigma_\E$]{};

\end{scope}
\end{scope}
\end{tikzpicture}
    \caption{\footnotesize Trajectories of the left screen on $\S_\Le$ (purple) and the right screen on $\S_\Ri$ (orange) in the $(\varepsilon_L,\varepsilon_R)=(+1,+1)$ bubble geometries. The interior regions of the pode and the antipode are the purple and orange shaded regions, respectively. The exterior region is white.}
    \label{fig:screen_trajectories_++}
\end{figure}
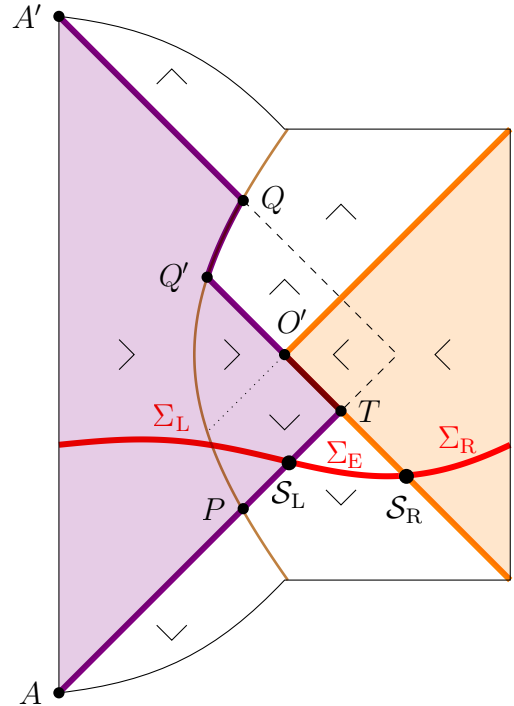

Given the orientation of the Bousso wedges, one can readily verify that the set of codimension-$2$ spheres satisfying the two criteria above corresponds to the purple shaded region of \Fig{fig:screen_trajectories_++},
which lies inside the causal patch of the pode. The boundary of this region is depicted by the thick purple line. It consists of the following four segments: i) The lightlike line segments $AP$ and $PT$ along the past cosmological horizon of the pode, ending at the intersection point $T$ with the past cosmological horizon of the antipode. Notice that $PT$ does not lie along the apparent horizon of the pode. In fact, along $PT$ the area of the spherical sections decreases from $\omega_{n-1}R_L^{n-1}$ to $\omega_{n-1}R_R^{n-1}$, as can be inferred from Eq.~(\ref{areapode}). ii) The lightlike segment $TQ'$ along the past cosmological horizon of the antipode, $\sigma_R=-\theta_R + \pi/2$. Along $TQ'$, the area of the spherical sections remains constant, equal to $\omega_{n-1}R_R^{n-1}$. iii) The timelike segment $Q'Q$ along the trajectory of the domain wall, for which the area of the spheres increases from $\omega_{n-1}R_R^{n-1}$ to $\omega_{n-1}R_L^{n-1}$. iv) The lightlike line segment $QA'$ along the future cosmological horizon of the pode. The area of the spheres remains constant equal to $\omega_{n-1}R_L^{n-1}$.   

Therefore, to maximize the size of $\Sigma_1$, we take $\S_1$ and the holographic screen to lie at the intersection of $\Sigma$ with the thick purple trajectory. Repeating this procedure for every Cauchy slice in the foliation $\F$, we get an evolution of the holographic screen along the thick purple trajectory. The purple region bounded by this trajectory, \Fig{fig:screen_trajectories_++}, will be referred to as the ``pode interior region'' (L). The maximal $\Sigma_1$ and its boundary $\S_1$ will be called, respectively, $\Sigma_\Le$ and $\S_\Le$. See \Fig{fig:screen_trajectories_++}.

A similar analysis can be carried out for a second holographic screen associated with the antipode observer. For these de Sitter bubbles with $(\varepsilon_L,\varepsilon_R)=(+1,+1)$, the construction is the same as in the eternal de Sitter case \cite{Franken:2023pni}: the trajectory of the right antipode screen coincides with the cosmological horizon of the antipode, depicted by the thick orange line in \Fig{fig:screen_trajectories_++}. Indeed, we can locate the right screen at the intersection of the Cauchy slice $\Sigma$ with the cosmological horizon of the antipode, the $(n-1)$-dimensional sphere $\S_\Ri$ in \Fig{fig:screen_trajectories_++}. The ``interior region of the antipode'' (R), bounded by the thick orange trajectory, 
is the full static patch of the antipode observer. It is shaded orange in \Fig{fig:screen_trajectories_++}. The spherical cap in $\Sigma$ bounded by $\S_\Ri$ and having the antipode as its pole is denoted by $\Sigma_\Ri$. The covariant entropy bound \cite{Bousso:1999xy,Bousso:2002ju} ensures that the state of $\Sigma_\Ri$ can be encoded on the holographic screen at the boundary sphere $\S_\Ri$. Indeed, $\Sigma_\Ri$ is the maximal such spherical cap that can be described holographically in terms of a quantum mechanical system on its boundary. 

The barrel-like part of $\Sigma$ bounded by $\S_\Le$ and $\S_\Ri$ is denoted by $\Sigma_\E$. As we follow the foliation $\F$, $\Sigma_\E$ spans the white region of \Fig{fig:screen_trajectories_++} between the purple and orange trajectories. This region will be called the ``exterior region'' (E).

Therefore, the holographic system consists of two subsystems on the screens at $\S_\Le$ and $\S_\Ri$, with enough degrees of freedom to describe the states on $\Sigma_\Le$ and $\Sigma_\Ri$. However, as we will see in the next section, the entanglement wedge of the two screen system is actually bigger; it covers not only $\Sigma_\Le \cup \Sigma_\Ri$ but the whole Cauchy slice $\Sigma$ (as follows from the geometrical, leading order analysis). Assuming entanglement wedge reconstruction, the two-screen system may be capable of encoding the state of $\Sigma$. As the right screen evolves along the orange trajectory, its area in Planck units remains constant. Therefore, the number of holographic degrees of freedom on it does not change. As the left screen evolves along the purple trajectory, the number of holographic degrees of freedom on it can change, since the area in Planck units is not always constant. Initially, the left screen carries more degrees of freedom than the right screen. Once the left screen crosses the domain wall trajectory and exits the bubble along the lightlike segment $PT$, where $T$ is the intersection point of the two past cosmological horizons (\Fig{fig:screen_trajectories_++}), the number of degrees of freedom on it decreases. Along the lightlike segment $TO'$ the location of the two screens on $\Sigma$ coincides, and they can exchange energy and information. Indeed, the two screens lie in the overlap of the pode and antipode causal patches in this case. Then it is interesting that along the segment $TO'$, the numbers of degrees of freedom on the two screens are equal. In fact, the number of degrees of freedom on the left screen remains constant and equal to that on the right screen along the larger lightlike segment $TQ'$. It starts to increase again along the segment $Q'Q$ of the wall trajectory, eventually reaching its initial value at $Q$. The evolution of the left screen system is not unitary and amounts to a sequence of mappings to Hilbert spaces of varying dimensionality. Similar behavior has been advocated in \cite{Cotler:2022weg,Cotler:2023xku} and has been shown to occur in the holographic examples of \cite{Franken:2023pni}, involving closed FRW cosmologies.  

Notice that along $TO'$, where the left and right screens coincide, $\Sigma_\Le \cup \Sigma_\Ri$ becomes equal to the complete Cauchy slice $\Sigma$. The two-screen system has enough degrees of freedom to describe the states on $\Sigma$ in this case. In fact, when the left and right screens are in the contracting phase of the bubble cosmology, it follows from the structure of the Bousso wedges in the region of trapped surfaces that the two-screen system also has adequate number of degrees of freedom to describe the states on $\Sigma$. Indeed, when the left screen is on the lightlike line segment $AT$ (or on $TO'$), its area in Planck units bounds the coarse-grained entropy on the complement of $\Sigma_\Le$,  $\Sigma -\Sigma_\Le$. Likewise, when the right screen is on the lightlike line segment $BT$ (or on $TO'$), its area in Planck units bounds the coarse-grained entropy on the complement of $\Sigma_\Ri$, $\Sigma -\Sigma_\Ri$. So, the number of degrees of freedom of the two-screen system bounds the sum of the coarse-grained entropies of these complements and the coarse-grained entropy of $\Sigma$, since $S_{\rm coarse \,grained}(\Sigma -\Sigma_\Le) + S_{\rm coarse \,grained}(\Sigma -\Sigma_\Ri) \ge S_{\rm coarse \,grained}\big( (\Sigma -\Sigma_\Le) \cup (\Sigma -\Sigma_\Ri)\big)=S_{\rm coarse \,grained}(\Sigma) $. In other words, the Bousso covariant entropy conjecture implies that the contracting phase of the bubble cosmology can be described in terms of a two-screen system. Since the expanding phase results from the contracting phase, we expect the two-screen system to be capable of describing it. Indeed, we will see that the bilayer entropy proposal for holographic entanglement entropy computation implies that the entanglement wedge of the two-screen system comprises complete Cauchy slices, even in the expanding phase, and as the screens evolve, it covers the entire bubble geometry.

In the $(\varepsilon_L,\varepsilon_R)=(-1,-1)$ cases on the other hand, the union $\Sigma_\Le \cup \Sigma_\Ri$ never coincides with a complete Cauchy slice and the pode and antipode causal patches are disconnected, as will be seen in \Sect{sect:--_geometry}. In no case can we argue that the union of the pode and antipode screens has enough degrees of freedom to describe the states on a complete Cauchy slice. This feature prevents the global description of the latter cases in terms of a holographic two-screen system, and the application of the bilayer proposal of the next section for holographic entropy computations.

\noindent {\it 2) The $(\varepsilon_L,\varepsilon_R)=(+1,-1)$ case}

\begin{figure}[h!]
%\begin{subfigure}[t]{0.48\linewidth}
    \centering
\begin{tikzpicture}
\begin{scope}[transparency group]
\begin{scope}[blend mode=multiply]
\path
       +(-3,5.25) coordinate (IItophat)
       +(0,3) coordinate (IItophatright)
       +(3,3)  coordinate (IIsquaretopright)
       +(3,-3) coordinate (IIsquarebotright)
       +(0,-3) coordinate (IIbothatright)
       +(-3,-5.25) coordinate (IIbothat)
       +(-3,3) coordinate (IIsquaretopleft)
       +(-3,-3) coordinate(IIsquarebotleft)
      
       ;
%\draw (IItophat) to
%          [bend left=20] (IItophatright) --
%          (IIsquaretopright) --       (IIsquarebotright) --
%          (IIbothatright) to 
%          [bend left=20] (IIbothat)--
%          (IItophat)--cycle;

%\draw[name path=A,line width= 1 pt,brown] plot[variable=\t,samples=90,domain=-44.8:44.8] ({-2*sec(\t)+2.83},{3*tan(\t)});

\draw[line width= 1 pt,brown] plot[variable=\t,samples=30,domain=30:45] ({-2*sec(\t)+2.83},{3*tan(\t)});
\draw[line width= 1 pt,brown] plot[variable=\t,samples=30,domain=-45:-15] ({-2*sec(\t)+2.83},{3*tan(\t)});

\draw[name path=C](IIbothatright) to [bend left=20] (IIbothat) -- (IItophat) to [bend left=20] (IItophatright);
\draw[name path=D](0,-3) -- (3,-3) -- (3,3) -- (0,3);

%\tikzfillbetween[of=A and C]{blue, opacity=0.1};
%\tikzfillbetween[of=A and D]{red, opacity=0.1};

%\draw[dotted] (IItophatright) -- (3,0);
%\draw[dotted] (IIbothatright) -- (3,0);
%\draw[dotted] (IItophatright) -- (-3,0);
%\draw[dotted] (IIbothatright) -- (-3,0);

\draw[dashed] (0.8,0.8) -- (0,0) -- (0.8,-0.8);
\draw[dashed] (0.52,1.7) -- (2.25,0) -- (1.1,-1.1);

\draw (1.05,2.3) -- (1.25,2.5) -- (1.45,2.3);
%\draw (-0.2,0.8) -- (0,1) -- (0.2,0.8);
\draw (-2.2,0.2) -- (-2,0) -- (-2.2,-0.2);
\draw (0.4,0.2) -- (0.6,0) -- (0.4,-0.2);
\draw (-1.2,3.6) -- (-1,3.8) -- (-0.8,3.6);
\draw (-1.2,-3.6) -- (-1,-3.8) -- (-0.8,-3.6);
\draw (2.7,0.2) -- (2.5,0) -- (2.7,-0.2);
\draw (1.35,0.2) -- (1.15,0) -- (1.35,-0.2);
\draw (1.05,-2.3) -- (1.25,-2.5) -- (1.45,-2.3);
\draw (0.75,-1.1) -- (0.85,-1.2) -- (0.95,-1.1);
\draw (0.75,1.1) -- (0.85,1.2) -- (0.95,1.1);
%\draw (-0.2,-0.8) -- (0,-1) -- (0.2,-0.8);

%trajectory of the left screen
\draw[violet,line width=0.8mm] (-2.99,-5.24) -- (1.1,-1.1) -- (0.75,-0.75) to [bend right=8] (0.75,0.75) to [bend right=5] (0.52,1.7) -- (-2.99,5.24);
%\draw[violet,line width=0.8mm] plot[variable=\t,samples=18,domain=18.7:34.5] ({3*sec(\t)-4.2},{3*tan(\t)});
%\draw[violet,line width=0.8mm] (-0.55,2.05) -- (IItophat); 

\fill[fill=violet!20] (IIbothat) -- (1.1,-1.1) -- (0.75,-0.75) to [bend right=8] (0.52,1.7) -- (IItophat) -- cycle;

%trajectory of the right screen
\draw[orange,line width=0.8mm] (IIsquaretopright) -- (0.75,0.75) to [bend left=10] (0.75,-0.75) -- (IIsquarebotright);
\fill[fill=orange!20] (IIsquaretopright) -- (0.75,0.75) to [bend left=10] (0.75,-0.75) -- (IIsquarebotright) -- cycle;

\node at (1.105,-1.105) [circle,fill,inner sep=1.5pt, label = right:$T$]{};

\node at (-3,-5.25) [label=left:$A$]{};
\node at (-3,5.25) [label=left:$A'$]{};
\node at (3,-3) [label=right:$B$]{};
\node at (3,3) [label=right:$B'$]{};

\node at (0.52,-1.7) [circle,fill,inner sep=1.5pt,label=left:$P$]{};
\node at (0.52,1.7) [circle,fill,inner sep=1.5pt]{};
\node at (0.7,1.4) [label=left:$Q$]{};

\node at (0.76,-0.76) [circle,fill,inner sep=1.5pt,label=left:$Q'$]{};
\node at (0.75,0.75) [circle,fill,inner sep=1.5pt,label=left:$P'$]{};

\node at (2.25,0) [circle,fill,inner sep=1.5pt, label = above:$O$]{};
\node at (0,0) [circle,fill,inner sep=1.5pt]{};
\node at (0.1,0) [label = left:$O'$]{};

%draw cauchy slice and label on the slice
\draw[line width=0.8mm,red,smooth] plot[variable=\t,samples=100,domain=0:3] ({\t-3},{0.1*\t*cos(45*\t)+2.4});
\draw[line width=0.8mm,red, smooth] plot[variable=\t,samples=100,domain=3:4.52] ({\t-3},{0.1*\t*cos(45*\t)+2.4});
\draw[line width=0.8mm,red, smooth] plot[variable=\t,samples=100,domain=4.52:6] ({\t-3},{0.1*\t*cos(45*\t)+2.4});
\node at (0.06,2.15) [circle,fill,inner sep=2pt]{};
\node at (-0.2,2.3) [label=below:$\S_\Le$]{};
\node at (2.05,2.05) [circle,fill,inner sep=2pt]{};
\node at (2.1,2.1) [label=below:$\S_\Ri$]{};
\node at (-1.5,1.65) [label = $\color{red} \Sigma_\Le$]{};
\node at (2.7,1.4) [label = $\color{red} \Sigma_\Ri$]{};
\node at (1.1,1.25) [label = $\color{red} \Sigma_\E$]{};

\end{scope}
\end{scope}
\end{tikzpicture}
    \caption{\footnotesize Trajectories of the left screen on $\S_\Le$ (purple) and the right screen on $\S_\Ri$ (orange) in the $(\varepsilon_L,\varepsilon_R)=(+1,-1)$ bubble geometries. The interior regions of the pode and the antipode are the purple and orange shaded regions, respectively. The exterior region is white.} \label{fig:screen_trajectories_+-}
\end{figure}
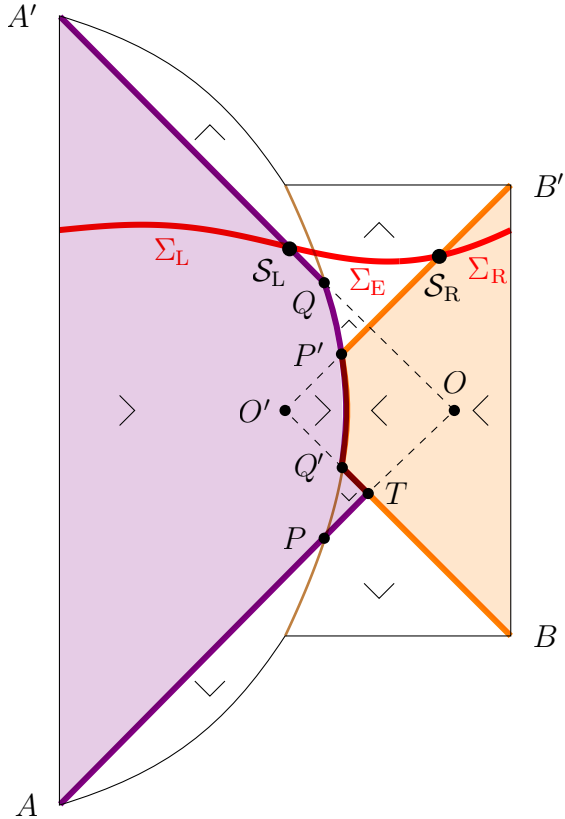

The procedure above can also be performed for the bubble geometries with $(\varepsilon_L,\varepsilon_R)=(+1,-1)$, leading to the screen trajectories depicted in the Penrose diagram of \Fig{fig:screen_trajectories_+-}.
One important difference with the $(\varepsilon_L,\varepsilon_R)=(+1,+1)$ bubbles is that now, the right screen on $\S_\Ri$ does not always follow the cosmological horizon of the antipode. As a consequence, the interior region of the antipode is no longer the full causal patch of the antipode observer, but only a subregion of it. The trajectory of the right screen, which corresponds to the thick orange line in \Fig{fig:screen_trajectories_+-}, consists of the following three segments: i) The lightlike line segment $BQ'$ along the past cosmological horizon of the antipode. Along $BQ'$, the area of the spherical sections remains constant, equal to $\omega_{n-1}R_R^{n-1}$. ii) The timelike segment $Q'P'$ along the trajectory of the domain wall. Along $Q'P'$, the area of the spheres decreases from $\omega_{n-1}R_R^{n-1}$ to $\omega_{n-1}R_B^{n-1}$, and then increases back to $\omega_{n-1}R_R^{n-1}$. iii) The lightlike line segment $P'B'$ along the future cosmological horizon of the antipode, where the area of the spheres remains constant, equal to $\omega_{n-1}R_R^{n-1}$. The trajectory of the left screen, which corresponds to the thick purple line in \Fig{fig:screen_trajectories_+-}, is similar to that of the $(\varepsilon_L,\varepsilon_R)=(+1,+1)$ bubbles described above.

As both screens evolve along their trajectories, the number of holographic degrees of freedom on them can change, since their areas in Planck units are not always constant. The evolution of the two-screen system is not unitary, suggesting a sequence of mappings involving Hilbert spaces of different dimensionalities.

%%%%%%%%%%%%%%%%%%%%%%%%%%%%%%%%%%%%%%%%%%%%%%%%%%%%%%%%%%%%%%%%%
\subsection{Bilayer holographic proposal}

We now present our prescription for computing the fine-grained von Neumann entropy of subsystems of the two-screen system associated with the bubble geometries with $(\varepsilon_L,\varepsilon_R)=(+1,+1)$ and $(\varepsilon_L,\varepsilon_R)=(+1,-1)$ described above, generalizing the analysis of \cite{Franken:2023jas,Franken:2023pni} for eternal de Sitter spacetime and closed FRW cosmologies and the bilayer proposal of \cite{Susskind:2021esx,Shaghoulian:2021cef,Shaghoulian:2022fop}. 

A generic SO$(n)$-symmetric Cauchy slice $\Sigma$ intersects the purple and orange trajectories of \Figs{fig:screen_trajectories_++} and~\ref{fig:screen_trajectories_+-} at the two S$^{n-1}$ spheres $\S_\Le$ and $\S_\Ri$, where the holographic screens are located. As discussed above, this construction divides such a generic slice $\Sigma$ into three parts, $\Sigma=\Sigma_\Le\cup\Sigma_\E\cup\Sigma_\Ri$. $\Sigma_\Le$ ($\Sigma_\Ri$) is the part of $\Sigma$ between $\S_\Le$ and the pode (antipode), while $\Sigma_\E$ is the part of $\Sigma$ between $\S_\Le$ and $\S_\Ri$. For $i\in\{\Le,\E,\Ri \}$, we denote by $\J_i$ the causal diamond of $\Sigma_i$. We will assume unitary evolution among the Cauchy slices of each $\J_i$. This is consistent with the holographic description since the gravitational subsystems associated with the Cauchy slices of $\J_i$ can be reconstructed from the same subsystems of the same pair of holographic screens \cite{Franken:2023jas}. 

Next, let us consider a holographic subsystem of the two-screen system associated with a subregion $A$ of $\S_\Le \cup \S_\Ri$.\footnote{Notice that we consider subsystems of the holographic dual associated with subregions of the two screens, but we do not assume that the holographic theory is local. One possibility is to envision a system of a number of qbits, one per Planckian area on each screen and allow for the possibility of non-local interactions among the qbits.} For any such subregion $A$ (and the associated subsystem), we define
\begin{equation}\label{parts of A}
A_\Le=A\cap \S_\Le, \qquad A_\E=A, \qquad A_\Ri=A\cap \S_\Ri.
\end{equation}
For $i\in\{\Le,\E,\Ri \}$, we denote by $\chi_i$ a codimension-$2$ surface of {\it minimal extremal area}, which is homologous to $A_i$ and lies in the causal diamond region $\J_i$, including its boundary. The homology condition means that there exists a codimension-$1$ surface $\mathcal{C}_i$ in some Cauchy slice $\hat{\Sigma}_i$ of $\J_i$ bounded by $\chi_i$ and $A_i$, \ie satisfying $\partial\mathcal{C}_i=\chi_i\cup A_i$, $\forall i\in\{\Le,\E,\Ri \}$. Notice that the extremization of the area functional of surfaces homologous to $A_i$ requires the inclusion of Lagrange multipliers and auxiliary fields enforcing the extremal surfaces to lie in $\J_i$, including its boundary \cite{Franken:2023jas,Franken:2023pni}. If more than one extremal surfaces $\chi_i$ exist, we choose the one with the minimal area.  The bilayer holographic entanglement entropy proposal then states that:
\begin{enumerate}
\item The classical geometrical contribution to the von Neumann entropy of $A$ is given by:
\begin{equation}
S(A)=\frac{{\rm Area}(\chi_\Le)+{\rm Area}(\chi_\E)+{\rm Area}(\chi_\Ri)}{4G\hbar}+\mathcal{O}\left((G\hbar)^0\right),
\label{classicalentropy}
\end{equation}
at leading order in $G\hbar$. 
\item The entanglement wedge of $A$ corresponds to the union of the causal diamonds of $\mathcal{C}_\Le,\mathcal{C}_\E,\mathcal{C}_\Ri$. Assuming entanglement wedge reconstruction \cite{Dong:2016eik,Cotler:2017erl}, the state on $\mathcal{C}_\Le\cup\mathcal{C}_\E\cup\mathcal{C}_\Ri$ is dual to the state of $A$. Given this, it follows that the von Neumann entropy of the gravitational subsystem associated with  $\mathcal{C}_\Le\cup\mathcal{C}_\E\cup\mathcal{C}_\Ri$ is equal to the von Neumann entropy of the holographic subsystem associated with $A$. 
\end{enumerate}

A proposal to include quantum corrections is discussed in \Appendix{app:quantum_corrections}, generalizing similar proposals in \cite{Franken:2023jas,Franken:2023pni}. It involves extremizing a generalized entropy with both geometrical and semiclassical entropy contributions. See also \cite{Almheiri:2020cfm} and references therein. In particular, we add to the leading geometrical contributions of Eq.~(\ref{classicalentropy}) the semiclassical entropy of the quantum field system (including the contributions from gravitons) on the $\mathcal{C}_\Le\cup\mathcal{C}_\E\cup\mathcal{C}_\Ri$ slice of the background bubble geometry, and then we extremize the resulting functional with respect to the $\chi_i$'s. See Eqs~(\ref{eq:generalized entropy}) and (\ref{eq:QES}). The leading contributions to the semiclassical entropy are of order $(G\hbar)^0$ and are computed by applying quantum field theory methods on curved backgrounds, in the semiclassical approximation \cite{Almheiri:2020cfm}. 

In the following two sections, we apply the bilayer proposal to compute the von Neumann entropy of the two-screen system $\S_\Le\cup\S_\Ri$ and the von Neumann entropies of the single-screen systems $\S_\Le$ and $\S_\Ri$ at order $(G\hbar)^{-1}$.
%%%%%%%%%%%%%%%%%%%%%%%%%%%%%%%%%%%%%%%%%%%%%%%%%%%%%%%%%%%%%%%%
\subsection{The two-screen system in the $(\varepsilon_L,\varepsilon_R)=(+1,\pm 1)$ bubbles}
\begin{figure}[h!]
    %\centering
    \begin{subfigure}[t]{0.48\linewidth}
\centering
\begin{tikzpicture}
\begin{scope}[transparency group]
\begin{scope}[blend mode=multiply]
%Gives a name to the 6 corners of the diagram
\path
       +(-3,4.5) coordinate (IItophat)
       +(0,3) coordinate (IItophatright)
       +(3,3)  coordinate (IIsquaretopright)
       +(3,-3) coordinate (IIsquarebotright)
       +(0,-3) coordinate (IIbothatright)
       +(-3,-4.5) coordinate (IIbothat)
       +(-3,3) coordinate (IIsquaretopleft)
       +(-3,-3) coordinate(IIsquarebotleft)
      
       ;

\fill[fill=blue!50] (0.33,0.33) -- (-0.84,1.5) -- (0.13,2.47) -- (1.3,1.3) -- cycle;

\fill[fill=blue!50] (-3,-0.66) -- (-0.84,1.5) -- (-3,3.66) --  cycle;

\fill[fill=blue!50] (3,-0.4) -- (1.3,1.3) -- (3,3) --  cycle;

%Draws the 6 boundaries of the diagram       
\draw (IItophat) to
          [bend left=20] (IItophatright) --
          (IIsquaretopright) --       (IIsquarebotright) --
          (IIbothatright) to 
          [bend left=20] (IIbothat)--
          (IItophat)--cycle;

%\draw (IIbothatright) to 
%          [bend left=20] (IItophatright);

%Draws the 4 dotted lines delimiting the interior diamond
%\draw[dotted] (IItophatright) -- (3,0);
%\draw[dotted] (IIbothatright) -- (3,0);
%\draw[dotted] (IItophatright) -- (-3,0);
%\draw[dotted] (IIbothatright) -- (-3,0);

\draw[dashed] (IIsquaretopright) -- (0,0) -- (IIsquarebotright);
\draw[dashed] (IItophat) -- (1.5,0) -- (IIbothat);
\draw[dotted] (-1,-1) -- (0,0) -- (-1,1);

\draw[line width= 1 pt,brown] plot[variable=\t,samples=90,domain=-45:45] ({3*sec(\t)-4.2},{3*tan(\t)});

\node at (-0.84,1.5) [circle,fill,inner sep=1.5pt, label = left:$\S_\Le$]{};
\node at (1.3,1.3) [circle,fill,inner sep=1.5pt, label = right:$\S_\Ri$]{};

\end{scope}
\end{scope}
\end{tikzpicture}
%\caption{\footnotesize  \label{}}
\end{subfigure}
\quad \,
\begin{subfigure}[t]{0.48\linewidth}
\centering
\begin{tikzpicture}
\begin{scope}[transparency group]
\begin{scope}[blend mode=multiply]
%Gives a name to the 6 corners of the diagram
\path
       +(-3,4.5) coordinate (IItophat)
       +(0,3) coordinate (IItophatright)
       +(3,3)  coordinate (IIsquaretopright)
       +(3,-3) coordinate (IIsquarebotright)
       +(0,-3) coordinate (IIbothatright)
       +(-3,-4.5) coordinate (IIbothat)
       +(-3,3) coordinate (IIsquaretopleft)
       +(-3,-3) coordinate(IIsquarebotleft)
      
       ;

\fill[fill=blue!50] (1.5/2,1.5/2) -- (-2,3.5) -- (-1.35,4.05) to [bend left=8] (0,3) -- (1,3) -- (2,2) -- (1.3,1.3) -- cycle;

\fill[fill=blue!50] (-3,2.5) -- (-2,3.5) -- (-3,4.5) --  cycle;

\fill[fill=blue!50] (3,1) -- (2,2) -- (3,3) --  cycle;

%Draws the 6 boundaries of the diagram       
\draw (IItophat) to
          [bend left=20] (IItophatright) --
          (IIsquaretopright) --       (IIsquarebotright) --
          (IIbothatright) to 
          [bend left=20] (IIbothat)--
          (IItophat)--cycle;

%\draw (IIbothatright) to 
%          [bend left=20] (IItophatright);

%Draws the 4 dotted lines delimiting the interior diamond
%\draw[dotted] (IItophatright) -- (3,0);
%\draw[dotted] (IIbothatright) -- (3,0);
%\draw[dotted] (IItophatright) -- (-3,0);
%\draw[dotted] (IIbothatright) -- (-3,0);

\draw[dashed] (IIsquaretopright) -- (0,0) -- (IIsquarebotright);
\draw[dashed] (IItophat) -- (1.5,0) -- (IIbothat);
\draw[dotted] (-1,-1) -- (0,0) -- (-1,1);

\draw[line width= 1 pt,brown] plot[variable=\t,samples=90,domain=-45:45] ({3*sec(\t)-4.2},{3*tan(\t)});

\node at (-2,3.5) [circle,fill,inner sep=1.5pt, label = left:$\S_\Le$]{};
\node at (2,2) [circle,fill,inner sep=1.5pt, label = right:$\S_\Ri$]{};

\end{scope}
\end{scope}
\end{tikzpicture}
    %\caption{\footnotesize  \label{}}
\end{subfigure}
    \caption{\footnotesize Entanglement wedge (blue shaded region) of the two-screen system $\S_\Le\cup\S_\Ri$, depicted at different conformal times for each screen, for the bubbles with $(\varepsilon_L,\varepsilon_R)=(+1,+1)$. It coincides with the union of the three regions $\J_i$, $i\in\{\Le,\E,\Ri \}$. For $i\in\{\Le,\E,\Ri \}$, the minimal extremal surface $\chi_i$ is the empty set: $\chi_i=\varnothing$. As the screens evolve in time, the entanglement wedge covers the full spacetime.}
    \label{fig:two_screen_system}
\end{figure}
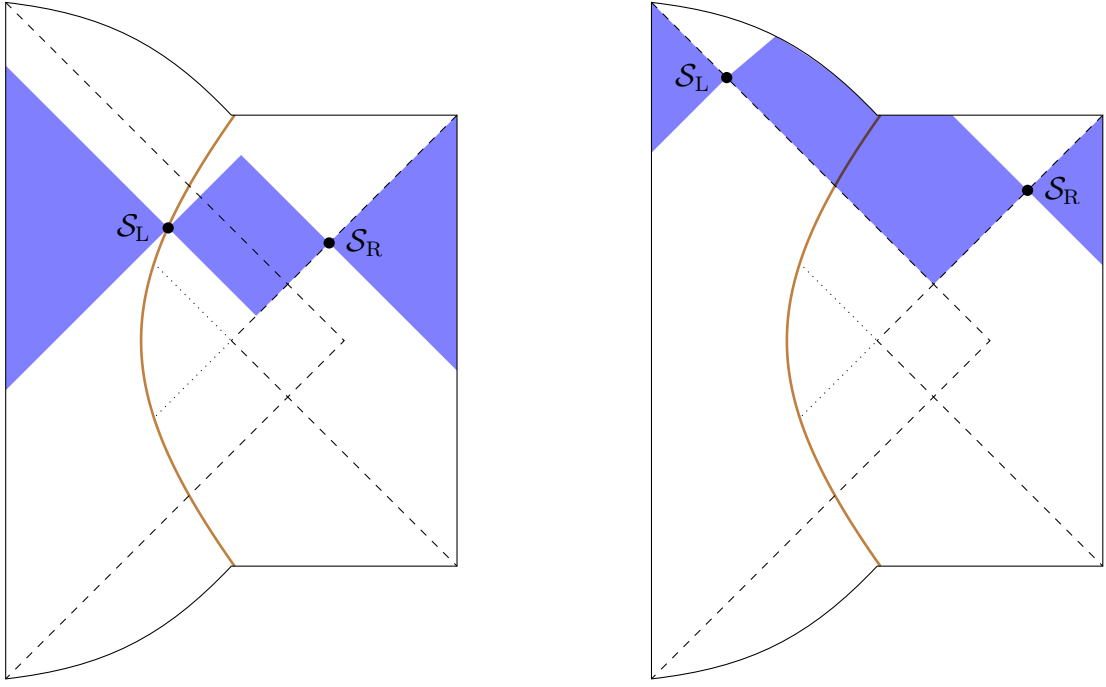

Let us first consider the full two-screen system $A=\S_\Le\cup\S_\Ri$. Eq.~\eqref{parts of A} yields
\begin{equation}
    A_\Le=\S_\Le, \qquad A_\E=\S_\Le\cup\S_\Ri , \qquad A_\Ri=\S_\Ri.
\end{equation}
For $i\in\{\Le,\E,\Ri \}$, we look for a surface $\chi_i$ of minimal extremal area homologous to $A_i$ and lying in $\J_i$. The three causal diamonds $\J_i$ are shown in blue in \Fig{fig:two_screen_system}, for the  $(\varepsilon_L,\varepsilon_R)=(+1,+1)$ cases. The situation is analogous for the bubble geometries with $(\varepsilon_L,\varepsilon_R)=(+1,-1)$. The blue rectangle in the middle is $\J_\E$, the causal diamond of $\Sigma_\E$. Along the lightlike segment $TO'$
where the screens coincide, it shrinks to a point, but becomes nontrivial when the screen trajectories separate. Its non-zero extent signals the entanglement between the two screen systems.

It is easy to verify that $\chi_i=\varnothing \,\,\, \forall i$, which yields vanishing geometrical contributions to the von Neumann entropy of the two-screen system. Indeed, the empty surface is of minimal extremal vanishing area (since it cannot be deformed) and satisfies the homology condition for $\mathcal{C}_i=\Sigma_i$. Therefore, to leading order the von Neumann entropy of the two-screen system is vanishing:
\begin{equation}
    S(\S_\Le\cup\S_\Ri)=0+\mathcal{O}(G\hbar)^0.
\end{equation}

Since $\mathcal{C}_i=\Sigma_i\,\, \forall i$, the entanglement wedge of the two-screen system is the blue-shaded region in \Fig{fig:two_screen_system}, or the union of the causal diamonds $\J_\Le\cup \J_\E\cup \J_\Ri$. Notice that the entanglement wedge contains all complete Cauchy slices passing through the two screens. Assuming entanglement wedge reconstruction, we expect the two-screen system to be able to holographically encode the state on any such Cauchy slice. Notice that when the two screens coincide along $TO'$, the union $\Sigma_\Le\cup\Sigma_\Ri$ amounts to a complete Cauchy slice, and so the description of the state on such slices was expected from the Bousso covariant entropy conjecture. The bilayer proposal implies that even when the screens are apart and $\Sigma_\E$ has a non-trivial extent, the two-screen system is capable of encoding the state on $\Sigma$. As the two screens evolve in time, the entanglement wedge sweeps the full spacetime, and so every slice of the foliation can be encoded during the evolution.

The quantum corrections to the entropy of the two-screen system can be obtained by extremizing a generalized entropy formula. As we have already remarked, for this purpose we add to the geometrical contributions the semiclassical entropy of the quantum field system on $\mathcal{C}_\Le\cup\mathcal{C}_\E\cup\mathcal{C}_\Ri$ and then we extremize with respect to the $\chi_i$'s. Consider the case for which the bulk state on $\Sigma$ is pure. Taking into account the fact that $S_{\rm semi}(\mathcal{C}_\Le\cup\mathcal{C}_\E\cup\mathcal{C}_\Ri)$ is nonnegative, it follows that the full generalized entropy is also minimized for $\chi_i=\varnothing \,\,\, \forall i$, since then $\mathcal{C}_\Le\cup\mathcal{C}_\E\cup\mathcal{C}_\Ri = \Sigma$ and both the geometrical contributions and the semiclassical entropy vanish. Therefore, if our prescription for computing quantum corrections is correct, the fine-grained entropy of the two-screen system vanishes to all orders in the $G\hbar$ expansion, suggesting that the state of the two-screen system is pure when the bulk state on $\Sigma$ is pure. This is compatible with entanglement wedge reconstruction. It also follows that the entropies of the single-screen systems are equal. The extremization problem is more complicated when the state on $\Sigma$ is not pure.

%%%%%%%%%%%%%%%%%%%%%%%%%%%%%%%%%%%%%%%%%%%%%%%%%%%%%%%%%%%%%%%%%%%%%%
\subsection{The single-screen system in the $(\varepsilon_L,\varepsilon_R)=(+1,+1)$ bubbles}
\label{sect:single_screen_++}
Let us now consider the single-screen system $A=\S_\Le$\footnote{The analysis for $A=\S_\Ri$ can be carried out in a similar way.} in the $(\varepsilon_L,\varepsilon_R)=(+1,+1)$ geometries. Eq.~\eqref{parts of A} gives
\begin{equation}
    A_\Le=\S_\Le, \qquad A_\E=\S_\Le, \qquad A_\Ri=\varnothing.
\end{equation}
As for the two-screen system described above, the minimal extremal surface $\chi_\Le$ in $\J_\Le$ homologous to $A_\Le=\S_\Le$ is the empty set, $\chi_\Le=\varnothing$. The geometrical contribution to the von Neumann entropy of $\S_\Le$ from the left interior region vanishes. The homology condition is satisfied for $\mathcal{C}_\Le=\Sigma_\Le$, and so the entanglement wedge in the left interior region is the full causal diamond $\J_\Le$.

Since $A_\Ri=\varnothing$, we also obtain $\chi_\Ri=\varnothing$. So the geometrical contributions to the entanglement entropy from the right antipode interior region is also vanishing. Since $\mathcal{C}_\Ri$ is trivially the empty set, the entanglement wedge of $\S_\Le$ does not extend in the antipode interior region.

The extremization problem \eqref{classicalentropy} is nontrivial in the exterior causal diamond $\J_\E$. For example, when the screen $\S_\Le$ lies on the lightlike segment $AT$ of the pode cosmological horizon and on the $Q'Q$ segment of the domain wall trajectory, $\J_\E$ corresponds to the red-shaded regions depicted in \Fig{fig:single_screen_contracting_phase} and in \Fig{fig:single_screen_expanding_phase_b}, respectively.

We will look for SO$(n)$-symmetric minimal extremal homologous surfaces, \ie S$^{n-1}$ spheres, which are represented by points on the Penrose diagram. As we remarked, the extremization problem includes suitable Lagrange multipliers and auxiliary fields that enforce the homologous spherical surfaces $\chi_\E$ to lie in the exterior causal diamond $\J_\E$, including its boundary \cite{Franken:2023jas,Franken:2023pni}. Essentially, minimal area spheres on the boundary of $\J_\E$ are extremal surfaces of the modified area functional. The extremization analysis is carried out in the Appendix \ref{app:area_extremization}. In the following, we summarize the main results and conclusions.  

Recall that while the area of $\S_\Ri$ remains constant when $(\varepsilon_L,\varepsilon_R)=(+1,+1)$, ${\rm Area}(\S_\Ri)=\omega_{n-1}R_R^{n-1}$, the area of $\S_\Le$ can change during its evolution. The trajectory of $\S_\Le$ splits into five different segments, as shown in \Fig{fig:screen_trajectories_++}, yielding different results regarding the calculation of the minimal extremal surfaces $\chi_\E$. Along each segment, the area of $\S_\Le$ is given by:
\renewcommand{\arraystretch}{1.4}
\begin{equation}\label{eq:area_S_L_++}
{\rm Area}(\S_\Le) = \omega_{n-1}\times  
\left\{\begin{array}{ll}
    \displaystyle R_L^{n-1} \, , &\mbox{if } \S_\Le \in AP  \\
    \displaystyle \left(\frac{R_R\sin(\sigma_R+\theta_{RP}-\sigma_{RP})}{\cos\sigma_R}\right)^{n-1} \, , &\mbox{if } \S_\Le \in PT \\
    \displaystyle R_R^{n-1} \, , &\mbox{if } \S_\Le \in TQ' \\
    \displaystyle \left(R_R^2\tan^2\sigma_R+R_B^2\right)^{\frac{n-1}{2}} \, , &\mbox{if } \S_\Le \in Q'Q \\
    \displaystyle R_L^{n-1} \, , &\mbox{if } \S_\Le \in QA'
    \end{array}\right. ,
\end{equation}
\renewcommand{\arraystretch}{1}where $\theta_{RP}$ and $\sigma_{RP}$ are given in \Eqs{eq:rp_rq_coordinates_++}. It is always larger than or equal to the area of $\S_\Ri$. 

Depending on the location of $\S_\Le$, the location of the classical minimal extremal surfaces $\chi_\E$ changes:
\begin{itemize}
\item When $\S_\Le$ lies on the lightlike line segments $AP$ and $PT$, along the past cosmological horizon of the pode, $\S_\Ri$ is on the lightlike segment $BT$, along the past cosmological horizon of the antipode. This is because $\S_\Le$ and $\S_\Ri$ are spacelike separated. As in the eternal de Sitter case \cite{Franken:2023pni}, there is a degeneracy of classical minimal extremal homologous surfaces. Indeed, any sphere on the segment $\S_\Ri T$ of the past cosmological horizon of the antipode is a minimal extremal homologous surface and a candidate for $\chi_\E$ at the classical level. In \Appendix{app:area_extremization} it is shown that these minimal area spheres are extrema of the modified area functional. $\S_\Ri T$ is depicted in red in \Fig{fig:single_screen_contracting_phase}.
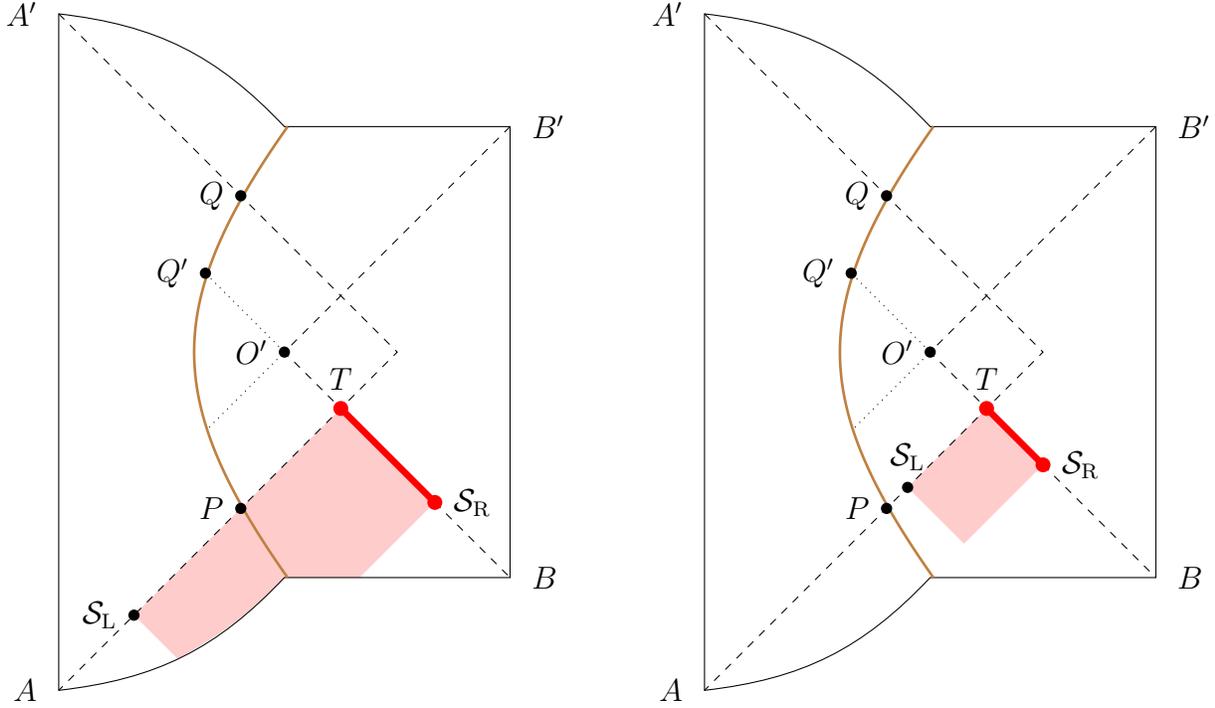
\begin{figure}[h!]
    %\centering
\begin{subfigure}[t]{0.48\linewidth}
\centering
\begin{tikzpicture}
%\begin{scope}[transparency group]
%\begin{scope}[blend mode=multiply]
%Gives a name to the 6 corners of the diagram
\path
       +(-3,4.5) coordinate (IItophat)
       +(0,3) coordinate (IItophatright)
       +(3,3)  coordinate (IIsquaretopright)
       +(3,-3) coordinate (IIsquarebotright)
       +(0,-3) coordinate (IIbothatright)
       +(-3,-4.5) coordinate (IIbothat)
       +(-3,3) coordinate (IIsquaretopleft)
       +(-3,-3) coordinate(IIsquarebotleft)
      
       ;

\fill[fill=red!20] (1.5/2,-1.5/2) -- (-2,-3.5) -- (-1.42,-4.08) to [bend right=8] (0,-3) -- (1,-3) -- (2,-2) -- (1.3,-1.3) -- cycle;

%\fill[fill=blue!50] (-3,2.5) -- (-2,3.5) -- (-3,4.5) --  cycle;

%Draws the 6 boundaries of the diagram       
\draw (IItophat) to
          [bend left=20] (IItophatright) --
          (IIsquaretopright) --       (IIsquarebotright) --
          (IIbothatright) to 
          [bend left=20] (IIbothat)--
          (IItophat)--cycle;

%\draw (IIbothatright) to 
%          [bend left=20] (IItophatright);

%Draws the 4 dotted lines delimiting the interior diamond
%\draw[dotted] (IItophatright) -- (3,0);
%\draw[dotted] (IIbothatright) -- (3,0);
%\draw[dotted] (IItophatright) -- (-3,0);
%\draw[dotted] (IIbothatright) -- (-3,0);

\draw[line width=2.5pt, red] (0.75,-0.75)--(2,-2);
%\node at (0.75,0.75) [circle,fill,inner sep=1.5pt, label = right:$\sigma'$]{};
\draw[dashed] (2,-2) -- (3,-3);
\draw[dashed] (3,3) -- (0,0) -- (0.75,-0.75) ;
\draw[dashed] (IItophat) -- (1.5,0) -- (IIbothat);
\draw[dotted] (-1,-1) -- (0,0) -- (-1,1);

\draw[line width= 1 pt,brown] plot[variable=\t,samples=90,domain=-45:45] ({3*sec(\t)-4.2},{3*tan(\t)});

\node at (-2,-3.5) [circle,fill,inner sep=1.5pt, label = left:$\S_\Le$]{};
\node at (2,-2) [circle,fill,inner sep=2pt, red, label = right:$\S_\Ri$]{};

\node at (-3,-4.5) [label=left:$A$]{};
\node at (-3,4.5) [label=left:$A'$]{};
\node at (3,-3) [label=right:$B$]{};
\node at (3,3) [label=right:$B'$]{};

%\node at (-0.5,-2.08) [label=$P$]{};
\node at (-0.58,-2.08) [circle,fill,inner sep=1.5pt, label=left:$P$]{};
%\node at (-0.60,2) [label=$Q$]{};
\node at (-0.58,2.08) [circle,fill,inner sep=1.5pt, label=left:$Q$]{};

\node at (-1.05,1.05) [circle,fill,inner sep=1.5pt,label=left:$Q'$]{};

\node at (0,0) [circle,fill,inner sep=1.5pt, label = left:$O'$]{};

\node at (0.75,-0.75) [circle,fill,inner sep=2pt, red, label = above:$T$]{};

%\end{scope}
%\end{scope}
\end{tikzpicture}
        %\caption{\footnotesize{A different choice of $\S_\Le$ and $\S_\Ri$. Again, the exterior surface of minimal area lies on the right horizon. } } \label{}
\end{subfigure}
\quad \,
\begin{subfigure}[t]{0.48\linewidth}
\centering
\begin{tikzpicture}
%\begin{scope}[transparency group]
%\begin{scope}[blend mode=multiply]
%Gives a name to the 6 corners of the diagram
\path
       +(-3,4.5) coordinate (IItophat)
       +(0,3) coordinate (IItophatright)
       +(3,3)  coordinate (IIsquaretopright)
       +(3,-3) coordinate (IIsquarebotright)
       +(0,-3) coordinate (IIbothatright)
       +(-3,-4.5) coordinate (IIbothat)
       +(-3,3) coordinate (IIsquaretopleft)
       +(-3,-3) coordinate(IIsquarebotleft)
      
       ;

\fill[fill=red!20] (0.75,-0.75) -- (1.5,-1.5) -- (0.9/2,-5.1/2) -- (-0.3,-1.8) -- cycle;

%\fill[fill=blue!50] (-3,-0.66) -- (-0.84,1.5) -- (-3,3.66) --  cycle;

%Draws the 6 boundaries of the diagram       
\draw (IItophat) to
          [bend left=20] (IItophatright) --
          (IIsquaretopright) --       (IIsquarebotright) --
          (IIbothatright) to 
          [bend left=20] (IIbothat)--
          (IItophat)--cycle;

\draw[dashed] (1.5,-1.5) -- (3,-3);
\draw[dashed] (3,3) -- (0,0) -- (0.75,-0.75);
\draw[dashed] (IItophat) -- (1.5,0) -- (IIbothat);
\draw[dotted] (-1,-1) -- (0,0) -- (-1,1);

\draw[line width= 1 pt,brown] plot[variable=\t,samples=90,domain=-45:45] ({3*sec(\t)-4.2},{3*tan(\t)});

%\node at (-0.84,1.5) [circle,fill,inner sep=1.5pt]{};
\node at (-0.3,-1.8) [circle,fill,inner sep=1.5pt, label = above:$\S_\Le$]{};
\node at (1.5,-1.5) [circle,fill,inner sep=2pt, red, label = right:$\S_\Ri$]{};

\draw[line width=2.5pt, red] (0.75,-0.75)--(1.5,-1.5);
%\node at (0.33,0.33) [circle,fill,inner sep=1.5pt, label = right:$\sigma'$]{};

\node at (-3,-4.5) [label=left:$A$]{};
\node at (-3,4.5) [label=left:$A'$]{};
\node at (3,-3) [label=right:$B$]{};
\node at (3,3) [label=right:$B'$]{};

%\node at (-0.5,-2.08) [label=$P$]{};
\node at (-0.58,-2.08) [circle,fill,inner sep=1.5pt, label=left:$P$]{};
%\node at (-0.60,2) [label=$Q$]{};
\node at (-0.58,2.08) [circle,fill,inner sep=1.5pt, label=left:$Q$]{};

\node at (-1.05,1.05) [circle,fill,inner sep=1.5pt,label=left:$Q'$]{};

\node at (0,0) [circle,fill,inner sep=1.5pt, label = left:$O'$]{};

\node at (0.75,-0.75) [circle,fill,inner sep=2pt, red, label = above:$T$]{};

%\end{scope}
%\end{scope}
\end{tikzpicture}
\end{subfigure}
    \caption{\footnotesize{Penrose diagrams when the screens are in the contracting phase of the bubble geometry. The causal diamond $\J_\E$ is red shaded, and corresponds to the region where we look for the minimal extremal homologous surface $\chi_\E$. Each point of the thick red line segment has the same (minimal extremal) area, and is thus a candidate surface for $\chi_\E$ at order $(G\hbar)^{-1}$, homologous to the single-screen system $A=\S_\Le$.}} \label{fig:single_screen_contracting_phase}
\end{figure}
Since the area is degenerate, each of these classical extremal surfaces yields the same contribution to the leading (of order $(G\hbar)^{-1}$) fine-grained entropy of $\S_\Le$. However, they lead to different entanglement wedges in the exterior region. For $\chi_\E = \S_\Ri$, the entanglement wedge in the exterior region becomes maximal, covering the whole $\J_\E$. See \Fig{fig:EW_SL}. 
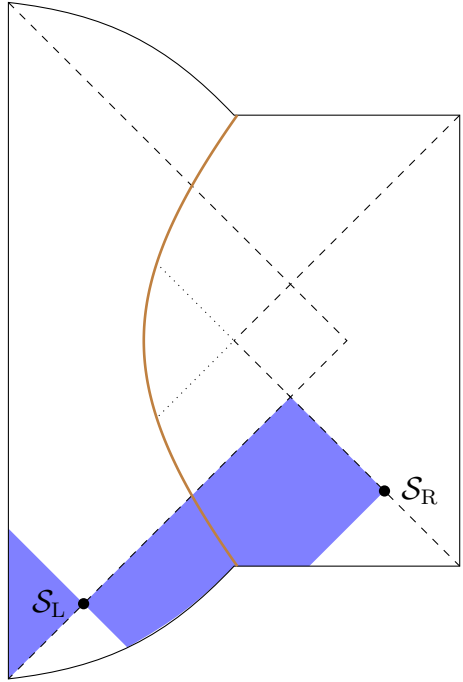
\begin{figure}[t]
\centering
\begin{tikzpicture}
%\begin{scope}[transparency group]
%\begin{scope}[blend mode=multiply]
%Gives a name to the 6 corners of the diagram
\path
       +(-3,4.5) coordinate (IItophat)
       +(0,3) coordinate (IItophatright)
       +(3,3)  coordinate (IIsquaretopright)
       +(3,-3) coordinate (IIsquarebotright)
       +(0,-3) coordinate (IIbothatright)
       +(-3,-4.5) coordinate (IIbothat)
       +(-3,3) coordinate (IIsquaretopleft)
       +(-3,-3) coordinate(IIsquarebotleft)
      
       ;

\fill[fill=blue!50] (1.5/2,-1.5/2) -- (-2,-3.5) -- (-1.42,-4.08) to [bend right=8] (0,-3) -- (1,-3) -- (2,-2) -- (1.3,-1.3) -- cycle;

\fill[fill=blue!50] (-3,-4.5) -- (-2,-3.5) -- (-3,-2.5) --  cycle;

%Draws the 6 boundaries of the diagram       
\draw (IItophat) to
          [bend left=20] (IItophatright) --
          (IIsquaretopright) --       (IIsquarebotright) --
          (IIbothatright) to 
          [bend left=20] (IIbothat)--
          (IItophat)--cycle;

%\draw (IIbothatright) to 
%          [bend left=20] (IItophatright);

%Draws the 4 dotted lines delimiting the interior diamond
%\draw[dotted] (IItophatright) -- (3,0);
%\draw[dotted] (IIbothatright) -- (3,0);
%\draw[dotted] (IItophatright) -- (-3,0);
%\draw[dotted] (IIbothatright) -- (-3,0);

%\draw[line width=2.5pt, red] (0.75,-0.75)--(2,-2);
%\node at (0.75,0.75) [circle,fill,inner sep=1.5pt, label = right:$\sigma'$]{};
\draw[dashed] (0.7,-0.7) -- (3,-3);
\draw[dashed] (3,3) -- (0,0) -- (0.75,-0.75) ;
\draw[dashed] (IItophat) -- (1.5,0) -- (IIbothat);
\draw[dotted] (-1,-1) -- (0,0) -- (-1,1);

\draw[line width= 1 pt,brown] plot[variable=\t,samples=90,domain=-45:45] ({3*sec(\t)-4.2},{3*tan(\t)});

\node at (-2,-3.5) [circle,fill,inner sep=1.5pt, label = left:$\S_\Le$]{};
\node at (2,-2) [circle,fill,inner sep=1.5pt, label = right:$\S_\Ri$]{};

%\node at (-3,-4.5) [label=left:$A$]{};
%\node at (-3,4.5) [label=left:$A'$]{};
%\node at (3,-3) [label=right:$B$]{};
%\node at (3,3) [label=right:$B'$]{};

%\node at (-0.5,-2.08) [label=$P$]{};
%\node at (-0.58,-2.08) [circle,fill,inner sep=1.5pt, label=left:$P$]{};
%\node at (-0.60,2) [label=$Q$]{};
%\node at (-0.58,2.08) [circle,fill,inner sep=1.5pt, label=left:$Q$]{};

%\node at (-1.05,1.05) [circle,fill,inner sep=1.5pt,label=left:$Q'$]{};

%\node at (0,0) [circle,fill,inner sep=1.5pt, label = left:$O'$]{};

%\node at (0.75,-0.75) [circle,fill,inner sep=1.5pt, label = above:$S$]{};

%\end{scope}
%\end{scope}
\end{tikzpicture}
        \caption{\footnotesize{The entanglement wedge of the left screen for $\chi_{\E}$ located on the right screen. In this case, the entanglement wedge of $\S_{\Le}$ contains the exterior region. This is a reasonable result, considering that $\S_{\Le}$ has more degrees of freedom than $\S_{\Ri}$.} } \label{fig:EW_SL}
\end{figure}
As we approach the intersection point $T$, this component of the entanglement wedge becomes smaller and smaller. For $\chi_\E = T$, it shrinks to the lightlike segment $\S_\Le T$ along the past cosmological horizon of the pode. The degeneracy at the classical, geometrical level can be lifted by quantum corrections. In \cite{Franken:2023pni} it was argued that for the eternal de Sitter case, the first order corrected entropy, which is given by the quantum area of $\chi_\E$\footnote{To obtain the quantum area of a surface, we add to the geometrical area term the entanglement entropy through the surface multiplied by $4G\hbar$ \cite{Strominger:2003br}.}, increases as we move from $\S_\Ri$ to $T$. This is because we expect the entanglement entropy through the sphere $\chi_\E$ to increase in the contracting phase as the conformal time increases, since more and more entangled particles can become separated in the causal patches of the pode and antipode without being able to reunite. Based on this, it is expected that for the bubble geometry as well, the quantum extremal surface $\chi_\E$ that minimizes the generalized entropy is close to the right screen $\S_\Ri$, and the entanglement wedge covers a large portion of $\J_\E$. Of course a more rigorous analysis of the generalized entropy at the next to leading order, including backreaction on the classical background, is needed in order to obtain the quantum extremal surface. In the bubble geometry case, we do not have a precise calculation for the first order corrected entropy associated with the classical minimal extremal spheres, which would provide us with an estimate of the effect of the quantum corrections \cite{Faulkner:2013ana}.

\item $\S_\Le$ and $\S_\Ri$ coincide along the lightlike line segment $TO'$ of the past cosmological horizon of the antipode. The exterior causal diamond region $\J_\E$ reduces to a point, corresponding to the sphere where the two screens are located. Therefore, $\chi_\E=\S_\Le=\S_\Ri$. The entanglement wedge of $\S_\Le$ ($\S_\Ri$) is the causal diamond region $\J_\Le$ ($\J_\Ri$), which contains $\Sigma_\Le$ ($\Sigma_\Ri$). Notice that in this case $\Sigma = \Sigma_\Le \cup \Sigma_\Ri$ and, so, the Bousso covariant entropy bound ensures that the two-screen system has enough degrees of freedom to describe the full Cauchy slice $\Sigma$. 

\item When $\S_\Le$ lies on the lightlike line segment $O'Q'$, $\S_\Ri$ is on the lightlike line segment $O'B'$ of the future cosmological horizon of the antipode. The two screens have the same area $\omega_{n-1}R_R^{n-1}$. In this case as well, there is a degeneracy of classical minimal extremal homologous surfaces $\chi_\E$. Any sphere on the lightlike line segments $O'\S_\Le$ and $O'\S_\Ri$, which are drawn in red in \Fig{fig:single_screen_expanding_phase_a},
\begin{figure}[h!]
    %\centering
    \begin{subfigure}[t]{0.48\linewidth}
\centering
\begin{tikzpicture}
%\begin{scope}[transparency group]
%\begin{scope}[blend mode=multiply]
%Gives a name to the 6 corners of the diagram
\path
       +(-3,4.5) coordinate (IItophat)
       +(0,3) coordinate (IItophatright)
       +(3,3)  coordinate (IIsquaretopright)
       +(3,-3) coordinate (IIsquarebotright)
       +(0,-3) coordinate (IIbothatright)
       +(-3,-4.5) coordinate (IIbothat)
       +(-3,3) coordinate (IIsquaretopleft)
       +(-3,-3) coordinate(IIsquarebotleft)
      
       ;

\fill[fill=red!20] (0.4,1.6) -- (-0.6,0.6) -- (0,0) -- (1,1) -- cycle;

%\fill[fill=blue!50] (-3,-0.66) -- (-0.84,1.5) -- (-3,3.66) --  cycle;

%Draws the 6 boundaries of the diagram       
\draw (IItophat) to
          [bend left=20] (IItophatright) --
          (IIsquaretopright) --       (IIsquarebotright) --
          (IIbothatright) to 
          [bend left=20] (IIbothat)--
          (IItophat)--cycle;

\draw[dashed] (0,0) -- (3,-3);
\draw[dashed] (1,1)--(3,3);
\draw[dashed] (IItophat) -- (1.5,0) -- (IIbothat);
\draw[dotted] (-1,1) -- (-0.6,0.6);
\draw[dotted] (0,0) -- (-1,-1);

\draw[line width= 1 pt,brown] plot[variable=\t,samples=90,domain=-45:45] ({3*sec(\t)-4.2},{3*tan(\t)});

\node at (-0.6,0.6) [circle,fill,inner sep=2pt,red]{};
\node at (-0.85,0.9) [label = below:$\S_\Le$]{};
\node at (1,1) [circle,fill,inner sep=2pt, red, label = right:$\S_\Ri$]{};

\draw[line width=2.5pt, red] (0,0)--(1,1);
\draw[line width=2.5pt, red] (0,0)--(-0.6,0.6);
%\node at (0.33,0.33) [circle,fill,inner sep=1.5pt, label = right:$\sigma'$]{};

\node at (-3,-4.5) [label=left:$A$]{};
\node at (-3,4.5) [label=left:$A'$]{};
\node at (3,-3) [label=right:$B$]{};
\node at (3,3) [label=right:$B'$]{};

%\node at (-0.5,-2.08) [label=$P$]{};
\node at (-0.58,-2.08) [circle,fill,inner sep=1.5pt, label=left:$P$]{};
%\node at (-0.60,2) [label=$Q$]{};
\node at (-0.58,2.08) [circle,fill,inner sep=1.5pt, label=left:$Q$]{};

\node at (-1.05,1.05) [circle,fill,inner sep=1.5pt,label=left:$Q'$]{};

\node at (0,0) [circle,fill,inner sep=2pt, red, label = below:$O'$]{};

\node at (0.75,-0.75) [circle,fill,inner sep=1.5pt, label = below:$T$]{};

%\end{scope}
%\end{scope}
\end{tikzpicture}
\caption{\footnotesize{For $\S_\Le\in O'Q'$, there is a classical degeneracy of minimal extremal homologous spheres along the two bottom edges of $\J_\E$.}}  \label{fig:single_screen_expanding_phase_a}
\end{subfigure}
\quad \,
%%%%%%%%%%%%%%%%%%%%%%%%%%%%%%%%%%%%%%%%%%%%%%%%%%%%%%%%%%%%%%
\begin{subfigure}[t]{0.48\linewidth}
\centering
\begin{tikzpicture}
%\begin{scope}[transparency group]
%\begin{scope}[blend mode=multiply]
%Gives a name to the 6 corners of the diagram
\path
       +(-3,4.5) coordinate (IItophat)
       +(0,3) coordinate (IItophatright)
       +(3,3)  coordinate (IIsquaretopright)
       +(3,-3) coordinate (IIsquarebotright)
       +(0,-3) coordinate (IIbothatright)
       +(-3,-4.5) coordinate (IIbothat)
       +(-3,3) coordinate (IIsquaretopleft)
       +(-3,-3) coordinate(IIsquarebotleft)
      
       ;

\fill[fill=red!20] (0.33,0.33) -- (-0.84,1.5) -- (0.13,2.47) -- (1.3,1.3) -- cycle;

%\fill[fill=blue!50] (-3,-0.66) -- (-0.84,1.5) -- (-3,3.66) --  cycle;

%Draws the 6 boundaries of the diagram       
\draw (IItophat) to
          [bend left=20] (IItophatright) --
          (IIsquaretopright) --       (IIsquarebotright) --
          (IIbothatright) to 
          [bend left=20] (IIbothat)--
          (IItophat)--cycle;

\draw[dashed] (0.32,0.32)--(0,0) -- (3,-3);
\draw[dashed] (1.3,1.3)--(3,3);
\draw[dashed] (IItophat) -- (1.5,0) -- (IIbothat);
\draw[dotted] (-1,-1) -- (0,0) -- (-1,1);

\draw[line width= 1 pt,brown] plot[variable=\t,samples=90,domain=-45:45] ({3*sec(\t)-4.2},{3*tan(\t)});

\node at (-0.84,1.5) [circle,fill,inner sep=1.5pt]{};
\node at (-0.7,1.65) [label = left:$\S_\Le$]{};
\node at (1.3,1.3) [circle,fill,inner sep=2pt, red, label = right:$\S_\Ri$]{};

\draw[line width=2.5pt, red] (0.32,0.32)--(1.3,1.3);
%\node at (0.33,0.33) [circle,fill,inner sep=1.5pt, label = right:$\sigma'$]{};

\node at (-3,-4.5) [label=left:$A$]{};
\node at (-3,4.5) [label=left:$A'$]{};
\node at (3,-3) [label=right:$B$]{};
\node at (3,3) [label=right:$B'$]{};

%\node at (-0.5,-2.08) [label=$P$]{};
\node at (-0.58,-2.08) [circle,fill,inner sep=1.5pt, label=left:$P$]{};
%\node at (-0.60,2) [label=$Q$]{};
\node at (-0.58,2.08) [circle,fill,inner sep=1.5pt, label=left:$Q$]{};

\node at (-1.05,1.05) [circle,fill,inner sep=1.5pt,label=left:$Q'$]{};

\node at (0,0) [circle,fill,inner sep=1.5pt, label = left:$O'$]{};

\node at (0.75,-0.75) [circle,fill,inner sep=1.5pt, label = below:$T$]{};

%\end{scope}
%\end{scope}
\end{tikzpicture}
\caption{\footnotesize{For $\S_\Le\in Q'Q$, the classical degeneracy for $\chi_\E$ along the bottom left edge of $\J_\E$ is raised. There remains a classical degeneracy along the bottom right edge of $\J_\E$.}}  \label{fig:single_screen_expanding_phase_b}
\end{subfigure}
    \caption{\footnotesize{Penrose diagram when the screens are in the expanding phase. The exterior causal diamond $\J_\E$ is red shaded. Each point of the thick red lines has the same (minimal extremal) area, and is thus a candidate surface for $\chi_\E$ homologous to the single-screen system $A=\S_\Le$.}}
    \label{fig:single_screen_expanding_phase}
\end{figure}
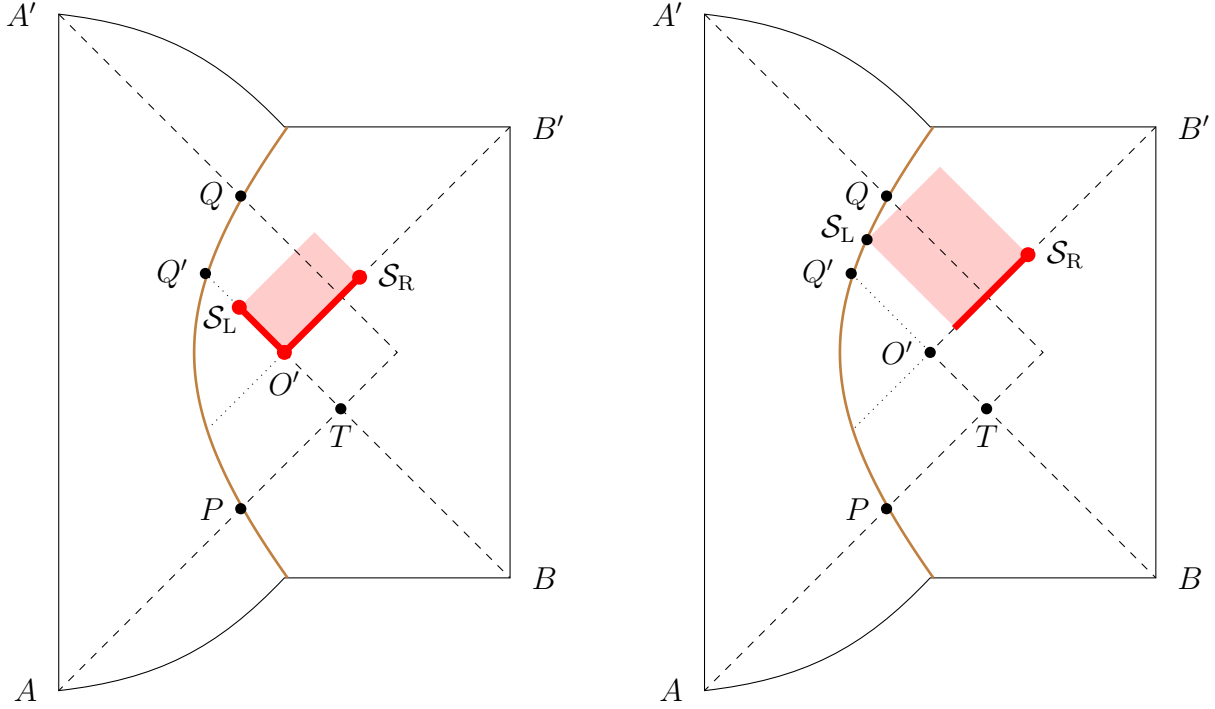
is a minimal extremal homologous surface at leading order $(G\hbar)^{-1}$. This situation is analogous to the eternal de Sitter case studied in \cite{Franken:2023pni}. All such surfaces yield the same contribution to the leading fine-grained entropy of $\S_\Le$. But, they lead to different entanglement wedge structures. If $\chi_\E$ lies on the lightlike line segment $O'\S_\Le$, the entanglement wedge of $\S_\Le$ in the exterior reduces to this lightlike line segment, $O'\S_\Le$. On the other hand, if $\chi_\E$ lies on the future horizon of the antipode, the entanglement wedge of $\S_\Le$ has a larger extent in the exterior region. It becomes equal to $\J_\E$ when $\chi_\E = \S_\Ri$. In the eternal de Sitter case, it was argued in \cite{Franken:2023pni} that quantum corrections will lift the degeneracy at the classical level, favoring the horizon sphere with the smallest quantum area. 
In fact, in the expanding phase of the cosmology, we expect the entanglement entropy through a horizon sphere to decrease with the conformal time. This is because entangled particles in the causal patches of the pode and antipode can reunite in the exterior region, leading to a decrease in the entanglement entropy, and hence the quantum area, of a horizon sphere. Hence, at the quantum level one expects the quantum minimal extremal homologous surface to be close to the screen of larger conformal time. So, if these arguments apply for the bubble geometry as well, and if $\sigma_{R\S_\Ri} > \sigma_{R\S_\Le}$, $\chi_\E=\S_\Ri$ and the exterior component of the entanglement wedge of $\S_\Le$ is the full $\J_\E$. If, on the other hand, $\sigma_{R\S_\Ri} < \sigma_{R\S_\Le}$, $\chi_\E = \S_{\Le}$ and the entanglement wedge of $\S_\Le$ in the exterior region reduces to a point. In this case, it is the exterior component of the entanglement wedge of $\S_{\Ri}$ that is equal to $\J_\E$, signaling interesting phase transitions when $\sigma_{R\S_\Ri} = \sigma_{R\S_\Le}$.    

\item When $\S_\Le$ lies on the timelike segment $Q'Q$ of the domain wall trajectory, $\S_\Ri$ is on the lightlike line segment $O'B'$ of the future cosmological horizon of the antipode. In this case, the classical degeneracy for $\chi_\E$ confines on the bottom right edge of $\J_\E$, along the future cosmological horizon of the antipode, as depicted by the red segment on \Fig{fig:single_screen_expanding_phase_b}. As in the other cases discussed above, we expect quantum corrections to lift the classical degeneracy, determining for  $\chi_\E$ a sphere close to the right screen $\S_\Ri$. This would imply that the entanglement wedge in the exterior region is the full $\J_\E$. Similar conclusions hold when the left screen is on the lightlike line segment $QA'$ along the future cosmological horizon of the pode.
\end{itemize}

Therefore, in all the above cases, the classical, geometrical contributions to the von Neumann entropy of $\S_\Le$ is constant, irrespectively of the fact that the area of this screen can change with time. This leading entropy is given by the area of $\S_\Ri$ divided by $4G\hbar$:
\begin{equation}\label{eq:vN_entropy_SL_++}
    S(\S_\Le)=\frac{\omega_{n-1}R_R^{n-1}}{4G\hbar}+\mathcal{O}(G\hbar)^0.
\end{equation}
Notice that this is equal to the Gibbons-Hawking entropy of the ``parent'' de Sitter space with the bigger cosmological constant. The leading geometrical entropy of $\S_\Ri$ is also equal to this. In fact, if the two-screen system is in a pure state, the von Neumann entropies of the single-screen systems are equal to all orders in $G\hbar$. The component of the entanglement wedge in the exterior region of $\S_\Le$, which carries more degrees of freedom is in general bigger than that of $\S_\Ri$. The only exception is when $\S_\Le$ is on the lightlike line segment $O'Q'$, in which case both systems carry the same number of degrees of freedom.

\begin{figure}[h!]
        \centering
	\includegraphics[height=70mm]{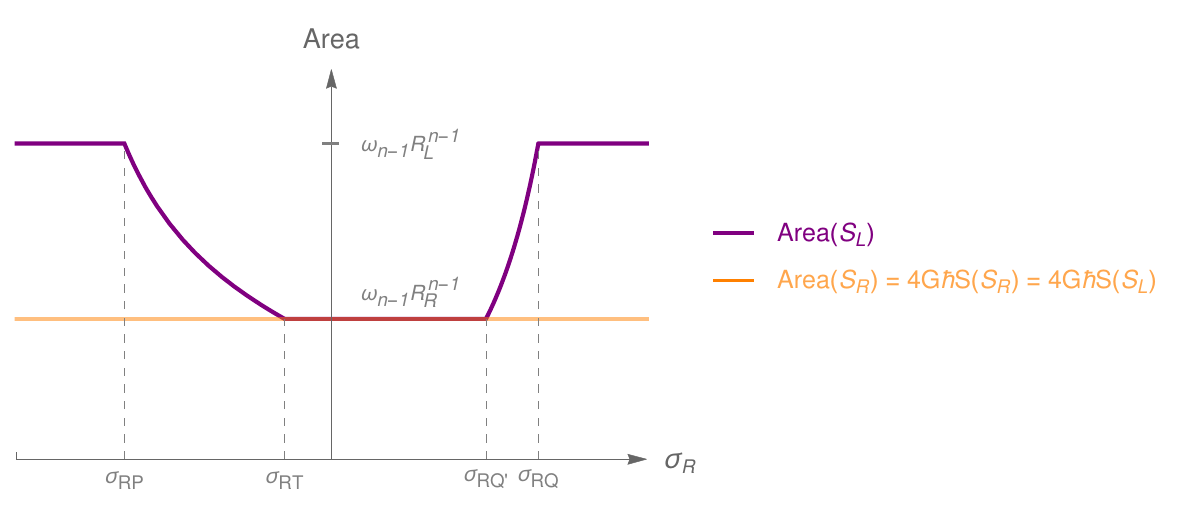}
 \caption{\footnotesize Evolution of the area and entanglement entropy of $\S_\Le$ and $\S_\Ri$ as a function of the conformal time $\sigma_R$, in the bubble geometries with $(\varepsilon_L,\varepsilon_R)=(+1,+1)$. } \label{fig:plot_area_++}
\end{figure}

The evolution of the area of $\S_\Le$ (\Eq{eq:area_S_L_++}) and $\S_\Ri$, as well as the entanglement entropy $S(\S_\Le)$ and $S(\S_\Ri)$, are plotted as functions of the conformal time $\sigma_R$ in \Fig{fig:plot_area_++}.

\subsection{The single-screen system in the $(\varepsilon_L,\varepsilon_R)=(+1,-1)$ bubbles}
\label{sect:single_screen_+-}
Next, we consider the single-screen system $A=\S_\Le$ in the bubble geometries with $(\varepsilon_L,\varepsilon_R)=(+1,-1)$. The screen trajectories are drawn on the Penrose diagram of \Fig{fig:screen_trajectories_+-}. In contrast to the previous case, the areas of both the left and the right screen can change during evolution. The trajectory of the left screen is divided into five segments, along which the area of the screen is given by
\renewcommand{\arraystretch}{1.4}
\begin{equation}\label{eq:area_S_L_+-}
{\rm Area}(\S_\Le) = \omega_{n-1}\times  
\left\{\begin{array}{ll}
    \displaystyle R_L^{n-1} \, , &\mbox{if } \S_\Le \in AP  \\
    \displaystyle \left(\frac{R_R\sin(\sigma_R+\theta_{RP}-\sigma_{RP})}{\cos\sigma_R}\right)^{n-1} \, , &\mbox{if } \S_\Le \in PT \\
    \displaystyle R_R^{n-1} \, , &\mbox{if } \S_\Le \in TQ' \\
    \displaystyle \left(R_R^2\tan^2\sigma_R+R_B^2\right)^{\frac{n-1}{2}} \, , &\mbox{if } \S_\Le \in Q'Q \\
    \displaystyle R_L^{n-1} \, , &\mbox{if } \S_\Le \in QA'
    \end{array}\right. ,
\end{equation}
\renewcommand{\arraystretch}{1}where $\theta_{RP}$ and $\sigma_{RP}$ are given in \Eqs{eq:rp_rq_coordinates_+-}. The trajectory of $\S_\Ri$ is divided into three segments, along which the area of the screen is given by:
\renewcommand{\arraystretch}{1.4}
\begin{equation}\label{eq:area_S_R_+-}
{\rm Area}(\S_\Ri) = \omega_{n-1}\times   
\left\{\begin{array}{ll}
    \displaystyle R_R^{n-1} \, , &\mbox{if } \S_\Ri \in BQ'  \\
    \displaystyle \left(R_R^2\tan^2\sigma_R+R_B^2\right)^{\frac{n-1}{2}} \, , &\mbox{if } \S_\Ri \in Q'P' \\
    \displaystyle R_R^{n-1} \, , &\mbox{if } \S_\Ri \in P'B' 
    \end{array}\right. .
\end{equation}
\renewcommand{\arraystretch}{1}

The minimal extremal surface $\chi_\Le$ in $\J_\Le$ homologous to $A_\Le=\S_\Le$ is the empty set, $\chi_\Le=\varnothing$. Since $A_\Ri=\varnothing$, we also obtain $\chi_\Ri=\varnothing$. So, the leading geometrical contributions to the von Neumann entropy of $\S_\Le$ from both the left and right interior regions are vanishing. The part of the entanglement wedge of $\S_\Le$ in the left interior region is the full causal diamond $\J_\Le$. The entanglement wedge does not have a part in the antipode interior region.

The geometrical contributions to the von Neumann entropy  arise from the exterior region. The locations of the classical minimal extremal homologous surfaces $\chi_\E$ depend on the location of $\S_\Le$ along its trajectory:
\begin{itemize}
\item When $\S_\Le$ is on the lightlike line segments $AP$ and $PT$, along the past cosmological horizon of the pode, the situation is analogous to the $(\varepsilon_L,\varepsilon_R)=(+1,+1)$ bubble geometries described in \Sect{sect:single_screen_++}. There is a degeneracy of classical extremal homologous surfaces $\chi_\E$ which lies on the line segment $\S_\Ri T$ of the past cosmological horizon of the antipode. See \Fig{fig:screen_trajectories_+-}. The area of $\chi_\E$ is given by $\omega_{n-1}R_R^{n-1}$, irrespective of the location of $\S_\Le$ on $AP$ or $PT$. We expect the classical degeneracy to be lifted by quantum corrections, with the location of the quantum extremal homologous surface determining the precise extent of the entanglement wedge in the exterior region.
\item $\S_\Le$ and $\S_\Ri$ coincide on the lightlike line segment $TQ'$. The exterior causal diamond region $\J_\E$ amounts to a point, and therefore, $\chi_\E=\S_\Le=\S_\Ri$ with area $\omega_{n-1}R_R^{n-1}$.
\item The two screens coincide also on the timelike segment $Q'P'$ of the domain wall trajectory. Since the exterior causal diamond region $\J_\E$ reduces to a point, $\chi_\E=\S_\Le=\S_\Ri$ with area given by $\omega_{n-1}\left(R_R^2\tan^2\sigma_R+R_B^2\right)^{\frac{n-1}{2}}$, see \Eq{eq:area_domain_wall}. The area decreases from $\sigma_{RQ'}$ until $\sigma_R=0$, and then increases again up to $\sigma_{RP'}$. 
%In both situations when $\S_\Le$ and $\S_\Ri$ coincide along $TQ'$ and $Q'P'$, $\Sigma = \Sigma_\Le \cup \Sigma_\Ri$ and, so, the Bousso covariant entropy bound ensures that the two-screen system has enough degrees of freedom to describe the full Cauchy slice $\Sigma$. 
\item When $\S_\Le$ lies on the timelike segment $P'Q$ of the domain wall trajectory, $\S_\Ri$ is on the lightlike line segment $P'B'$ of the future cosmological horizon of the antipode. There is a degeneracy of classical minimal extremal surfaces $\chi_\E$ along the lightlike line segment $P'\S_\Ri$ of the future cosmological horizon of the antipode. The area of these extremal surfaces is equal to $\omega_{n-1}R_R^{n-1}$. Similar conclusions hold when $\S_\Le$ is on the lightlike line segment $QA'$ of the future cosmological horizon of the pode.
\end{itemize}
Gathering these results, we find that the classical, geometrical contributions to the von Neumann entropy of $\S_\Le$ are given by:
\renewcommand{\arraystretch}{1.4}
\begin{equation}\label{eq:vN_entropy_SL_+-}
S(\S_\Le) =   
\left\{\begin{array}{ll}
    \displaystyle \frac{\omega_{n-1}R_R^{n-1}}{4G\hbar}+\mathcal{O}(G\hbar)^0 \, , &\mbox{if } \S_\Le \in AT,~ \S_\Le \in TQ'  \\
    \displaystyle \frac{\omega_{n-1}\left(R_R^2\tan^2\sigma_R+R_B^2\right)^{\frac{n-1}{2}}}{4G\hbar}+ \mathcal{O}((G\hbar)^0)\, , &\mbox{if } \S_\Le \in Q'P' \\
    \displaystyle \frac{\omega_{n-1}R_R^{n-1}}{4G\hbar}+\mathcal{O}(G\hbar)^0 \, , &\mbox{if } \S_\Le \in P' Q,~ \S_\Le \in QA'  
    \end{array}\right. .
\end{equation}
\renewcommand{\arraystretch}{1}It is easy to verify that the leading geometrical entropy of $\S_\Ri$ is also given by this equation. A difference from the $(\varepsilon_L,\varepsilon_R)=(+1,+1)$ case is that now the entanglement entropy of the single-screen system is time-dependent, when $\S_\Le$ is on the segment $Q'P'$ of the domain wall trajectory. Using the expressions of $\sigma_{RQ'}$ and $\sigma_{RP'}$ given in \Eq{eq:rp'_rq'_coordinates_+-}, one sees that the entropy $S(\S_\Le)$ is continuous at the points $Q'$ and $P'$, and thus throughout the trajectory of $\S_\Le$.

\begin{figure}[h!]
        \centering
	\includegraphics[height=70mm]{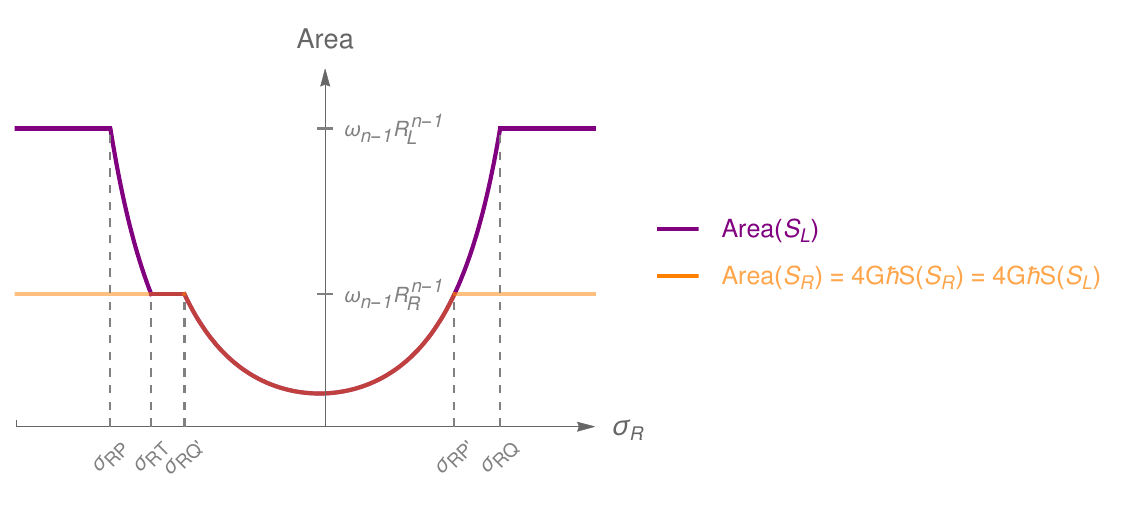}
 \caption{\footnotesize Evolution of the area and entanglement entropy of $\S_\Le$ and $\S_\Ri$ as a function of the conformal time $\sigma_R$, in the bubble geometries with $(\varepsilon_L,\varepsilon_R)=(+1,-1)$.} \label{fig:plot_area_+-}
\end{figure}

The evolution of the area of $\S_\Le$ (\Eq{eq:area_S_L_+-}) and $\S_\Ri$ (\Eq{eq:area_S_R_+-}), as well as the entanglement entropy $S(\S_\Le)$ and $S(\S_\Ri)$, are plotted as functions of the conformal time $\sigma_R$ in \Fig{fig:plot_area_+-}.

\section{Bubble geometry with $(\varepsilon_L,\varepsilon_R)=(-1,-1)$}
\label{sect:--_geometry}
We now proceed to discuss the de Sitter bubble geometries with $(\varepsilon_L,\varepsilon_R)=(-1,-1)$, for which the domain wall has negative tension. As we mentioned in \Sect{sect:lorentzian_geometry}, in these cases, the polar coordinates $\theta_L$ and $\theta_R$ along the trajectories of the left and right domain walls are larger than $\pi/2$. As a result, the causal patch of the pode observer is fully contained inside the de Sitter bubble (left) region. In other words, this observer is not aware of a ``parent'' de Sitter region or the bubble wall that separates the two regions. The causal patch of the antipode observer extends in both de Sitter regions, but as depicted in the Penrose diagrams of \Fig{fig:Penrose_diag_--},
\begin{figure}[!h]
 \begin{subfigure}[t]{0.48\linewidth}
    \centering
\begin{tikzpicture}
%\begin{scope}[transparency group]
%\begin{scope}[blend mode=multiply]

%wall
\draw[name path=B, line width= 1 pt,brown] plot[variable=\t,samples=91,domain=-45:45] ({-4.3*sec(\t)+6.1},{3*tan(\t)});
\draw[name path=A, line width= 1 pt,brown] plot[variable=\t,samples=91,domain=-45:45] ({-3.5*sec(\t)+4.95},{3*tan(\t)});
\tikzfillbetween[of=A and B]{gray, opacity=0.3}; 

%\draw[dashed] (0,3)--(0,-3);

\draw[name path=C](0,-3) -- (-3,-3) -- (-3,3) -- (0,3);
\draw[name path=D](0,-3) -- (3,-3) -- (3,3) -- (0,3);

\tikzfillbetween[of=A and C]{blue, opacity=0.1};
\tikzfillbetween[of=B and D]{red, opacity=0.1};

\draw[dashed] (-3,3)--(0,0)--(-3,-3);
\draw[dashed] (3,3)--(1.4,1.4);
\draw[dashed] (3,-3)--(1.4,-1.4);
\draw[dashed] (1.35,-0.6)--(0.75,0)--(1.35,0.6);
\draw[dotted] (0,0)--(1.15,1.15);
\draw[dotted] (0,0)--(1.15,-1.15);

%Bousso's wedges
\draw (1.3,2.3) -- (1.5,2.5) -- (1.7,2.3);
\draw (-0.2,1.5) -- (0.0,1.7) -- (0.2,1.5);
\draw (-0.2,-1.5) -- (0,-1.7) -- (0.2,-1.5);
\draw (-1.7,0.2) -- (-1.5,0) -- (-1.7,-0.2);
%\draw (-0.6,0.2) -- (-0.4,0) -- (-0.6,-0.2);
%\draw (-1.5,2.2) -- (-1.3,2.4) -- (-1.1,2.2);
%\draw (-1.5,-2.2) -- (-1.3,-2.4) -- (-1.1,-2.2);
\draw (2.5,0.2) -- (2.3,0) -- (2.5,-0.2);
\draw (1.3,0.2) -- (1.1,0) -- (1.3,-0.2);
\draw (0.6,0.2) -- (0.4,0) -- (0.6,-0.2);
\draw (1.3,-2.3) -- (1.5,-2.5) -- (1.7,-2.3);

\node at (-3,-3) [label=left:$A$]{};
\node at (-3,3) [label=left:$A'$]{};
\node at (3,-3) [label=right:$B$]{};
\node at (3,3) [label=right:$B'$]{};

\node at (1.38,1.38) [circle,fill,inner sep=1.5pt,label=right:$P'$]{};
\node at (1.35,0.6) [circle,fill,inner sep=1.5pt,label=right:$P'$]{};
\node at (1.38,-1.38) [circle,fill,inner sep=1.5pt,label=right:$Q'$]{};
\node at (1.35,-0.6) [circle,fill,inner sep=1.5pt,label=right:$Q'$]{};

\node at (0,0) [circle,fill,inner sep=1.5pt, label = above:$O$]{};
\node at (0.75,0) [circle,fill,inner sep=1.5pt]{};
\node at (0.76,0) [label = above:$O'$]{};

\node at (1.18,1.18) [circle,fill,inner sep=1.5pt,label=left:$P$]{};
\node at (1.18,-1.18) [circle,fill,inner sep=1.5pt,label=left:$Q$]{};

%axes
\node at (-3, 0) [label={[rotate=90]$\theta_L = 0$}] {};
\node at (3.6, 0) [label={[rotate=90]$\theta_R = \pi$}] {};
\node at (0,-3.75) [label={$\sigma_L=\sigma_R=-\pi/2$}] {};
\node at (0,2.75) [label={$\sigma_L=\sigma_R=\pi/2$}] {};
%\node at (1.7, 1.5) [label={[rotate=45] \scriptsize $\theta_R-\sigma_R = \pi/2$}] {};
%\node at (-2,1.8) [label={[rotate=-45]\scriptsize $\theta_L+\sigma_L = \pi/2$}] {};
%\node at (-1.6, -2.4) [label={[rotate=45]\scriptsize $ \theta_L-\sigma_L = \pi/2$}] {};
%\node at (1.6, -1.8) [label={[rotate=-45]\scriptsize $ \theta_R+\sigma_R = \pi/2$}] {};

%\end{scope}
%\end{scope}
\end{tikzpicture}
    \caption{\footnotesize Two different conformal coordinate systems $(\theta_L,\sigma_L)$ and $(\theta_R,\sigma_R)$ covering, respectively, the blue and red de Sitter regions (with different cosmological constants). Each region is bounded by a domain wall trajectory (brown lines). The gray shaded region is not a part of the spacetime. Points on the left and right domain wall trajectories are identified appropriately.} \label{fig:Penrose_diag_--_disconnected}
    \end{subfigure}
\quad \
\begin{subfigure}[t]{0.48\linewidth}
 \centering
 \raisebox{0.65cm}{
\begin{tikzpicture}
\begin{scope}[transparency group]
\begin{scope}[blend mode=multiply]
\path
       +(-3,1.5) coordinate (IItophat)
       +(0,3) coordinate (IItophatright)
       +(3,3)  coordinate (IIsquaretopright)
       +(3,-3) coordinate (IIsquarebotright)
       +(0,-3) coordinate (IIbothatright)
       +(-3,-1.5) coordinate (IIbothat)
       +(-3,3) coordinate (IIsquaretopleft)
       +(-3,-3) coordinate(IIsquarebotleft)
      
       ;
%\draw (IItophat) to
%          [bend left=20] (IItophatright) --
%          (IIsquaretopright) --       (IIsquarebotright) --
%          (IIbothatright) to 
%          [bend left=20] (IIbothat)--
%          (IItophat)--cycle;

\draw[name path=A,line width= 1 pt,brown] plot[variable=\t,samples=90,domain=-44.8:44.8] ({-3*sec(\t)+4.23},{3.02*tan(\t)});

\draw[name path=C](0,-3) to [bend right=20] (-3,-1.5) -- (-3,1.5) to [bend right=20] (0,3);
\draw[name path=D](0,-3) -- (3,-3) -- (3,3) -- (0,3);

\tikzfillbetween[of=A and C]{blue, opacity=0.1};
\tikzfillbetween[of=A and D]{red, opacity=0.1};

%\draw[dotted] (IItophatright) -- (3,0);
%\draw[dotted] (IIbothatright) -- (3,0);
%\draw[dotted] (IItophatright) -- (-3,0);
%\draw[dotted] (IIbothatright) -- (-3,0);

\draw[dashed] (IIsquaretopright) -- (0,0) -- (IIsquarebotright);
\draw[dashed] (IItophat) -- (-1.5,0) -- (IIbothat);
\draw[dotted] (0.55,-2.05) -- (-1.5,0) -- (0.55,2.05);

%\draw (0,-3) to[bend left=44] (0,3);

\draw (1.3,2.3) -- (1.5,2.5) -- (1.7,2.3);
\draw (-2.45,0.2) -- (-2.25,0) -- (-2.45,-0.2);
\draw (-0.55,0.2) -- (-0.75,0) -- (-0.55,-0.2);
\draw (-1.7,1) -- (-1.5,1.2) -- (-1.3,1);
\draw (-1.7,-1) -- (-1.5,-1.2) -- (-1.3,-1);
\draw (2.2,0.2) -- (2,0) -- (2.2,-0.2);
\draw (0.85,0.2) -- (0.65,0) -- (0.85,-0.2);
\draw (1.3,-2.3) -- (1.5,-2.5) -- (1.7,-2.3);

\node at (-3,-1.5) [label=left:$A$]{};
\node at (-3,1.5) [label=left:$A'$]{};
\node at (3,-3) [label=right:$B$]{};
\node at (3,3) [label=right:$B'$]{};

\node at (1.05,1.05) [circle,fill,inner sep=1.5pt,label=right:$P'$]{};
\node at (1.05,-1.05) [circle,fill,inner sep=1.5pt,label=right:$Q'$]{};

\node at (-1.5,0) [circle,fill,inner sep=1.5pt, label = above:$O$]{};
\node at (0,0) [circle,fill,inner sep=1.5pt, label = above:$O'$]{};

\node at (0.58,2.08) [circle,fill,inner sep=1.5pt,label=left:$P$]{};
\node at (0.58,-2.08) [circle,fill,inner sep=1.5pt,label=left:$Q$]{};

\end{scope}
\end{scope}
\end{tikzpicture}}
    \caption{\footnotesize A single global conformal coordinate system $(\theta_G,\sigma_G)$ covering the whole spacetime. The brown curve depicts the domain wall trajectory.}
\label{fig:Penrose_diag_--_connected}
\end{subfigure}
\caption{\footnotesize The Penrose diagrams for the $(\varepsilon_L,\varepsilon_R)=(-1,-1)$ bubble geometry. The blue (red) region is the interior (exterior) of the bubble, which is a part of dS$_{n+1}$ with a smaller (larger) cosmological constant. Appropriate boundary conditions enforce the identification of the points on the left and right domain wall trajectories bounding the two de Sitter regions. The  causal patches of the pode and antipode observers are bounded by future and past cosmological horizons, depicted by the thick dashed lines. Bousso wedges are drawn in each region of the diagrams.
}
\label{fig:Penrose_diag_--}
\end{figure}
it has no overlap with the causal patch of the pode observer. 

In the right de Sitter region, the future and past cosmological horizons delimiting the causal patch of the antipode observer are given by the null lines of \eqref{eq:right_horizons_right}, as described in \Sect{sect:++_+-_geometry} for the $(\varepsilon_L,\varepsilon_R)=(+1,-1)$ cases. The right  conformal coordinates of the endpoints $P'$ and $Q'$ of these horizons on the trajectory of the domain wall are given by Eq.~(\ref{eq:rp'_rq'_coordinates_+-}). The left conformal coordinates are given by
\begin{equation}
\theta_{LP'}=\theta_{LQ'}=\pi-\arctan\frac{R_R}{\sqrt{R_L^2-R_B^2}}, \quad \sigma_{LP'}=-\sigma_{LQ'}=\arctan\frac{\sqrt{R_R^2-R_B^2}}{R_L}.
\label{}
\end{equation}
The segments of the antipode horizons in the left de Sitter region are described by Eq.~\eqref{right_horizon_left}. In this case, however, the coordinates of the bifurcation point $O'$ are given by 
\begin{equation}
\theta_{LO'}=\pi-\arctan\frac{R_L R_R+\sqrt{R_R^2-R_B^2}\sqrt{R_L^2-R_B^2}}{R_L\sqrt{R_L^2-R_B^2}-R_R\sqrt{R_R^2-R_B^2}}, \,\,\,\,\, \sigma_{LO'}=0. 
\end{equation}
Thus, the area of the horizon sphere at $O'$ is given by
\begin{equation}
{\rm Area}(O')=\omega_{n-1}R_L^{n-1}\left(\frac{R_L R_R+\sqrt{R_R^2-R_B^2}\sqrt{R_L^2-R_B^2}}{R_R^2+R_L^2-R_B^2}\right)^{n-1}.
\end{equation}

The future and past cosmological horizons of the observer at the pode are described, respectively, by the equations:
\begin{equation}
 \left\{\!\begin{array}{ll}
\dis \sigma_L+\theta_L=\frac{\pi}{2}\, , &\mbox{if } \dis 0\le \sigma_L <\frac{\pi}{2}\, , \\
\dis \sigma_L-\theta_L=-\frac{\pi}{2}\, , &\mbox{if } \dis -\frac{\pi}{2}<\sigma_L \le 0\, .\esp
\end{array}\right.
\end{equation}
The bifurcation point $O$ occurs at $(\sigma_L,\theta_L)=(0,\pi/2)$ with ${\rm Area}(O)=\omega_{n-1}R_L^{n-1}$. The bubble wall surrounds the bifurcation horizon at $O$, and its beyond reach of the observer at the pode.

The causal patches of the pode and antipode observers are non-overlapping and causally disconnected, in sharp contrast with the bubble geometries with $(\varepsilon_L,\varepsilon_R)=(+1,\pm 1)$ described in \Sect{sect:++_+-_geometry}. A connected Penrose diagram can be built from the conformal transformation described in \Appendix{app:global_coord}, as depicted on \Fig{fig:Penrose_diag_--_connected}. The area of the bubble wall at $\sigma_L=\sigma_R=0$, given by $\omega_{n-1}R_B^{n-1}$, is smaller than the area of both bifurcate horizons at $O'$ and $O$. This is thus an example of a subhorizon bubble.

\begin{figure}[h!]
%\begin{subfigure}[t]{0.48\linewidth}
    \centering
\begin{tikzpicture}
\begin{scope}[transparency group]
\begin{scope}[blend mode=multiply]
\path
       +(-3,1.5) coordinate (IItophat)
       +(0,3) coordinate (IItophatright)
       +(3,3)  coordinate (IIsquaretopright)
       +(3,-3) coordinate (IIsquarebotright)
       +(0,-3) coordinate (IIbothatright)
       +(-3,-1.5) coordinate (IIbothat)
       +(-3,3) coordinate (IIsquaretopleft)
       +(-3,-3) coordinate(IIsquarebotleft)
      
       ;
%\draw (IItophat) to
%          [bend left=20] (IItophatright) --
%          (IIsquaretopright) --       (IIsquarebotright) --
%          (IIbothatright) to 
%          [bend left=20] (IIbothat)--
%          (IItophat)--cycle;

\draw[name path=A,line width= 1 pt,brown] plot[variable=\t,samples=90,domain=-44.8:44.8] ({-3*sec(\t)+4.23},{3.02*tan(\t)});

\draw[name path=C](0,-3) to [bend right=20] (-3,-1.5) -- (-3,1.5) to [bend right=20] (0,3);
\draw[name path=D](0,-3) -- (3,-3) -- (3,3) -- (0,3);

%\tikzfillbetween[of=A and C]{blue, opacity=0.1};
%\tikzfillbetween[of=A and D]{red, opacity=0.1};

%\draw[dotted] (IItophatright) -- (3,0);
%\draw[dotted] (IIbothatright) -- (3,0);
%\draw[dotted] (IItophatright) -- (-3,0);
%\draw[dotted] (IIbothatright) -- (-3,0);

%\draw[dashed] (IIsquaretopright) -- (0,0) -- (IIsquarebotright);
%\draw[dashed] (IItophat) -- (-1.5,0) -- (IIbothat);
\draw[dotted] (0.55,-2.05) -- (-1.5,0) -- (0.55,2.05);

%\draw (0,-3) to[bend left=44] (0,3);

\draw (1.3,2.3) -- (1.5,2.5) -- (1.7,2.3);
\draw (-2.45,0.2) -- (-2.25,0) -- (-2.45,-0.2);
\draw (-0.55,0.2) -- (-0.75,0) -- (-0.55,-0.2);
\draw (-1.7,1) -- (-1.5,1.2) -- (-1.3,1);
\draw (-1.7,-1) -- (-1.5,-1.2) -- (-1.3,-1);
\draw (2.2,0.2) -- (2,0) -- (2.2,-0.2);
\draw (0.85,0.2) -- (0.65,0) -- (0.85,-0.2);
\draw (1.3,-2.3) -- (1.5,-2.5) -- (1.7,-2.3);

%trajectory of left screen
\draw[violet,line width=0.8mm] (IIbothat) -- (-1.5,0) --  (IItophat);
\fill[fill=violet!20] (IIbothat) -- (IIbothat) -- (-1.5,0) --  (IItophat) -- cycle;

%draw right horizon
\draw[orange,line width=0.8mm] (IIsquaretopright) -- (0,0) -- (IIsquarebotright);
\fill[fill=orange!20] (IIsquaretopright) -- (0,0) -- (IIsquarebotright) -- cycle;

%draw cauchy slice and label on the slice
\draw[line width=0.8mm,red,smooth] plot[variable=\t,samples=100,domain=0:3] ({\t-3},{0.1*\t*cos(45*\t)-0.4});
\draw[line width=0.8mm,red, smooth] plot[variable=\t,samples=100,domain=3:4.52] ({\t-3},{0.1*\t*cos(45*\t)-0.4});
\draw[line width=0.8mm,red, smooth] plot[variable=\t,samples=100,domain=4.5:6] ({\t-3},{0.1*\t*cos(45*\t)-0.4});
\node at (-1.8,-0.33) [circle,fill,inner sep=2pt, label=below:$\S_\Le$]{};
\node at (0.8,-0.79) [circle,fill,inner sep=2pt]{};
\node at (0.7,-0.65) [label=below:$\S_\Ri$]{};
\node at (-2.7,-1.2) [label = $\color{red} \Sigma_\Le$]{};
\node at (2.3,-1.6) [label = $\color{red} \Sigma_\Ri$]{};
\node at (-0.3,-1.3) [label = $\color{red} \Sigma$]{};

%x+, x- axes
\draw[->] (-3,0)--(-2.5,0.5);
\draw[->] (-3,0)--(-3.5,0.5);
\node at (-3.1,0.7) [label = right:$x^+$]{};
\node at (-3,0.7) [label = left:$x^-$]{};

\end{scope}
\end{scope}
\end{tikzpicture}
    \caption{\footnotesize Trajectories of the left screen on $\S_\Le$ (purple) and the right screen on $\S_\Ri$ (orange) in the bubble geometries with $(\varepsilon_L,\varepsilon_R)=(-1,-1)$. The causal patches of the pode and antipode observers, which are non-overlapping, are the purple and orange shaded regions, respectively.} \label{fig:screen_trajectories_--}
\end{figure}
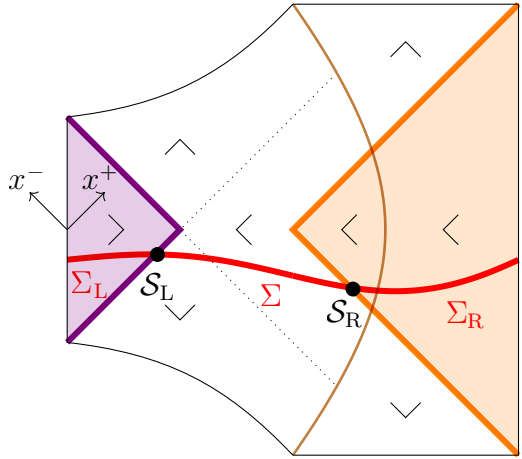

Repeating the procedure described in \Sect{sect:screen_trajectories}, we consider a foliation $\F$ of spacetime in terms of SO$(n)$-symmetric Cauchy slices $\Sigma$, and locate two spherical holographic screens at the intersections of $\Sigma$ with the cosmological horizons of the pode and the antipode, $\S_\Le$ and $\S_\Ri$, respectively. The Bousso covariant entropy bound ensures that each screen has an adequate number of degrees of freedom to holographically encode the state on $\Sigma_\Le$ and $\Sigma_\Ri$, the parts of $\Sigma$ between the pode and $\S_\Le$ and the antipode and $\S_\Ri$, respectively. See \Fig{fig:screen_trajectories_--}.
The main difference with the bubble geometries with $(\varepsilon_L,\varepsilon_R)=(+1,\pm 1)$ described in \Sect{sect:screen_trajectories} is that now the screens remain out of causal contact as they evolve, and their trajectories never intersect. Crucially, the union of the Cauchy slices $\Sigma_\Le$ and $\Sigma_\Ri$ never amounts to a complete Cauchy slice. Another sharp contrast with the $(\varepsilon_L,\varepsilon_R)=(+1,\pm 1)$ cases is the following. In the contracting phase of the cosmology, the two-screen system does not have the capacity to describe the state on any of the full Cauchy slices $\Sigma$ of the foliation $\F$. Indeed, consider, for example, the sphere $\S_\Le$, where the left screen is located, in \Fig{fig:screen_trajectories_--}. A future directed lightsheet of non-positive expansion, which runs parallel to the $x^+$ axis, emanates from it. Naively, this seems to suggest that the area of $\S_\Le$ divided by $4G\hbar$ also bounds the coarse-grained entropy on the complement of $\Sigma_\Le$, $\Sigma - \Sigma_\Le$. Notice, however, that this lightsheet terminates on the domain wall and not on the trajectory of the antipode. Therefore, it is possible for massless particles on $\Sigma - \Sigma_\Le$ to reach future null infinity without having to pass or cross this lightsheet of non-positive expansion. In other words, a spacelike projection theorem cannot be established in order to obtain an upper bound for the coarse-grained entropy on $\Sigma - \Sigma_\Le$. Therefore, there is no reason to expect that the degrees of freedom of the two-screen system can fully encode the state on any of the full Cauchy slices $\Sigma$ of these bubble geometries. Rather, each screen system provides an effective holographic description of the corresponding patch. Since more than two screens are needed to holographically describe the full bubble cosmology, we cannot apply the bilayer prescription for holographic entanglement entropy calculations in the $(\varepsilon_L,\varepsilon_R)=(-1,-1)$ cases. 

This situation is reminiscent of a particular class of Big Bang/Big Crunch closed FRW cosmologies, with perfect fluid index in the range $4/n -1< w \le 1$ (where $n$ is the number of spatial dimensions). For these closed FRW cosmologies, the Penrose diagrams are wider than tall \cite{Franken:2023jas} and the causal patches of the pode and antipode observers are completely disconnected. For this class of cosmologies as well, the two-screen system does not have the capacity to holographically describe the full cosmology, and so the bilayer proposal cannot be applied \cite{Franken:2023jas}.

\section{Flat Minkowski bubbles}
\label{sect:flat_bubbles}
In this section, we consider an $(n+1)$-dimensional de Sitter spacetime containing a flat Minkowski bubble. Such geometries arise in the limit $R_L\rightarrow+\infty$ of the de Sitter bubble geometries with $(\varepsilon_L,\varepsilon_R)=(+1,\pm 1)$, described in \Sect{sect:++_+-_geometry}. For both cases, the tension $T$ of the domain wall is positive, given by 
\begin{equation}
    8\pi GT/(n-1)=1/R_B\mp\sqrt{R_B^{-2}-2\Lambda_R/(n(n-1))}.
\end{equation}
The $(\varepsilon_L,\varepsilon_R)=(+1, +1)$ cases are subcritical, while the $(\varepsilon_L,\varepsilon_R)=(+1,-1)$ are supercritical.

We will see that the holographic prescription of \Sect{sect:holography} can be applied straightforwardly to these spacetimes as well. In four spacetime dimensions, our construction can be linked to a conjectured duality between the Milne patch of the Minkowski bubble and a $2$-dimensional CFT on its asymptotic spatial boundary, originally proposed in \cite{Freivogel:2006xu}.

Recall that the metric of the $(n+1)$-dimensional sphere S$^{n+1}$ of radius $R_L$ is given by:
\begin{equation}
    \d s^2=R_L^2\left(\d\chi_L^2+\sin^2\chi_L (\d\theta_L^2+\sin^2\theta_L\d\Omega^2_{n-1}\right).
\end{equation}
We then define new coordinates $x_L$ and $r_L$ via 
\begin{equation}\label{eq:xL_rL}
  \chi_L=\frac{\pi}{2}+\frac{x_L}{R_L},\quad \theta_L=\frac{r_L}{R_L},  
\end{equation}
and take the limit
\begin{equation}\label{eq:flat_limit}
   R_L\rightarrow+\infty, \qquad  x_L \mbox{ and } r_L \text{ fixed}.
\end{equation}
We obtain the flat Euclidean metric in $n+1$ dimensions
\begin{equation}\label{eq:euclidean_metric}
    \d s^2=\d x_L^2+\d r_L^2+r_L^2\d\Omega^2_{n-1},
\end{equation}
in spherical coordinates.

The geometric properties of flat bubbles can be directly obtained from the finite-$R_L$ case described in \Sect{sect:bubble_geometry}, by applying the coordinate transformation \eqref{eq:xL_rL} and taking the flat space limit \eqref{eq:flat_limit}. For instance, the equation of the domain wall \eqref{eq:euclid_dom_wall} in the left coordinate system becomes 
\begin{equation}
    x_L^2+r_L^2=R_B^2, 
\end{equation}
and the matching condition \eqref{eq:euclid_matching_cond} between the left and right coordinates on the domain wall reduces to
\begin{equation}
    x_L=-R_R\cos\chi_R.
\end{equation}

Via the analytic continuation $x_L\rightarrow i\tau_L$, the Euclidean metric \eqref{eq:euclidean_metric} gives the metric of $(n+1)$-dimensional Minkowski spacetime, in spherical coordinates: 
\begin{equation}\label{eq:Mink_metric}
\d s^2=-\d \tau_L^2+\d r_L^2+r_L^2\d\Omega^2_{n-1}.
\end{equation}
The equation of the domain wall trajectory reads
\begin{equation}\label{eq:domain_wall_eq}
-\tau_L^2 + r_L^2 = R_B^2,
\end{equation}
while the matching condition between the left and right coordinate systems along the domain wall is given by
\begin{equation}
\tau_L = R_R \sinh \tau_R.
\end{equation}

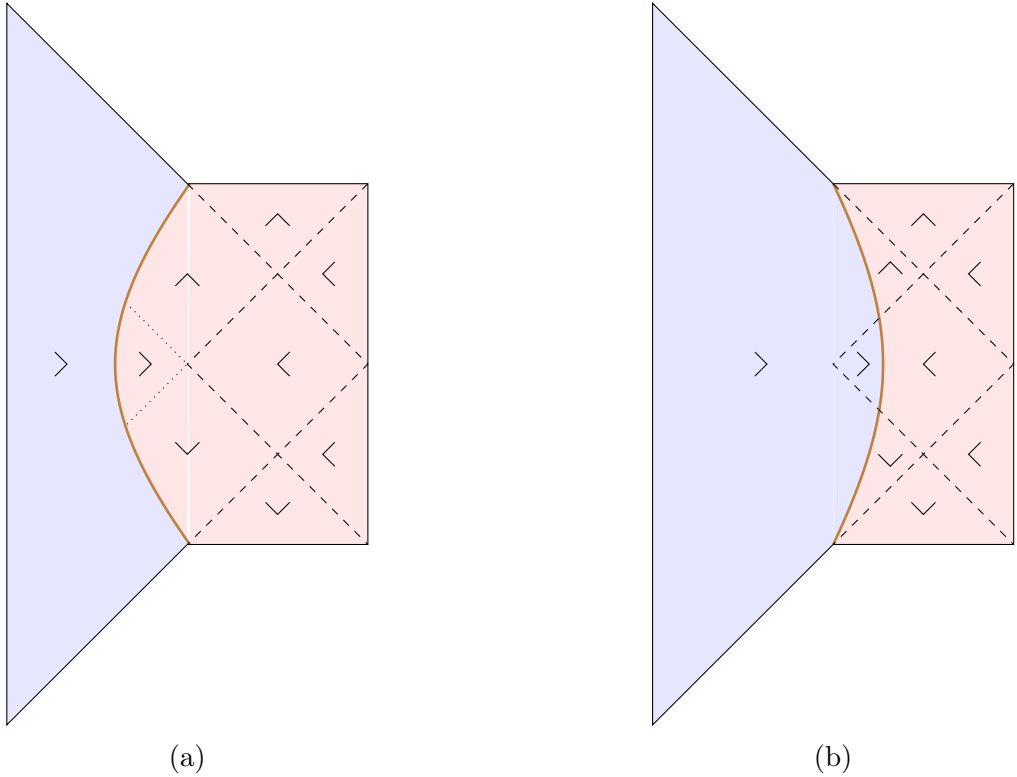
\begin{figure}[h!]
    %\centering
    \begin{subfigure}[t]{0.48\linewidth}
\centering
\begin{tikzpicture}[scale=0.8]
\begin{scope}[transparency group]
\begin{scope}[blend mode=multiply]
\path
       +(-3,6) coordinate (IItophat)
       +(0,3) coordinate (IItophatright)
       +(3,3)  coordinate (IIsquaretopright)
       +(3,-3) coordinate (IIsquarebotright)
       +(0,-3) coordinate (IIbothatright)
       +(-3,-6) coordinate (IIbothat)
       +(-3,3) coordinate (IIsquaretopleft)
       +(-3,-3) coordinate(IIsquarebotleft)
      
       ;
%\draw (IItophat) -- (IItophatright) --
%          (IIsquaretopright) --       (IIsquarebotright) --
%          (IIbothatright) -- (IIbothat)--
%          (IItophat)--cycle;

\draw[name path=A,line width= 1 pt,brown] plot[variable=\t,samples=91,domain=-45:45] ({3*sec(\t)-4.24},{3*tan(\t)});

\draw[name path=C](0,-3) -- (-3,-6) -- (-3,6) -- (0,3);
\draw[name path=D](0,-3) -- (3,-3) -- (3,3) -- (0,3);
%\draw[name path=E,red, opacity=0.1](0,-3) -- (0,3);

\tikzfillbetween[of=A and C]{blue, opacity=0.1};
\tikzfillbetween[of=A and D]{red, opacity=0.1};
%\tikzfillbetween[of=D and E]{red, opacity=0.1};

\draw[dashed] (IIsquaretopright) -- (0,0) -- (IIsquarebotright);
\draw[dashed] (IItophatright) -- (3,0) -- (IIbothatright);
\draw[dotted] (-1,-1) -- (0,0) -- (-1,1);

%\draw (-2.2,4+0.2) -- (-2,4) -- (-2.2,4-0.2);
\draw (1.5-0.2,2.3) -- (1.5,2.5) -- (1.5+0.2,2.3);
\draw (-0.2,1.5-0.2) -- (0,1.5) -- (0.2,1.5-0.2);
\draw (-2.2,0.2) -- (-2,0) -- (-2.2,-0.2);
\draw (-0.8,0.2) -- (-0.6,0) -- (-0.8,-0.2);
%\draw (2.2,0.2) -- (2,0) -- (2.2,-0.2);
\draw (1.5+0.2,0.2) -- (1.5,0) -- (1.5+0.2,-0.2);
\draw (1.5-0.2,-2.3) -- (1.5,-2.5) -- (1.5+0.2,-2.3);
\draw (-0.2,-1.5+0.2) -- (0,-1.5) -- (0.2,-1.5+0.2);

\draw (2.25+0.2,1.5-0.2) -- (2.25,1.5) -- (2.25+0.2,1.5+0.2);

\draw (2.25+0.2,-1.5-0.2) -- (2.25,-1.5) -- (2.25+0.2,-1.5+0.2);

%\draw (-2.2,-4+0.2) -- (-2,-4) -- (-2.2,-4-0.2);

%draw cauchy slice and label on the slice
%\draw[line width=0.8mm,red,smooth] plot[variable=\t,samples=100,domain=0:1.6] ({\t-3},{0.1*\t*cos(45*\t)-1.5});
%\draw[red, smooth] plot[variable=\t,samples=100,domain=1.6:6] ({\t-3},{0.1*\t*cos(45*\t)-1.5});
%\node at (-1.4,-1.45) [circle,fill,inner sep=1.5pt, label=below:$\S_1$]{};
%\node at (-2.3,-1.6) [label = $\color{red} \Sigma_1$]{};
%\node at (0.4,-1.95) [label = $\color{red} \Sigma$]{};

\end{scope}
\end{scope}
\end{tikzpicture}
\caption{ }
    \label{fig:Penrose_diag_Mink_+}
    \end{subfigure}
\quad \,
%%%%%%%%%%%%%%%%%%%%%%%%%%%%%%%%%%%%%%%%%%%%%%%%%%%%%%%%%%%%%%
\begin{subfigure}[t]{0.48\linewidth}
\centering
\begin{tikzpicture}[scale=0.8]
\begin{scope}[transparency group]
\begin{scope}[blend mode=multiply]
\path
       +(-3,6) coordinate (IItophat)
       +(0,3) coordinate (IItophatright)
       +(3,3)  coordinate (IIsquaretopright)
       +(3,-3) coordinate (IIsquarebotright)
       +(0,-3) coordinate (IIbothatright)
       +(-3,-6) coordinate (IIbothat)
       +(-3,3) coordinate (IIsquaretopleft)
       +(-3,-3) coordinate(IIsquarebotleft)
      
       ;
%\draw (IItophat) -- (IItophatright) --
 %         (IIsquaretopright) --       (IIsquarebotright) --
 %         (IIbothatright) -- (IIbothat)--
 %         (IItophat)--cycle;

\draw[name path=A, line width= 1 pt,brown] plot[variable=\t,samples=91,domain=-45:45] ({-2*sec(\t)+2.83},{3*tan(\t)});

\draw[name path=C](IIbothatright) -- (IIbothat) -- (IItophat) -- (IItophatright);
\draw[name path=D](0,-3) -- (3,-3) -- (3,3) -- (0,3);

\tikzfillbetween[of=A and C]{blue, opacity=0.1};
\tikzfillbetween[of=A and D]{red, opacity=0.1};

\draw[dashed] (IIsquaretopright) -- (0,0) -- (IIsquarebotright);
\draw[dashed] (IItophatright) -- (3,0) -- (0,-3);

%\draw[line width= 1 pt,brown] plot[variable=\t,samples=91,domain=-45:45] ({3*sec(\t)-4.2},{3*tan(\t)});

%\draw (-2.2,4+0.2) -- (-2,4) -- (-2.2,4-0.2);
\draw (1.5-0.2,2.3) -- (1.5,2.5) -- (1.5+0.2,2.3);
\draw (0.75,1.5) -- (0.95,1.7) -- (1.15,1.5);
\draw (-1.3,0.2) -- (-1.1,0) -- (-1.3,-0.2);
\draw (0.4,0.2) -- (0.6,0) -- (0.4,-0.2);
\draw (1.5+0.2,0.2) -- (1.5,0) -- (1.5+0.2,-0.2);
\draw (1.5-0.2,-2.3) -- (1.5,-2.5) -- (1.5+0.2,-2.3);
\draw (0.75,-1.5) -- (0.95,-1.7) -- (1.15,-1.5);
\draw (2.25+0.2,1.5-0.2) -- (2.25,1.5) -- (2.25+0.2,1.5+0.2);
\draw (2.25+0.2,-1.5-0.2) -- (2.25,-1.5) -- (2.25+0.2,-1.5+0.2);

%draw cauchy slice and label on the slice
%\draw[line width=0.8mm,red,smooth] plot[variable=\t,samples=100,domain=0:1.6] ({\t-3},{0.1*\t*cos(45*\t)-1.5});
%\draw[red, smooth] plot[variable=\t,samples=100,domain=1.6:6] ({\t-3},{0.1*\t*cos(45*\t)-1.5});
%\node at (-1.4,-1.45) [circle,fill,inner sep=1.5pt, label=below:$\S_1$]{};
%\node at (-2.3,-1.6) [label = $\color{red} \Sigma_1$]{};
%\node at (0.4,-1.95) [label = $\color{red} \Sigma$]{};

\end{scope}
\end{scope}

\end{tikzpicture}
\caption{}  \label{fig:Penrose_diag_Mink_-}
\end{subfigure}
    \caption{\footnotesize The Penrose diagrams for a de Sitter spacetime containing a Minkowski bubble. The blue region is the interior of the bubble with vanishing cosmological constant $\Lambda_L=0$, and the red region is part of dS$_{n+1}$ with cosmological constant $\Lambda_R>0$. The two regions are separated by a thin domain wall trajectory (brown line). The left and right vertical edges correspond to the worldlines of an observer sitting at the center of the bubble, and at the pole of the de Sitter spherical cap, respectively. The causal patches are bounded by future and past cosmological horizons depicted by the thick dashed lines. The observer in the Minkowski bubble is always causally aware of the ``parent'' de Sitter space. However, the causal patch of the de Sitter observer can either be causally disconnected from the Minkowski region (case $(a)$), or overlapping with it (case $(b)$). Bousso wedges are drawn in each region of the diagrams.}
    \label{fig:Penrose_diag_Mink}
\end{figure}

Performing the conformal transformation 
\begin{equation}\label{eq:conformal_transfo}
 \left\{\!\begin{array}{ll}
\dis \tau_L+r_L=\tan\dfrac{\tilde\sigma_L+\tilde\theta_L}{2}\\
\dis \tau_L-r_L=\tan\dfrac{\tilde\sigma_L-\tilde\theta_L}{2}
\end{array}\right.
\quad
\Longleftrightarrow
\quad 
\left\{\!\begin{array}{ll}
\dis \tau_L=\dfrac{\sin\tilde\sigma_L}{\cos\tilde\sigma_L+\cos\tilde\theta_L} \\[2ex]
\dis r_L=\dfrac{\sin\tilde\theta_L}{\cos\tilde\sigma_L+\cos\tilde\theta_L}
\end{array}\right. ,
\end{equation}
Eq.~\eqref{eq:Mink_metric} takes the following form:
\begin{equation}
\d s^2=\frac{1}{(\cos\tilde\sigma_L+\cos\tilde\theta_L)^2}\left(-\d \tilde\sigma_L^2+\d \tilde\theta_L^2+\sin^2 \tilde\theta_L~\d\Omega^2_{n-1}\right).
\end{equation}
In the absence of the domain wall, the new coordinates take values in a finite range: $0\leq \tilde\theta_L<\pi$, $-\pi+\tilde\theta_L<\tilde\sigma_L<\pi-\tilde\theta_L$. In terms of $(\tilde\theta_L,\tilde\sigma_L)$, Eq.~\eqref{eq:domain_wall_eq} can be written as\footnote{We use the fact that $\dis R_B=r_L(\tilde\sigma_L=0)=\frac{\sin\tilde\theta_{L0}}{1+\cos\tilde\theta_{L0}}$.}:
\begin{equation}
    \cos\tilde\theta_L=\cos\tilde\theta_{L0} \cos\tilde\sigma_L,
\end{equation}
where $\tilde\theta_{L0}$ is the domain wall polar angle at $\tilde\sigma_L=0$. Thus, depending on whether $\tilde\theta_{L0}$ lies in $[0,\pi/2)$ or $(\pi/2,\pi]$, the coordinate $\tilde\theta_L$ along the domain wall trajectory is always smaller or larger than $\pi/2$. The two classes of geometries are depicted in the Penrose diagrams of \Figs{fig:Penrose_diag_Mink_+} and~\ref{fig:Penrose_diag_Mink_-}, respectively.
Physically, the two cases correspond to whether an observer located at the pole of the de Sitter spherical cap has causal access to a part of the Minkowski bubble. From the point of view of the right de Sitter region, the equation of the domain wall in the right conformal coordinate system $(\theta_R,\sigma_R)$ is still given by:
\begin{equation}
    \cos\theta_{R}=\varepsilon_R\sqrt{1-\frac{R_B^2}{R^2_R}}\cos \sigma_R,
\end{equation}
with $\varepsilon_R=\pm 1$. 

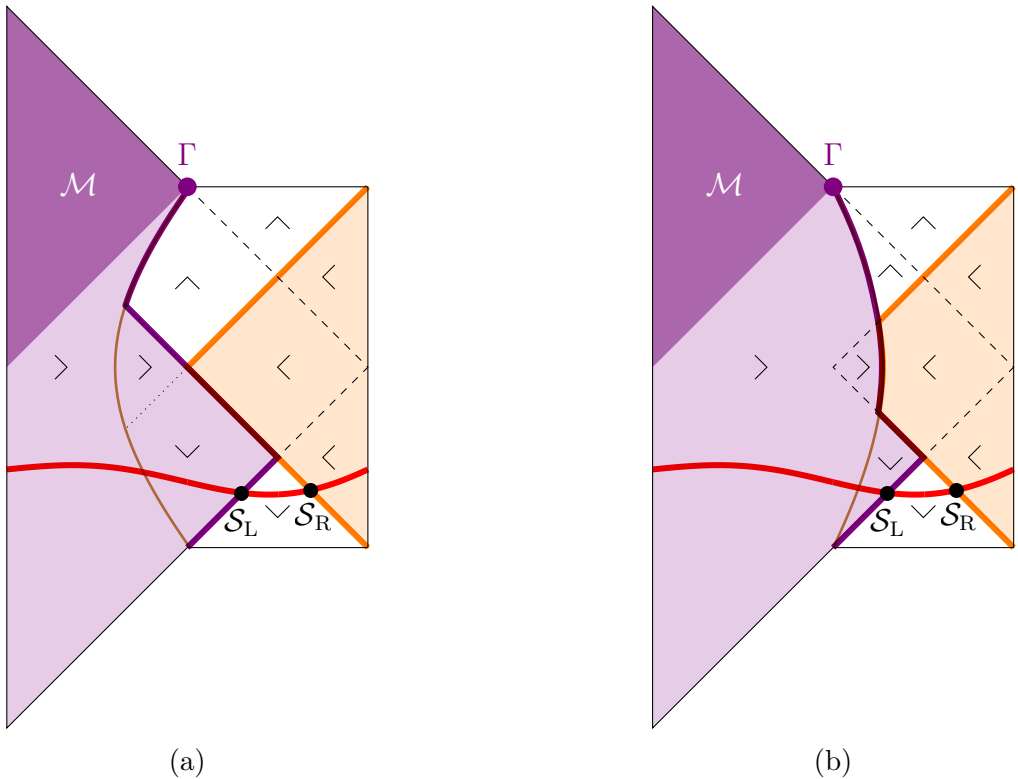
\begin{figure}[h!]
    %\centering
    \begin{subfigure}[t]{0.48\linewidth}
\centering
\begin{tikzpicture}[scale=0.8]
\begin{scope}[transparency group]
\begin{scope}[blend mode=multiply]
\path
       +(-3,6) coordinate (IItophat)
       +(0,3) coordinate (IItophatright)
       +(3,3)  coordinate (IIsquaretopright)
       +(3,-3) coordinate (IIsquarebotright)
       +(0,-3) coordinate (IIbothatright)
       +(-3,-6) coordinate (IIbothat)
       +(-3,3) coordinate (IIsquaretopleft)
       +(-3,-3) coordinate(IIsquarebotleft)
      
       ;
\draw (IItophat) -- (IItophatright) --
          (IIsquaretopright) --       (IIsquarebotright) --
          (IIbothatright) -- (IIbothat)--
          (IItophat)--cycle;

%\draw[dotted] (IItophatright) -- (3,0);
%\draw[dotted] (IIbothatright) -- (3,0);
%\draw[dotted] (IItophatright) -- (-3,0);
%\draw[dotted] (IIbothatright) -- (-3,0);

%\draw[dashed] (IIsquaretopright) -- (0,0) -- (IIsquarebotright);
\draw[dashed] (IItophatright) -- (3,0) -- (1.5,-1.5);
\draw[dotted] (-1,-1) -- (0,0) -- (-1,1);

\draw[line width= 1 pt,brown] plot[variable=\t,samples=91,domain=-45:45] ({3*sec(\t)-4.2},{3*tan(\t)});

%\draw (-2.2,4+0.2) -- (-2,4) -- (-2.2,4-0.2);
\draw (1.5-0.2,2.3) -- (1.5,2.5) -- (1.5+0.2,2.3);
\draw (-0.2,1.5-0.2) -- (0,1.5) -- (0.2,1.5-0.2);
\draw (-2.2,0.2) -- (-2,0) -- (-2.2,-0.2);
\draw (-0.8,0.2) -- (-0.6,0) -- (-0.8,-0.2);
%\draw (2.2,0.2) -- (2,0) -- (2.2,-0.2);
\draw (1.5+0.2,0.2) -- (1.5,0) -- (1.5+0.2,-0.2);
\draw (1.5-0.2,-2.3) -- (1.5,-2.5) -- (1.5+0.2,-2.3);
\draw (-0.2,-1.5+0.2) -- (0,-1.5) -- (0.2,-1.5+0.2);

\draw (2.25+0.2,1.5-0.2) -- (2.25,1.5) -- (2.25+0.2,1.5+0.2);

\draw (2.25+0.2,-1.5-0.2) -- (2.25,-1.5) -- (2.25+0.2,-1.5+0.2);

%\draw (-2.2,-4+0.2) -- (-2,-4) -- (-2.2,-4-0.2);

\draw[violet,line width=0.8mm] (IIbothatright) -- (1.5,-1.5) --  (-1.05,1.05);
\draw[violet,line width=0.8mm] plot[variable=\t,samples=18,domain=18.7:45] ({3*sec(\t)-4.2},{3*tan(\t)});
\fill[fill=violet!20] (IIbothat) -- (1.5,-1.5) -- (-1.05,1.05)  to [bend left=8] (0,3) -- (IItophat) -- cycle;
\fill[fill=violet!50] (IItophat) -- (0,3) -- (-3,0) -- cycle;
\draw[orange,line width=0.8mm] (IIsquaretopright) -- (0,0) -- (IIsquarebotright);
\fill[fill=orange!20] (IIsquaretopright) -- (0,0) -- (IIsquarebotright) -- cycle;

%draw cauchy slice and label on the slice
\draw[line width=0.8mm,red,smooth] plot[variable=\t,samples=100,domain=0:3] ({\t-3},{0.1*\t*cos(45*\t)-1.7});
\draw[line width=0.8mm,red, smooth] plot[variable=\t,samples=100,domain=3:4.52] ({\t-3},{0.1*\t*cos(45*\t)-1.7});
\draw[line width=0.8mm,red, smooth] plot[variable=\t,samples=100,domain=4.52:6] ({\t-3},{0.1*\t*cos(45*\t)-1.7});
\node at (0.9,-2.1) [circle,fill,inner sep=2pt]{};
\node at (0.9,-2) [label=below:$\S_\Le$]{};
\node at (2.05,-2.05) [circle,fill,inner sep=2pt]{};
\node at (2.1,-1.9) [label=below:$\S_\Ri$]{};
%\node at (-1.5,1.65) [label = $\color{red} \Sigma_\Le$]{};
%\node at (2.7,1.4) [label = $\color{red} \Sigma_\Ri$]{};
%\node at (1.1,1.25) [label = $\color{red} \Sigma_\E$]{};

\end{scope}
\end{scope}

\node at (0,3) [violet, circle,fill,inner sep=2.5pt]{};
\node at (0,3) [label=above:$\color{violet} \Gamma$]{};
%\node at (0,-3) [violet, circle,fill,inner sep=2.5pt]{};
\node at (-1.8,2.5) [label=above:$\color{white} \mathcal{M}$]{};

\end{tikzpicture}
    \caption{ }
    \label{fig:screen_trajectories_Mink_+}
    \end{subfigure}
\quad \,
%%%%%%%%%%%%%%%%%%%%%%%%%%%%%%%%%%%%%%%%%%%%%%%%%%%%%%%%%%%%%%
\begin{subfigure}[t]{0.48\linewidth}
\centering
\begin{tikzpicture}[scale=0.8]
\begin{scope}[transparency group]
\begin{scope}[blend mode=multiply]
\path
       +(-3,6) coordinate (IItophat)
       +(0,3) coordinate (IItophatright)
       +(3,3)  coordinate (IIsquaretopright)
       +(3,-3) coordinate (IIsquarebotright)
       +(0,-3) coordinate (IIbothatright)
       +(-3,-6) coordinate (IIbothat)
       +(-3,3) coordinate (IIsquaretopleft)
       +(-3,-3) coordinate(IIsquarebotleft)
      
       ;
\draw (IItophat) -- (IItophatright) --
          (IIsquaretopright) --       (IIsquarebotright) --
          (IIbothatright) -- (IIbothat)--
          (IItophat)--cycle;

%\draw[dotted] (IItophatright) -- (3,0);
%\draw[dotted] (IIbothatright) -- (3,0);
%\draw[dotted] (IItophatright) -- (-3,0);
%\draw[dotted] (IIbothatright) -- (-3,0);

%\draw[dashed] (IIsquaretopright) -- (0,0) -- (IIsquarebotright);
\draw[dashed] (IItophatright) -- (3,0) -- (1.5,-1.5);
\draw[dashed] (0.8,0.8) -- (0,0) -- (0.8,-0.8);

\draw[name path=A, line width= 1 pt,brown] plot[variable=\t,samples=91,domain=-45:45] ({-2*sec(\t)+2.83},{3*tan(\t)});

%\draw[line width= 1 pt,brown] plot[variable=\t,samples=91,domain=-45:45] ({3*sec(\t)-4.2},{3*tan(\t)});

%\draw (-2.2,4+0.2) -- (-2,4) -- (-2.2,4-0.2);
\draw (1.5-0.2,2.3) -- (1.5,2.5) -- (1.5+0.2,2.3);
\draw (0.75,1.5) -- (0.95,1.7) -- (1.15,1.5);
\draw (-1.3,0.2) -- (-1.1,0) -- (-1.3,-0.2);
\draw (0.4,0.2) -- (0.6,0) -- (0.4,-0.2);
\draw (1.5+0.2,0.2) -- (1.5,0) -- (1.5+0.2,-0.2);
\draw (1.5-0.2,-2.3) -- (1.5,-2.5) -- (1.5+0.2,-2.3);
\draw (0.75,-1.5) -- (0.95,-1.7) -- (1.15,-1.5);
\draw (2.25+0.2,1.5-0.2) -- (2.25,1.5) -- (2.25+0.2,1.5+0.2);
\draw (2.25+0.2,-1.5-0.2) -- (2.25,-1.5) -- (2.25+0.2,-1.5+0.2);

%trajectory of the left screen
\draw[violet,line width=0.8mm] (0,-3) -- (1.5,-1.5) -- (0.75,-0.75) to [bend right=8] (0.75,0.75) to [bend right=8] (0,3) ;

\fill[fill=violet!20] (IIbothat) -- (1.5,-1.5) -- (0.75,-0.75) to [bend right=14] (0,3) -- (IItophat) -- cycle;

\fill[fill=violet!50] (IItophat) -- (0,3) -- (-3,0) -- cycle;

%trajectory of the right screen
\draw[orange,line width=0.8mm] (IIsquaretopright) -- (0.75,0.75) to [bend left=10] (0.75,-0.75) -- (IIsquarebotright);
\fill[fill=orange!20] (IIsquaretopright) -- (0.75,0.75) to [bend left=10] (0.75,-0.75) -- (IIsquarebotright) -- cycle;

%draw cauchy slice and label on the slice
\draw[line width=0.8mm,red,smooth] plot[variable=\t,samples=100,domain=0:3] ({\t-3},{0.1*\t*cos(45*\t)-1.7});
\draw[line width=0.8mm,red, smooth] plot[variable=\t,samples=100,domain=3:4.52] ({\t-3},{0.1*\t*cos(45*\t)-1.7});
\draw[line width=0.8mm,red, smooth] plot[variable=\t,samples=100,domain=4.52:6] ({\t-3},{0.1*\t*cos(45*\t)-1.7});
\node at (0.9,-2.1) [circle,fill,inner sep=2pt]{};
\node at (0.9,-2) [label=below:$\S_\Le$]{};
\node at (2.05,-2.05) [circle,fill,inner sep=2pt]{};
\node at (2.1,-1.9) [label=below:$\S_\Ri$]{};
%\node at (-1.5,1.65) [label = $\color{red} \Sigma_\Le$]{};
%\node at (2.7,1.4) [label = $\color{red} \Sigma_\Ri$]{};
%\node at (1.1,1.25) [label = $\color{red} \Sigma_\E$]{};

\end{scope}
\end{scope}

\node at (0,3) [violet, circle,fill,inner sep=2.5pt]{};
\node at (0,3) [label=above:$\color{violet} \Gamma$]{};
%\node at (0,-3) [violet, circle,fill,inner sep=2.5pt]{};
\node at (-1.8,2.5) [label=above:$\color{white} \mathcal{M}$]{};

\end{tikzpicture}
\caption{}  \label{fig:screen_trajectories_Mink_-}
\end{subfigure}
\caption{\footnotesize Trajectories of the left screen on $\S_\Le$ (purple) and the right screen on $\S_\Ri$ (orange) in flat Minkowski bubble geometries. The Milne patch $\mathcal{M}$ of the Minkowski bubble is shaded in dark purple. The interior regions of the pode corresponds to the union of $\mathcal{M}$ and the light purple region, and the interior region of the antipode is orange shaded. The exterior region is white. Eventually, $\S_\Le$ reaches the codimension-$2$ surface $\Gamma$, which corresponds to the spatial infinity of $\mathcal{M}$. Two cases can be distinguished, depending whether the Minkowski bubble geometry is obtained as the flat limit $R_L\rightarrow+\infty$ of the de Sitter bubble geometry with $(\varepsilon_L,\varepsilon_R)=(+1,+1)$ (case $(a)$), or for the de Sitter bubble $(\varepsilon_L,\varepsilon_R)=(+1,-1)$ (case $(b)$).}
    \label{fig:screen_trajectories_Mink}
\end{figure}

We now proceed to holographically describe these flat Minkowski bubble geometries using two holographic screens associated with a pair of observers: one sitting at the center of the flat bubble, and the other at the pole of the de Sitter spherical cap. The procedure described in \Sect{sect:screen_trajectories} can be repeated, leading to the screen trajectories shown in the Penrose diagrams of \Fig{fig:screen_trajectories_Mink}.
The trajectory and the area of the right screen $\S_\Ri$ are the same as in the de Sitter bubble geometries with $(\varepsilon_L,\varepsilon_R)=(+1,\pm 1)$. However, the area of the left screen $\S_\Le$ now increases to $+\infty$ as $\S_\Le$ approaches the far past or the far future. Despite the fact that the area of $\S_\Le$ grows to infinity, the classical contribution to the von Neumann entropy of the single-screen system $\S_\Le$ remains finite throughout the cosmological evolution. It is given by the same expressions as in the de Sitter bubble geometries with $(\varepsilon_L,\varepsilon_R)=(+1,+1)$ and $(\varepsilon_L,\varepsilon_R)=(+1,-1)$, see \Eqs{eq:vN_entropy_SL_++} and \eqref{eq:vN_entropy_SL_+-}, respectively. This follows from the fact that \Eqs{eq:vN_entropy_SL_++} and \eqref{eq:vN_entropy_SL_+-} depend on $R_R$ but not on $R_L$, and therefore remain the same in the flat space limit $R_L\rightarrow+\infty$.

Eventually, the left screen $\S_\Le$ reaches the purple dot denoted $\Gamma$ in \Fig{fig:screen_trajectories_Mink}. This codimension-$2$ surface has infinite area. It corresponds to the intersection between the spacelike future infinity of the de Sitter region and the future null infinity of the Minkowski region, and represents the spatial infinity of the Milne patch $\mathcal{M}$ of the Minkowski bubble. 

A holographic description of the bulk region $\mathcal{M}$ in four spacetime dimensions, in terms of a $2$-dimensional local Euclidean CFT on $\Gamma$, has been proposed in \cite{Freivogel:2006xu}. The starting point of their argument is the following. On the one hand, by performing the change of coordinates
\begin{equation}
\left\{\!\begin{array}{ll}
\dis \tau_L=T_L\cosh X_L \\[1ex]
\dis r_L=T_L\sinh X_L
\end{array}\right. ,
\end{equation}
with $T_L>0$, $X_L\geq 0$, the Minkowski metric \eqref{eq:Mink_metric} is brought into the form:
\begin{equation}\label{eq:Milne_metric}
    \d s^2=-\d T_L^2+T_L^2\left(\d X_L^2+\sinh^2 X_L\d \Omega^2_2\right).
\end{equation}
The coordinates $(T_L,X_L)$ cover the Milne patch $\mathcal{M}$ of the Minkowski region, corresponding to the dark purple shaded triangle in \Fig{fig:screen_trajectories_Mink}. From Eq.~\eqref{eq:Milne_metric}, one sees that a constant $T_L$ slice corresponds to a spacelike $3$-dimensional hyperbolic space of constant negative curvature, whose isometry group is SO$(1,3)$. On the other hand, SO$(1,3)$ acts as the $2$-dimensional Euclidean conformal group on $\Gamma$. The fact that every constant $T_L$ slice of $\mathcal{M}$ has the same symmetry suggests the existence of a holographic duality between $\mathcal{M}$ and a Euclidean CFT on $\Gamma$, in the spirit of the dS/CFT correspondence \cite{Strominger:2001gp}. In particular, both time and a spatial direction should emerge in such a correspondence \cite{Freivogel:2006xu}.

In the proposal of \cite{Freivogel:2006xu}, the dual CFT is defined implicitly through a holographic Wheeler--DeWitt construction. The starting point is the Wheeler--DeWitt wavefunction of the bulk region $\mathcal{M}$ with asymptotic boundary $\Gamma$. The geometric fluctuations of $\Gamma$ are described by a Liouville field $L$, while bulk matter fields induce boundary degrees of freedom denoted $f$. The fields appearing in the holographic description are thus labeled $(L,f)$. Projecting the bulk wavefunction onto these boundary variables gives a boundary Wheeler--DeWitt wavefunction $\Psi(L,f)$, which satisfies a boundary Wheeler--DeWitt equation \cite{Freivogel:2006xu}. The measure $\Psi^{\dagger}\Psi$ for computing expectation values of functions of $L$ and $f$ is then written as $\Psi^{\dagger}\Psi=e^{-S(L,f)}$, and $S(L,f)$ is interpreted as the Euclidean action of a $2$-dimensional CFT on $\Gamma$. The central charge of the CFT is related to the entropy of the ``parent'' de Sitter space \cite{Freivogel:2006xu}. 

In four spacetime dimensions, our holographic construction could naturally incorporate the holographic proposal of \cite{Freivogel:2006xu}, providing a candidate for the dual theory on the left screen $\S_\Le$ when it coincides with the asymptotic sphere $\Gamma$. It would be interesting to understand how the proposal of \cite{Freivogel:2006xu} could be used to decipher the nature of the holographic theory on $\S_\Le$ at an arbitrary location along its trajectory.

\section{Conclusion}
\label{sect:conclusions}
In this work, we generalize the de Sitter static patch holographic proposal and the bilayer holographic entanglement entropy prescription of \cite{Susskind:2021esx,Shaghoulian:2021cef,Shaghoulian:2022fop,Franken:2023jas,Franken:2023pni} to de Sitter geometries containing a bouncing bubble of smaller positive or vanishing cosmological constant. We study in detail three classes of de Sitter bubble geometries, labeled $(\varepsilon_L,\varepsilon_R)=(+1,+1)$, $(\varepsilon_L,\varepsilon_R)=(+1,-1)$ and $(\varepsilon_L,\varepsilon_R)=(-1,-1)$. We consider a pair of comoving observers, one at the center of the bubble, and another at the antipodal point of the spatial sphere. In the first two cases, there is a non-trivial overlap of the causal patches of the antipodal observers. In the third case, the causal patches are non-overlapping and causally disconnected. 

Each observer is associated with a holographic screen, which follows a non-spacelike trajectory in his/her causal patch. Application of the bilayer proposal to the two-screen system leads us to conjecture that the full spacetime of the bubble geometries with $(\varepsilon_L,\varepsilon_R)=(+1,+1)$ and $(\varepsilon_L,\varepsilon_R)=(+1,-1)$ can be holographically encoded on the two holographic screens. In these two cases, the dominant geometrical contribution to the entanglement entropy between the single-screen subsystems is given by the area of a minimal extremal homologous surface in the exterior region between the screens, in Planck units. This entanglement entropy is constant for the bubbles with $(\varepsilon_L,\varepsilon_R)=(+1,+1)$, equal to the Gibbons-Hawking entropy of the ``parent'' de Sitter space, but time-dependent in the case $(\varepsilon_L,\varepsilon_R)=(+1,-1)$. In the latter case, it remains bounded by the  Gibbons-Hawking entropy of the ``parent'' de Sitter space. The emergence of the exterior region between the two screens as a result of entanglement is a manifestation of the ER=EPR paradigm. We argue that in the bubbles with $(\varepsilon_L,\varepsilon_R)=(-1,-1)$ the bilayer proposal cannot be applied, and that more than two holographic screens will be needed to holographically encode the entire spacetime.

In this work, we do not provide a detailed derivation of the covariant version of the bilayer holographic entanglement entropy prescription for de Sitter bubble geometries. As part of future work, it would be particularly interesting to obtain evidence for this proposal from a replica bulk path-integral computation, via Lorentzian Schwinger-Keldysh path integrals. Furthermore, it would be interesting to explore the semiclassical contributions to the fine-grained entropy of the single-screen subsystems. Another major breakthrough in the holographic description of de Sitter space would be to understand the nature of a possible (non-gravitational) dual quantum theory, which is not known yet. It has recently been argued that the high temperature limit of the double-scaled Sachdev-Ye-Kitaev (SYK) model could provide a dual quantum description of two-dimensional de Sitter JT gravity \cite{Susskind:2021esx,Susskind:2022dfz,Lin:2022nss,Susskind:2022bia,Susskind:2023hnj,Narovlansky:2023lfz,Verlinde:2024znh,Verlinde:2024zrh,Rahman:2022jsf,Rahman:2023pgt,Rahman:2024vyg,Aguilar-Gutierrez:2024yzu}. It would be particularly interesting to investigate how this duality could be generalized to de Sitter JT gravity with bubbles of different cosmological constants. It would be very interesting to understand how to describe holographically de Sitter multi-bubble cosmologies in order to provide a holographic framework for the string theory landscape. 

\section*{Acknowledgements}
The authors thank Hervé Partouche for helpful discussions during the early stages of this investigation and Victor Franken for valuable comments on this article. F.R. and N.T. would like to thank the University of Cyprus and the Theory Group at Ecole Polytechnique, Paris for hospitality. This work was partially supported by the Cyprus Research and Innovation Foundation grant EXCELLENCE/0421/0362.

\newpage
\appendix
\numberwithin{equation}{section}

\section{Global conformal coordinate system}
\label{app:global_coord}
The aim of this appendix is to construct the connected Penrose diagrams of \Figs{fig:Penrose_diag_++_connected}, \ref{fig:Penrose_diag_+-_connected} and~\ref{fig:Penrose_diag_--_connected}, from the disconnected ones drawn in \Figs{fig:Penrose_diag_++_disconnected}, \ref{fig:Penrose_diag_+-_disconnected} and ~\ref{fig:Penrose_diag_--_disconnected}, respectively. In order to obtain a global conformal coordinate system $(\theta_G,\sigma_G)$ covering both the interior and the exterior of the bubble, we follow the strategy presented in \cite{Fabinger:2003gp}. In the exterior of the bubble, we take $(\theta_G,\sigma_G)=(\theta_R,\sigma_R)$, and perform a conformal transformation on the coordinates $(\theta_L,\sigma_L)$ in the interior the bubble. Using light-cone coordinates $x^{\pm}=\sigma\pm\theta$, this means that inside the bubble, we have $x_G^{\pm}=f^{\pm}(x_L^{\pm})$. The functions $f^\pm$ will be determined by imposing continuity of the global coordinates across the wall, \ie we impose $x_R^{\pm}=f^{\pm}(x_L^{\pm})$ along the wall.

Along the domain wall, the left and right conformal coordinates $(\theta_L,\sigma_L)$ and $(\theta_R,\sigma_R)$ satisfy the relations \eqref{eq:lorenzian_dom_walls}:
\begin{equation}
    \cos\theta_{L}=\frac{\Delta_L}{R_L}\cos \sigma_L, \qquad \cos\theta_R=\frac{\Delta_R}{R_R}\cos\sigma_R,
\end{equation}
where we have defined $\Delta_L\equiv \varepsilon_L\sqrt{R_L^2-R_B^2}$, $\Delta_R\equiv \varepsilon_R\sqrt{R_R^2-R_B^2}$, with $\varepsilon_L=\pm 1$ and $\varepsilon_R=\pm 1$. The light-cone coordinates $x_L^\pm=\sigma_L\pm\theta_L$ can take values in the intervals $x_L^+\in[-\pi/2,\pi]$, $x_L^-\in[-\pi,\pi/2]$.
 
Let us first consider a light ray at constant $x_L^+$. For $x_L^+\in[0,\pi]$, the light ray intersects the domain wall at a conformal time $\sigma_L$ satisfying:
\begin{equation}
    \cos(x_L^+-\sigma_L)=\frac{\Delta_L}{R_L}\cos \sigma_L.
\end{equation}
This can be rewritten as follows
\begin{equation}
    \cos x_L^++\sin x_L^+\tan\sigma_L=\frac{\Delta_L}{R_L}.
\end{equation}
From the matching condition \eqref{eq:matching_cond}, which relates $\tan\sigma_L$ with $\tan\sigma_R$, we find the conformal time $\sigma_R$ when the light ray intersects the domain wall
\begin{equation}
\tan\sigma_R=\frac{\Delta_L-R_L\cos x_L^+}{R_R\sin x_L^+}.
\end{equation}
Using the relation $\displaystyle \cos\left(\arctan x\right)=\frac{1}{\sqrt{1+x^2}}, \forall x\in\mathbb{R}$, this can be written as
\begin{equation}\label{eq:sigma_R}
    \sigma_R=\eta \arccos\frac{R_R\sin x_L^+}{\sqrt{R_R^2\sin^2 x_L^++\left(\Delta_L-R_L\cos x_L^+\right)^2}},
\end{equation}
with $\eta=+1$ if $\displaystyle x_L^+>\arccos\left(\Delta_L/R_L\right)$, and $\eta=-1$ if $\displaystyle x_L^+<\arccos\left(\Delta_L/R_L\right)$. The polar angle $\theta_R$ where the light ray intersects the domain wall is given by $\displaystyle \cos\theta_R=\frac{\Delta_R}{R_R}\cos\sigma_R$. Using the expression for $\sigma_R$ found in \eqref{eq:sigma_R} and the fact that $\theta_R\in[0,\pi]$, one gets
\begin{equation}\label{eq:theta_R}
\theta_R=\arccos\frac{\Delta_R\sin x_L^+}{\sqrt{R_R^2\sin^2 x_L^++\left(\Delta_L-R_L\cos x_L^+\right)^2}}.
\end{equation}
One can now combine the expressions \eqref{eq:sigma_R} and \eqref{eq:theta_R} to find \footnote{We use the trigonometric identity $\arccos x \pm \arccos y=\arccos\left(xy \mp \sqrt{1-x^2}\sqrt{1-y^2}\right), \forall x,y\in[-1,1]$.} $x_R^+=\sigma_R+\theta_R$. Since we have defined $x_G^+=x_R^+$, we get:
\begin{equation}\label{eq:x_g^+}
x_G^+=\arccos\frac{\Delta_R R_R\sin^2 x_L^+-\left(\Delta_L-R_L\cos x_L^+\right)\sqrt{\left(\Delta_L-R_L\cos x_L^+\right)^2+R_B^2\sin^2 x_L^+}}{R_R^2\sin^2 x_L^++\left(\Delta_L-R_L\cos x_L^+\right)^2},
\end{equation}
for $x_L^+\in[0,\pi]$. For $x_L^+\in[-\pi/2,0)$, the light ray does not intersect the wall, and $x_G^+$ is given in this case by minus the expression \eqref{eq:x_g^+}.

Let us now consider a light ray at constant $x_L^-\in[-\pi,\pi/2]$. For $x_L^-\in[-\pi,0]$, the light ray intersects the domain wall, and one can carry out a similar analysis as presented above. We get in this case:
\begin{equation}\label{eq:x_g^-}
x_G^-=-\arccos\frac{\Delta_R R_R\sin^2 x_L^- -\left(\Delta_L-R_L\cos x_L^-\right)\sqrt{\left(\Delta_L-R_L\cos x_L^-\right)^2+R_B^2\sin^2 x_L^-}}{R_R^2\sin^2 x_L^- +\left(\Delta_L-R_L\cos x_L^-\right)^2},
\end{equation}
for $x_L^-\in[-\pi,0]$. For $x_L^-\in(0,\pi/2]$, the light ray does not intersect the wall, and $x_G^-$ is given in this case by minus the expression \eqref{eq:x_g^-}.

To summarize, the global conformal coordinates $x_G^{\pm}=\sigma_G\pm\theta_G$ covering the whole de Sitter bubble geometry are defined by
\begin{equation}
 \left\{\!\begin{array}{ll}
\displaystyle x_G^{\pm}=x_R^{\pm} \, , &\mbox{on the right of the wall, outside the bubble} \, , \\
\displaystyle x_G^{\pm}=f^{\pm}(x_L^{\pm})\, , &\mbox{on the left of the wall, inside the bubble} \, ,\esp
\end{array}\right.
\end{equation}
where the functions $f^{\pm}$ read:
{\footnotesize
\begin{equation}
f^+(x_L^+)=
 \left\{\!\begin{array}{ll}
\dis +\arccos\frac{\Delta_R R_R\sin^2 x_L^+-\left(\Delta_L-R_L\cos x_L^+\right)\sqrt{\left(\Delta_L-R_L\cos x_L^+\right)^2+R_B^2\sin^2 x_L^+}}{R_R^2\sin^2 x_L^++\left(\Delta_L-R_L\cos x_L^+\right)^2} \, , &\mbox{if } \dis x_L^+\in\left[0,\pi\right]\, , \\
\dis -\arccos\frac{\Delta_R R_R\sin^2 x_L^+-\left(\Delta_L-R_L\cos x_L^+\right)\sqrt{\left(\Delta_L-R_L\cos x_L^+\right)^2+R_B^2\sin^2 x_L^+}}{R_R^2\sin^2 x_L^++\left(\Delta_L-R_L\cos x_L^+\right)^2} \, , &\mbox{if } \dis x_L^+\in\left[-\frac{\pi}{2},0\right]\, ,\esp
\end{array}\right.
\end{equation}
}and
{\footnotesize
\begin{equation}
f^-(x_L^-)=
 \left\{\!\begin{array}{ll}
\dis -\arccos\frac{\Delta_R R_R\sin^2 x_L^- -\left(\Delta_L-R_L\cos x_L^-\right)\sqrt{\left(\Delta_L-R_L\cos x_L^-\right)^2+R_B^2\sin^2 x_L^-}}{R_R^2\sin^2 x_L^- +\left(\Delta_L-R_L\cos x_L^-\right)^2} \, , &\mbox{if } \dis x_L^-\in\left[-\pi,0\right]\, , \\
\dis +\arccos\frac{\Delta_R R_R\sin^2 x_L^- -\left(\Delta_L-R_L\cos x_L^-\right)\sqrt{\left(\Delta_L-R_L\cos x_L^-\right)^2+R_B^2\sin^2 x_L^-}}{R_R^2\sin^2 x_L^- +\left(\Delta_L-R_L\cos x_L^-\right)^2} \, , &\mbox{if } \dis x_L^-\in\left[0,\frac{\pi}{2}\right]\, .\esp
\end{array}\right.
\end{equation}
}Notice that the functions $f^{\pm}$ are continuous at $x_L^{\pm}=0$. The eternal de Sitter limit is easily recovered by taking $R_L=R_R=R_B$ (and so $\Delta_L=\Delta_R=0$), which yields, as expected: $f^{\pm}(x_L^{\pm})=x_L^{\pm}$, for $x_L^+\in[-\pi/2,\pi]$ and $x_L^-\in[-\pi,\pi/2]$.

The light rays $x_L^+=0$ and $x_L^-=0$, which intersect the domain wall at $\theta_L=\theta_R=\theta_G=\pi/2$ and $\sigma_L=\sigma_R=\sigma_G=\pm\pi/2$, are left invariant by the conformal transformation, $f^{\pm}(x_L^{\pm}=0)=0$. The parts of the future and past cosmological horizons of the pode lying inside the bubble, given in the left coordinate system by $x_L^+=\pi/2$ and $x_L^-=-\pi/2$ respectively, are mapped to: 
\begin{equation}
f^{\pm}\left(x_L^{\pm}=\pm\frac{\pi}{2}\right)=\pm\arccos\frac{\Delta_R R_R-\Delta_L R_L}{R_R^2+R_L^2-R_B^2}.
\end{equation}
One can easily check that the cosmic line of the observer sitting at the pode, given in the left coordinate system by $x_L^+=x_L^-$, \ie $\theta_L=0$, is given in the global coordinate system by $x_G^+=x_G^-$, \ie $\theta_G=0$. Therefore, defining 
\begin{equation}\label{eq:sigma_infty}
    \sigma_G^{\infty}=\arccos\frac{\Delta_R R_R-\Delta_L R_L}{R_R^2+R_L^2-R_B^2}=\arccos\frac{\varepsilon_R R_R \sqrt{R_R^2-R_B^2}-\varepsilon_L R_L\sqrt{R_L^2-R_B^2}}{R_R^2+R_L^2-R_B^2},
\end{equation}
the conformal time $\sigma_G$ varies along the pode between $-\sigma_G^{\infty}$ at past infinity to $+\sigma_G^{\infty}$ at future infinity.

For the bubble geometries with $(\varepsilon_L,\varepsilon_R)=(+1,+1)$ and $(\varepsilon_L,\varepsilon_R)=(+1,-1)$, the argument of the $\arccos$ function in \eqref{eq:sigma_infty} is negative since $R_L>R_R$, and so $\sigma_G^{\infty}$ is between $\pi/2$ and $\pi$. This justifies the shape of the Penrose diagrams depicted in \Figs{fig:Penrose_diag_++_connected}-\ref{fig:Penrose_diag_+-_connected} and used in the holographic construction of \Sect{sect:holography}. For $R_L=R_R=R_B$, one recovers the eternal de Sitter case with $\sigma_G^{\infty}=\pi/2$, while for $R_L\rightarrow +\infty$, corresponding to a flat Minkowski bubble, we get $\sigma_G^{\infty}=\pi$, as shown in \Fig{fig:Penrose_diag_Mink}.  

For the bubble geometries with $(\varepsilon_L,\varepsilon_R)=(-1,-1)$, the argument of the $\arccos$ function in \eqref{eq:sigma_infty} is positive, and so $\sigma_G^{\infty}<\pi/2$. This justifies the shape of the Penrose diagram drawn in \Fig{fig:Penrose_diag_--_connected}.

%%%%%%%%%%%%%%%%%%%%%%%%%%%%%%%%%%%%%%%%%%%%%%%%%%%%%%%%%%%%%%%%%%%%%%%%%%%%%%%%%%%%%%%%%%%%%%%%%%%%%%%%%%%%%%%%%%%%%%%%%%%%5
\section{Quantum corrections}
\label{app:quantum_corrections}

In the context of the bilayer proposal, we can incorporate in the calculation of the entropy of a subsystem $A$ quantum corrections. To this end, we add to the area functionals of the homologous surfaces $\chi_i$ (divided by 4G$\hbar$) the \textit{semiclassical entropy} of the quantum fields, including gravitons, on a certain region of the semiclassical background geometry, in order to obtain a generalized entropy \cite{Almheiri:2020cfm}. The semiclassical contribution to the generalized entropy is of order $(G\hbar)^0$. Although we do not derive the precise form of the generalized entropy from a bulk replica path integral, the expression we propose gives consistent results, at least for the cases of the eternal de Sitter space and the closed FRW cosmologies studied in \cite{Franken:2023jas,Franken:2023pni}, and the cases we study in this paper.

Following \cite{Franken:2023jas,Franken:2023pni}, we define the generalized entropy associated with the subsystem $A$ as
\begin{align}
S_{\textrm{gen}}(\chi_\Le,\chi_\E,\chi_\Ri)=\frac{{\rm Area}(\chi_\Le)+{\rm Area}(\chi_\E)+{\rm Area}(\chi_\Ri)}{4G\hbar}+\S_{\textrm{semicl}}
(\mathcal{C}_\Le\cup\mathcal{C}_\E\cup\mathcal{C}_\Ri),
\label{eq:generalized entropy}
\end{align}
where for $i\in \{\Le,\E,\Ri\}$, $\chi_i$ is a codimension-$2$ surface homologous to $A_i$\footnote{Recall that $A_\Le=A\cap \S_\Le$, $A_\E=A$, $A_\Ri=A\cap \S_\Ri$.}, lying in the causal diamond region $\J_i$, and $\mathcal{C}_i$ is the bulk codimension-$1$ surface with boundary $\chi_i \cup A_i$. To obtain the fine-grained entropy of $A$, we need to extremize the generalized entropy with respect to $\chi_\Le, \chi_\E, \chi_\Ri$. If there are more than one extrema, we choose the one with the minimum generalized entropy:
\begin{align}\label{eq:QES}
S(A)=\textrm{min}\ \textrm{ext}\ S_{\textrm{gen}}(\chi_\Le,\chi_\E,\chi_\Ri).
\end{align}

Notice that to have a minimal generalized entropy, we do not add to the geometrical terms the sum of the semiclassical entropies of each $\mathcal{C}_i$ : $S_{\textrm{semicl}}(\mathcal{C}_\Le)+S_{\textrm{semicl}}
(\mathcal{C}_\E)+S_{\textrm{semicl}}
(\mathcal{C}_\Ri)$. Instead, we add to the geometrical terms the semiclassical entropy of the combined region $\mathcal{C}_\Le \cup \mathcal{C}_\E \cup \mathcal{C}_\Ri$. Indeed, entropy inequalities require that $S_{\textrm{semicl}}(\mathcal{C}_\Le)+S_{\textrm{semicl}}
(\mathcal{C}_\E)+S_{\textrm{semicl}}
(\mathcal{C}_\Ri)\ge S_{\textrm{semicl}}
(\mathcal{C}_\Le\cup\mathcal{C}_\E\cup\mathcal{C}_\Ri)$. Cauchy slices in the entanglement wedge of a generic subsystem $A$ of the holographic dual are expected to contain parts from all three regions left, exterior and right. The bulk gravitational subsystems associated with these slices are expected to be reconstructible from $A$, and their fine-grained entropy should be equal to the fine-grained entropy of $A$. So, bulk degrees of freedom in different regions (left, exterior, right) that are located in the entanglement wedge of $A$ and are entangled with each other should not contribute to the fine-grained entropy of $A$ and the dual bulk subsystems. 

Furthermore, in the literature, there are cases where the semiclassical term in the generalized entropy formula involves two separated regions, as in the island rule for black holes. In this case, the semiclassical entropy used is the one of the union of the island $I$ and the asymptotic surface $R$, where the radiation resides, and not the sum of the two semiclassical entropies. This is justified using the replica path integral, which involves a gluing between $n$ copies of the original geometry via the surface $I\cup R$. In the limit $n\rightarrow 1$, the above replica manifold is the one needed to calculate/define the semiclassical entropy of $I\cup R$, rather than the sum of $S_{\textrm{semicl}}(I)$ and $S_{\textrm{semicl}}(R)$.

\section{Area extremization in a causal diamond}
\label{app:area_extremization}
At fixed conformal coordinates $(\theta,\sigma)$, the metric of dS$_{n+1}$ given in \eqref{eq:dS_metric} reduces to the metric of S$^{n-1}$, whose area is given in \Eq{eq:area}. The first goal of this Appendix is to find the extrema of this area functional in the whole Penrose diagram. We will observe that for a generic Cauchy slice $\Sigma$, none of these extremal points lies in the causal diamond $\J_\E$ of $\Sigma_\E$. However, we will show that if we restrict the homologous surfaces to lie in $\J_\E$,  by supplementing the area functional with suitable Lagrange multipliers and auxiliary fields that enforce this restriction, certain points on the boundary of $\J_\E$ become extremal, as in \cite{Franken:2023jas,Franken:2023pni}. 

\subsection{Extrema of the area defined in the Penrose diagram}
In this subsection, we define the area on the whole Penrose diagram. It is a function of the conformal coordinates $(\theta_L,\sigma_L)$ and $(\theta_R,\sigma_R)$, in the left and right de Sitter regions, respectively. In both cases, it has an extremum if $\d {\rm Area}=0$ at this point. Using \Eq{eq:derivative_of_area}, we find that:

\noindent $\bullet$ When $n\geq 3$, extrema are located at $(\theta_L,\sigma_L)=(0,\sigma_L)$, $(\theta_R,\sigma_R)=(\pi,\sigma_R)$, where $-\pi/2<\sigma_{L,R}<\pi/2$.

\noindent $\bullet$ For the bubble geometries with $(\varepsilon_L,\varepsilon_R)=(+1,+1)$, depicted in the Penrose diagrams of \Fig{fig:Penrose_diag_++}, the area is also extremal at $(\theta_R,\sigma_R)=(\pi/2,0)$.

\noindent $\bullet$ For the bubble geometries with $(\varepsilon_L,\varepsilon_R)=(+1,-1)$, depicted in the Penrose diagrams of \Fig{fig:Penrose_diag_+-}, there are no other extrema.

\noindent $\bullet$ For the bubble geometries with $(\varepsilon_L,\varepsilon_R)=(-1,-1)$, depicted in the Penrose diagrams of \Fig{fig:Penrose_diag_--}, the area is also extremal at $(\theta_L,\sigma_L)=(\pi/2,0)$.

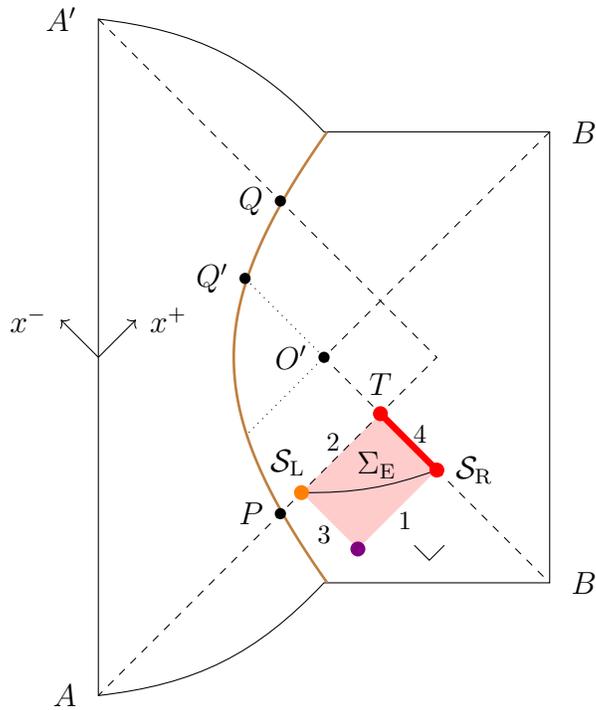
\begin{figure}[h!]
\centering
\begin{tikzpicture}
%\begin{scope}[transparency group]
%\begin{scope}[blend mode=multiply]
%Gives a name to the 6 corners of the diagram
\path
       +(-3,4.5) coordinate (IItophat)
       +(0,3) coordinate (IItophatright)
       +(3,3)  coordinate (IIsquaretopright)
       +(3,-3) coordinate (IIsquarebotright)
       +(0,-3) coordinate (IIbothatright)
       +(-3,-4.5) coordinate (IIbothat)
       +(-3,3) coordinate (IIsquaretopleft)
       +(-3,-3) coordinate(IIsquarebotleft)
      
       ;

\fill[fill=red!20] (0.75,-0.75) -- node[pos=0.7, above] {{\footnotesize $4$}}(1.5,-1.5) -- node[pos=0.4, below] {{\footnotesize $1$}} (0.9/2,-5.1/2) --  node[pos=0.6, below] {{\footnotesize $3$}}(-0.3,-1.8) -- node[pos=0.4, above] {{\footnotesize $2$}}(0.75,-0.75);

%\fill[fill=blue!50] (-3,-0.66) -- (-0.84,1.5) -- (-3,3.66) --  cycle;

%Draws the 6 boundaries of the diagram       
\draw (IItophat) to
          [bend left=20] (IItophatright) --
          (IIsquaretopright) --       (IIsquarebotright) --
          (IIbothatright) to 
          [bend left=20] (IIbothat)--
          (IItophat)--cycle;

\draw[dashed] (1.5,-1.5) -- (3,-3);
\draw[dashed] (3,3) -- (0,0) -- (0.75,-0.75);
\draw[dashed] (IItophat) -- (1.5,0) -- (IIbothat);
\draw[dotted] (-1,-1) -- (0,0) -- (-1,1);

\draw[line width= 1 pt,brown] plot[variable=\t,samples=90,domain=-45:45] ({3*sec(\t)-4.2},{3*tan(\t)});

\draw (-0.3,-1.8) to [bend right=10] (1.5,-1.5) ;
\node at (0.7,-1.9) [label=above:$\Sigma_\E$]{};

%\node at (-0.84,1.5) [circle,fill,inner sep=1.5pt]{};
\node at (-0.3,-1.8) [circle,fill,inner sep=2pt, orange]{};
\node at (-1,-1.4) [label = right:$\S_\Le$]{};
\node at (1.5,-1.5) [circle,fill,inner sep=2pt, red, label = right:$\S_\Ri$]{};

\draw[line width=2.5pt, red] (0.75,-0.75)--(1.5,-1.5);
%\node at (0.33,0.33) [circle,fill,inner sep=1.5pt, label = right:$\sigma'$]{};

\node at (-3,-4.5) [label=left:$A$]{};
\node at (-3,4.5) [label=left:$A'$]{};
\node at (3,-3) [label=right:$B$]{};
\node at (3,3) [label=right:$B'$]{};

%\node at (-0.5,-2.08) [label=$P$]{};
\node at (-0.58,-2.08) [circle,fill,inner sep=1.5pt, label=left:$P$]{};
%\node at (-0.60,2) [label=$Q$]{};
\node at (-0.58,2.08) [circle,fill,inner sep=1.5pt, label=left:$Q$]{};

\node at (-1.05,1.05) [circle,fill,inner sep=1.5pt,label=left:$Q'$]{};

\node at (0,0) [circle,fill,inner sep=1.5pt, label = left:$O'$]{};

\node at (0.75,-0.75) [circle,fill,inner sep=2pt, red, label = above:$T$]{};

\node at (0.9/2,-5.1/2) [circle,fill,inner sep=2pt, violet]{};

\draw (1.2,-2.5) -- (1.4,-2.7) -- (1.6,-2.5);

%x+, x- axes
\draw[->] (-3,0)--(-2.5,0.5);
\draw[->] (-3,0)--(-3.5,0.5);
\node at (-2.6,0.5) [label = right:$x^+$]{};
\node at (-3.4,0.5) [label = left:$x^-$]{};

%\end{scope}
%\end{scope}
\end{tikzpicture}
    \caption{\footnotesize{Causal diamond $\J_\E$ (red rectangle) of a Cauchy slice $\Sigma_\E$ located in the contracting phase of the cosmology, in the case $(\varepsilon_L,\varepsilon_R)=(+1,+1)$. When we restrict the domain of definition to be the $\J_\E$ causal diamond, the S$^{n-1}$-area functional has minima (red), a maximum (purple) and a saddle point (orange) on the boundary of the diamond.}} \label{fig:extremization}
\end{figure}

In this Appendix, we focus on the  $(\varepsilon_L,\varepsilon_R)=(+1,+1)$ bubble geometries. To be specific, we will also study the case where the left screen $\S_\Le$ is along the lightlike segment $PT$ and the right screen $\S_\Ri$ along the lightlike segment $BT$. See \Fig{fig:extremization}.
The analysis for different screen locations, and/or other classes of bubble geometries, can be carried out in an analogous way.

In this situation, both screens lie in the right de Sitter region. We will denote by $x_R^{\pm}(\S_\Le)$ and $x_R^{\pm}(\S_\Ri)$ the null coordinates of the screens ($x_R^{\pm}=\sigma_R\pm\theta_R$). Along the lightlike segments $PT$ and $BT$, we have:
\begin{eqnarray}
    x_R^-(\S_\Le) &=& -\arccos\frac{R_R\sqrt{R_R^2-R_B^2} - R_L\sqrt{R_L^2-R_B^2}}{R_R^2+R_L^2-R_B^2},\label{eq:x_R^-_S_L}\\
    x_R^+(\S_\Ri)&=&\frac{\pi}{2}.
\end{eqnarray}
The conformal time of the top vertex $T$ of $\J_\E$ is thus given by:
\begin{equation}
    \sigma_R(T)=\frac{\pi}{4}-\frac{1}{2}\arccos\frac{R_R\sqrt{R_R^2-R_B^2} - R_L\sqrt{R_L^2-R_B^2}}{R_R^2+R_L^2-R_B^2}.
\end{equation}
For $-\pi/2\leq (x_R^+(\S_\Le)+x_R^-(\S_\Ri))/2 < \sigma_R(T)$, $\J_\E$ is a rectangle corresponding to the domain
\begin{equation}
\left\{\!\!\begin{array}{l}
\dis x_R^+(\S_\Le)\le x_R^+\le \frac{\pi}{2}\\
\dis x_R^-(\S_\Ri)\le x_R^-\le x_R^-(\S_\Le) \esp
\end{array}\right.,
\label{eq:causal_diam}
\end{equation}
where $x_R^-(\S_\Le)$ is given by \eqref{eq:x_R^-_S_L}. It is shaded in red in the Penrose diagram of \Fig{fig:extremization}, and its boundary is composed of $4$ segments labelled as $1,\dots,4$. The derivatives of the area,
\begin{equation}\label{eq:deriv}
\frac{\partial {\rm Area}}{\partial x_R^{\pm}}(x_R^+,x_R^-)=\pm\omega_{n-1}R^{n-1}\frac{n-1}{2}\cos(x_R^\mp)\frac{(\sin\theta)^{n-2}}{(\cos\sigma)^n},
\end{equation}
do not vanish simultaneously in the diamond \eqref{eq:causal_diam}, so that no point of $\J_\E$ satisfies the extremization condition $\d{\rm Area}=0$.

\subsection{Minima, maxima and saddle points of the area defined in a causal diamond}
\label{sect:min_max_saddle}
Now, let us restrict the domain of definition of the area functional to be the causal diamond $\J_\E$ only. Following the orientation of the Bousso wedge in \Fig{fig:extremization}, one observes that:

- ${\rm Area}(x_R^+,x_R^-)$ initially at the boundary 4 of the diamond increases when $x_R^+$ decreases. We also know that all spheres located on the cosmological horizon of the antipode are degenerate. Each sphere on the boundary 4 thus corresponds to degenerate minima of the area functional in $\J_\E$.

- ${\rm Area}(x_R^+,x_R^-)$ initially at the lower vertex of $\J_\E$ decreases when $x_R^+$ increases and when $x_R^-$ increases: the lower vertex is thus a global maximum.

- ${\rm Area}(x_R^+,x_R^-)$ initially at the left vertex of $\J_\E$ decreases when $x_R^+$ increases, and increases when $x_R^-$ decreases: the left vertex is thus a saddle point.

\noindent Those minima, maximum and saddle point are respectively depicted in red, purple and orange in the Penrose diagram of \Fig{fig:extremization}. These points, however, are not extrema of the area functional, in the sense that $\d {\rm Area}\neq 0$ at these points. 

\subsection{Minima, maxima and saddle points as solutions of a constrained extremization problem}
Following the method introduced in \cite{Franken:2023jas,Franken:2023pni}, we show how those minima, maxima and saddle points of the area function defined in $\J_\E$ can be obtained as extremal points in a constrained extremization problem.

The inequalities \eqref{eq:causal_diam} restricting the homologous surfaces to lie in the causal diamond $\J_\E$ can be enforced by supplementing the area functional with terms proportional to Lagrange multipliers~$\nu_I$, $I\in\{1,2,3,4\}$,
\begin{eqnarray}
\widehat{{\rm Area}}(x_R^+,x_R^-,\nu_I,a_I)={\rm Area}(x_R^+,x_R^-)&+&\nu_1\left(x_R^--x_R^-(\S_\Ri)-a_1^2\right)+\nu_2\left(x_R^-(\S_\Le)-x_R^--a_2^2\right)\nonumber\\
&+&\nu_3\left(x_R^+-x_R^+(\S_\Le)-a_3^3\right)+\nu_4\left(\frac{\pi}{2}-x_R^+-a_4^2\right)\!.
\label{eq:hat_A}
\end{eqnarray}
Beside the multipliers, the $a_I$'s are extra variables whose squares are the ``positive distances from $(x_R^+,x_R^-)$ to the edges $I$ in the Penrose diagram'' in \Fig{fig:extremization}. The introduction of the $a_I$'s is necessary because the constraints \eqref{eq:causal_diam} are inequalities rather than equalities. The key point is that the functional $\widehat{{\rm Area}}$ is defined in a domain without boundary, as $x_R^{\pm}$, $\nu_I$, $a_I$ are all spanning $\R$. To find its extrema, \ie the points in $\R^{10}$ satisfying $\d\widehat{{\rm Area}}=0$, we first vary $\widehat{{\rm Area}}$ with respect to $x_R^{\pm}$, which yields:
\begin{subequations}\label{eq:x_variations}
\begin{align}
    \frac{\partial {\rm Area}}{\partial x_R^+}&=\nu_4-\nu_3,\label{eq:x+}\\
    \frac{\partial {\rm Area}}{\partial x_R^-}&=\nu_2-\nu_1\label{eq:x-}.
\end{align}
\end{subequations} 
Varying $\widehat{{\rm Area}}$ with respect to the Lagrange multipliers gives the equations
\begin{subequations}\label{eq:nu_variations}
\begin{align}
\label{eq:nu1}
a_1^2&= x_R^- - x_R^-(\S_\Ri),\\
a_2^2&=x_R^-(\S_\Le)-x_R^-,\label{eq:nu2}\\
a_3^2&=x_R^+-x_R^+(\S_\Le),\label{eq:nu3}\\
a_4^2&=\frac{\pi}{2}-x_R^+,\label{eq:nu4}
\end{align}
\end{subequations} 
which imply $\widehat{{\rm Area}}={\rm Area}$ when satisfied. Finally, the variations of $\widehat{{\rm Area}}$ with respect to the $a_I$'s give
\begin{subequations}
\label{eq:a_variations}
\begin{align}
\nu_1a_1&=0,\label{eq:a1}\\
\nu_2a_2&=0,\label{eq:a2}\\
\nu_3a_3&=0,\label{eq:a3}\\
\nu_4a_4&=0.\label{eq:a4}
\end{align}
\end{subequations}
To find all solutions of the system of 10 equations, let us organize our discussion from the location of $(x_R^+,x_R^-)\in\R^2$:
 
- When $(x_R^+,x_R^-)$ is not on the boundary of the diamond, \Eqs{eq:nu_variations} impose it to lie in the bulk of the diamond and determine the values of $a^2_I> 0$, $I\in\{1,2,3,4\}$. As a result, $\nu_I=0$ from \Eqs{eq:a_variations}. However, \Eqs{eq:x_variations} are not satisfied, as seen below \Eq{eq:deriv}. 

- Take now $(x_R^+,x_R^-)$ in the bulk of the boundary segment 1, \ie not at its endpoints. We have $x_R^-=x_R^-(\S_\Ri)$, and thus $a_1=0$ from \Eq{eq:nu1}. This implies that \Eq{eq:a1} is satisfied. \Eqs{eq:nu2}--(\ref{eq:nu4}) determine $a^2_{2,3,4}> 0$, which imposes $\nu_{2,3,4}=0$ from \Eqs{eq:a2}--(\ref{eq:a4}). However, since $\cos x_R^-=\cos(x_R^-(\S_\Ri))\neq 0$, \Eq{eq:x+} is not satisfied. In a similar way, there is no solution to the equations when $(x_R^+,x_R^-)$ lies in the interior of edges 2 and 3.

- When $(x_R^+,x_R^-)$ is in the bulk of the boundary segment 4, we have $x_R^+=\pi/2$, and thus $a_4=0$ from \Eq{eq:nu4}. This implies that \Eq{eq:a4} is satisfied. \Eqs{eq:nu1}--(\ref{eq:nu3}) determine $a^2_{1,2,3}> 0$, which imposes $\nu_{1,2,3}=0$ from \Eqs{eq:a1}--(\ref{eq:a3}). In that case, \Eq{eq:x+} fixes $\nu_4$ and \Eq{eq:x-} is solved for any $x_R^-$. We thus have found degenerate extrema of $\widehat{{\rm Area}}$. 

- If $(x_R^+,x_R^-)$ is at the lower tip of the diamond, \ie the intersection of the boundary segments 1 and 3, we have $x_R^--x_R^-(\S_\Ri)=0$, $x_R^+-x_R^+(\S_\Le)=0$ and thus $a_{1,3}=0$ from \Eqs{eq:nu1}, (\ref{eq:nu3}). \Eqs{eq:a1}, (\ref{eq:a3}) are thus satisfied. We also have $a^2_{2,4} > 0$, from \Eqs{eq:nu2}, (\ref{eq:nu4}), which implies $\nu_{2,4}=0$ from \Eqs{eq:a2}, (\ref{eq:a4}). \Eqs{eq:x+},(\ref{eq:x-}) then determine $\nu_{1,3}$. We thus have found an extremum of $\widehat{{\rm Area}}$. 

- Similarly, when $(x_R^+,x_R^-)$ is at another tip $\S_\Le$, $\S_\Ri$ or $T$ of the diamond, one finds an extremum of $\widehat{{\rm Area}}$.

\noindent To summarize, we have recovered all the minima, maximum and saddle point of ${\rm Area}$ found in \Sect{sect:min_max_saddle}, except that now the extremization condition $\d \widehat{{\rm Area}}=0$ is satisfied. The Lagrange multipliers are tuned in order to compensate the derivatives of ${\rm Area}$ and thus obtain extrema of $\widehat{{\rm Area}}$.

\newpage
%\normalem
\bibliographystyle{jhep}
\bibliography{bib}

@article{Coleman:1980aw,
    author = "Coleman, Sidney R. and De Luccia, Frank",
    title = "{Gravitational Effects on and of Vacuum Decay}",
    reportNumber = "SLAC-PUB-2463",
    doi = "10.1103/PhysRevD.21.3305",
    journal = "Phys. Rev. D",
    volume = "21",
    pages = "3305",
    year = "1980"
}

@article{Brown:1987dd,
    author = "Brown, J. David and Teitelboim, C.",
    title = "{Dynamical neutralization of the cosmological constant}",
    doi = "10.1016/0370-2693(87)91190-7",
    journal = "Phys. Lett. B",
    volume = "195",
    pages = "177--182",
    year = "1987"
}

@article{Brown:1988kg,
    author = "Brown, J. David and Teitelboim, C.",
    title = "{Neutralization of the cosmological constant by membrane creation}",
    doi = "10.1016/0550-3213(88)90559-7",
    journal = "Nucl. Phys. B",
    volume = "297",
    pages = "787--836",
    year = "1988"
}

@article{Dong:2011gx,
    author = "Dong, Xi and Harlow, Daniel",
    title = "{Analytic Coleman-De Luccia Geometries}",
    eprint = "1109.0011",
    archivePrefix = "arXiv",
    primaryClass = "hep-th",
    reportNumber = "SLAC-PUB-14555, SU-ITP-11-45",
    doi = "10.1088/1475-7516/2011/11/044",
    journal = "JCAP",
    volume = "11",
    pages = "044",
    year = "2011"
}

@article{Banks:2002nm,
    author = "Banks, T.",
    title = "{Heretics of the false vacuum: Gravitational effects on and of vacuum decay. 2.}",
    eprint = "hep-th/0211160",
    archivePrefix = "arXiv",
    reportNumber = "RUNHETC-2002-44, SCIPP-02-27",
    month = "11",
    year = "2002"
}

@article{Fabinger:2003gp,
    author = "Fabinger, Michal and Silverstein, Eva",
    title = "{D-Sitter space: Causal structure, thermodynamics, and entropy}",
    eprint = "hep-th/0304220",
    archivePrefix = "arXiv",
    reportNumber = "SLAC-PUB-9717, SU-ITP-03-08",
    doi = "10.1088/1126-6708/2004/12/061",
    journal = "JHEP",
    volume = "12",
    pages = "061",
    year = "2004"
}

@article{Susskind:2021omt,
    author = "Susskind, Leonard",
    title = "{De Sitter Holography: Fluctuations, Anomalous Symmetry, and Wormholes}",
    eprint = "2106.03964",
    archivePrefix = "arXiv",
    primaryClass = "hep-th",
    doi = "10.3390/universe7120464",
    journal = "Universe",
    volume = "7",
    number = "12",
    pages = "464",
    year = "2021"
}

@article{Banks:2000fe,
    author = "Banks, Tom",
    editor = "Duff, Michael J. and Liu, J. T. and Lu, J.",
    title = "{Cosmological breaking of supersymmetry?}",
    eprint = "hep-th/0007146",
    archivePrefix = "arXiv",
    reportNumber = "RUNHETC-2000-24, SCIPP-00-23",
    doi = "10.1142/S0217751X01003998",
    journal = "Int. J. Mod. Phys. A",
    volume = "16",
    pages = "910--921",
    year = "2001"
}

@inproceedings{Witten:2001kn,
    author = "Witten, Edward",
    title = "{Quantum gravity in de Sitter space}",
    booktitle = "{Strings 2001: International Conference}",
    eprint = "hep-th/0106109",
    archivePrefix = "arXiv",
    month = "6",
    year = "2001"
}

@article{Strominger:2001pn,
    author = "Strominger, Andrew",
    title = "{The dS / CFT correspondence}",
    eprint = "hep-th/0106113",
    archivePrefix = "arXiv",
    doi = "10.1088/1126-6708/2001/10/034",
    journal = "JHEP",
    volume = "10",
    pages = "034",
    year = "2001"
}

@article{Strominger:2001gp,
    author = "Strominger, Andrew",
    title = "{Inflation and the dS / CFT correspondence}",
    eprint = "hep-th/0110087",
    archivePrefix = "arXiv",
    doi = "10.1088/1126-6708/2001/11/049",
    journal = "JHEP",
    volume = "11",
    pages = "049",
    year = "2001"
}

@article{Franken:2023pni,
    author = "Franken, Victor and Partouche, Herv\'e and Rondeau, Fran\c{c}ois and Toumbas, Nicolaos",
    title = "{Bridging the static patches: de Sitter holography and entanglement}",
    eprint = "2305.12861",
    archivePrefix = "arXiv",
    primaryClass = "hep-th",
    reportNumber = "CPHT-RR018.042023",
    doi = "10.1007/JHEP08(2023)074",
    journal = "JHEP",
    volume = "08",
    pages = "074",
    year = "2023"
}

@article{Franken:2023jas,
    author = "Franken, Victor and Partouche, Herv\'e and Rondeau, Fran\c{c}ois and Toumbas, Nicolaos",
    title = "{Closed FRW holography: a time-dependent ER=EPR realization}",
    eprint = "2310.20652",
    archivePrefix = "arXiv",
    primaryClass = "hep-th",
    reportNumber = "CPHT-RR066.102023",
    doi = "10.1007/JHEP05(2024)219",
    journal = "JHEP",
    volume = "05",
    pages = "219",
    year = "2024"
}

@article{Bousso:1999xy,
    author = "Bousso, Raphael",
    title = "{A Covariant entropy conjecture}",
    eprint = "hep-th/9905177",
    archivePrefix = "arXiv",
    reportNumber = "SU-ITP-99-23",
    doi = "10.1088/1126-6708/1999/07/004",
    journal = "JHEP",
    volume = "07",
    pages = "004",
    year = "1999"
}

@article{Bousso:2002ju,
    author = "Bousso, Raphael",
    title = "{The Holographic principle}",
    eprint = "hep-th/0203101",
    archivePrefix = "arXiv",
    reportNumber = "NSF-ITP-02-17",
    doi = "10.1103/RevModPhys.74.825",
    journal = "Rev. Mod. Phys.",
    volume = "74",
    pages = "825--874",
    year = "2002"
}

@article{Dong:2016eik,
    author = "Dong, Xi and Harlow, Daniel and Wall, Aron C.",
    title = "{Reconstruction of Bulk Operators within the Entanglement Wedge in Gauge-Gravity Duality}",
    eprint = "1601.05416",
    archivePrefix = "arXiv",
    primaryClass = "hep-th",
    reportNumber = "NSF-KITP-16-005",
    doi = "10.1103/PhysRevLett.117.021601",
    journal = "Phys. Rev. Lett.",
    volume = "117",
    number = "2",
    pages = "021601",
    year = "2016"
}

@article{Bousso:1999cb,
    author = "Bousso, Raphael",
    title = "{Holography in general space-times}",
    eprint = "hep-th/9906022",
    archivePrefix = "arXiv",
    reportNumber = "SU-ITP-99-24",
    doi = "10.1088/1126-6708/1999/06/028",
    journal = "JHEP",
    volume = "06",
    pages = "028",
    year = "1999"
}

@article{Bak:1999hd,
    author = "Bak, Dongsu and Rey, Soo-Jong",
    title = "{Cosmic holography}",
    eprint = "hep-th/9902173",
    archivePrefix = "arXiv",
    reportNumber = "SNUST-99-002, UOSTP-99-004",
    doi = "10.1088/0264-9381/17/15/101",
    journal = "Class. Quant. Grav.",
    volume = "17",
    pages = "L83",
    year = "2000"
}

@article{McNamara:2020uza,
    author = "McNamara, Jacob and Vafa, Cumrun",
    title = "{Baby Universes, Holography, and the Swampland}",
    eprint = "2004.06738",
    archivePrefix = "arXiv",
    primaryClass = "hep-th",
    month = "4",
    year = "2020"
}

@article{Harlow:2025pvj,
    author = "Harlow, Daniel and Usatyuk, Mykhaylo and Zhao, Ying",
    title = "{Quantum mechanics and observers for gravity in a closed universe}",
    eprint = "2501.02359",
    archivePrefix = "arXiv",
    primaryClass = "hep-th",
    reportNumber = "MIT-CTP/5824",
    month = "1",
    year = "2025"
}

@article{Cotler:2017erl,
    author = "Cotler, Jordan and Hayden, Patrick and Penington, Geoffrey and Salton, Grant and Swingle, Brian and Walter, Michael",
    title = "{Entanglement Wedge Reconstruction via Universal Recovery Channels}",
    eprint = "1704.05839",
    archivePrefix = "arXiv",
    primaryClass = "hep-th",
    doi = "10.1103/PhysRevX.9.031011",
    journal = "Phys. Rev. X",
    volume = "9",
    number = "3",
    pages = "031011",
    year = "2019"
}

@article{tHooft:1993dmi,
    author = "'t Hooft, Gerard",
    title = "{Dimensional reduction in quantum gravity}",
    eprint = "gr-qc/9310026",
    archivePrefix = "arXiv",
    reportNumber = "THU-93-26",
    journal = "Conf. Proc. C",
    volume = "930308",
    pages = "284--296",
    year = "1993"
}

@article{Susskind:1994vu,
    author = "Susskind, Leonard",
    title = "{The World as a hologram}",
    eprint = "hep-th/9409089",
    archivePrefix = "arXiv",
    reportNumber = "SU-ITP-94-33",
    doi = "10.1063/1.531249",
    journal = "J. Math. Phys.",
    volume = "36",
    pages = "6377--6396",
    year = "1995"
}

@article{Maldacena:1997re,
    author = "Maldacena, Juan Martin",
    title = "{The Large N limit of superconformal field theories and supergravity}",
    eprint = "hep-th/9711200",
    archivePrefix = "arXiv",
    reportNumber = "HUTP-97-A097, HUTP-98-A097",
    doi = "10.4310/ATMP.1998.v2.n2.a1",
    journal = "Adv. Theor. Math. Phys.",
    volume = "2",
    pages = "231--252",
    year = "1998"
}

@article{Gubser:1998bc,
    author = "Gubser, S. S. and Klebanov, Igor R. and Polyakov, Alexander M.",
    title = "{Gauge theory correlators from noncritical string theory}",
    eprint = "hep-th/9802109",
    archivePrefix = "arXiv",
    reportNumber = "PUPT-1767",
    doi = "10.1016/S0370-2693(98)00377-3",
    journal = "Phys. Lett. B",
    volume = "428",
    pages = "105--114",
    year = "1998"
}

@article{Witten:1998qj,
    author = "Witten, Edward",
    title = "{Anti-de Sitter space and holography}",
    eprint = "hep-th/9802150",
    archivePrefix = "arXiv",
    reportNumber = "IASSNS-HEP-98-15",
    doi = "10.4310/ATMP.1998.v2.n2.a2",
    journal = "Adv. Theor. Math. Phys.",
    volume = "2",
    pages = "253--291",
    year = "1998"
}

@article{Susskind:2007pv,
    author = "Susskind, Leonard",
    title = "{The Census taker's hat}",
    eprint = "0710.1129",
    archivePrefix = "arXiv",
    primaryClass = "hep-th",
    month = "10",
    year = "2007"
}

@inproceedings{Seiberg:2006wf,
    author = "Seiberg, Nathan",
    title = "{Emergent spacetime}",
    booktitle = "{23rd Solvay Conference in Physics: The Quantum Structure of Space and Time}",
    eprint = "hep-th/0601234",
    archivePrefix = "arXiv",
    doi = "10.1142/9789812706768_0005",
    pages = "163--178",
    month = "1",
    year = "2006"
}

@article{Ryu:2006bv,
    author = "Ryu, Shinsei and Takayanagi, Tadashi",
    title = "{Holographic derivation of entanglement entropy from AdS/CFT}",
    eprint = "hep-th/0603001",
    archivePrefix = "arXiv",
    reportNumber = "NSF-KITP-06-11",
    doi = "10.1103/PhysRevLett.96.181602",
    journal = "Phys. Rev. Lett.",
    volume = "96",
    pages = "181602",
    year = "2006"
}

@article{Hubeny:2007xt,
    author = "Hubeny, Veronika E. and Rangamani, Mukund and Takayanagi, Tadashi",
    title = "{A Covariant holographic entanglement entropy proposal}",
    eprint = "0705.0016",
    archivePrefix = "arXiv",
    primaryClass = "hep-th",
    reportNumber = "DCPT-07-13, KUNS-2069",
    doi = "10.1088/1126-6708/2007/07/062",
    journal = "JHEP",
    volume = "07",
    pages = "062",
    year = "2007"
}

@article{VanRaamsdonk:2010pw,
    author = "Van Raamsdonk, Mark",
    title = "{Building up spacetime with quantum entanglement}",
    eprint = "1005.3035",
    archivePrefix = "arXiv",
    primaryClass = "hep-th",
    doi = "10.1142/S0218271810018529",
    journal = "Gen. Rel. Grav.",
    volume = "42",
    pages = "2323--2329",
    year = "2010"
}

@article{Lewkowycz:2013nqa,
    author = "Lewkowycz, Aitor and Maldacena, Juan",
    title = "{Generalized gravitational entropy}",
    eprint = "1304.4926",
    archivePrefix = "arXiv",
    primaryClass = "hep-th",
    doi = "10.1007/JHEP08(2013)090",
    journal = "JHEP",
    volume = "08",
    pages = "090",
    year = "2013"
}

@article{Faulkner:2013ana,
    author = "Faulkner, Thomas and Lewkowycz, Aitor and Maldacena, Juan",
    title = "{Quantum corrections to holographic entanglement entropy}",
    eprint = "1307.2892",
    archivePrefix = "arXiv",
    primaryClass = "hep-th",
    doi = "10.1007/JHEP11(2013)074",
    journal = "JHEP",
    volume = "11",
    pages = "074",
    year = "2013"
}

@article{Almheiri:2020cfm,
    author = "Almheiri, Ahmed and Hartman, Thomas and Maldacena, Juan and Shaghoulian, Edgar and Tajdini, Amirhossein",
    title = "{The entropy of Hawking radiation}",
    eprint = "2006.06872",
    archivePrefix = "arXiv",
    primaryClass = "hep-th",
    doi = "10.1103/RevModPhys.93.035002",
    journal = "Rev. Mod. Phys.",
    volume = "93",
    number = "3",
    pages = "035002",
    year = "2021"
}

@article{Penington:2019npb,
    author = "Penington, Geoffrey",
    title = "{Entanglement Wedge Reconstruction and the Information Paradox}",
    eprint = "1905.08255",
    archivePrefix = "arXiv",
    primaryClass = "hep-th",
    doi = "10.1007/JHEP09(2020)002",
    journal = "JHEP",
    volume = "09",
    pages = "002",
    year = "2020"
}

@article{Almheiri:2019psf,
    author = "Almheiri, Ahmed and Engelhardt, Netta and Marolf, Donald and Maxfield, Henry",
    title = "{The entropy of bulk quantum fields and the entanglement wedge of an evaporating black hole}",
    eprint = "1905.08762",
    archivePrefix = "arXiv",
    primaryClass = "hep-th",
    doi = "10.1007/JHEP12(2019)063",
    journal = "JHEP",
    volume = "12",
    pages = "063",
    year = "2019"
}

@article{Almheiri:2019hni,
    author = "Almheiri, Ahmed and Mahajan, Raghu and Maldacena, Juan and Zhao, Ying",
    title = "{The Page curve of Hawking radiation from semiclassical geometry}",
    eprint = "1908.10996",
    archivePrefix = "arXiv",
    primaryClass = "hep-th",
    doi = "10.1007/JHEP03(2020)149",
    journal = "JHEP",
    volume = "03",
    pages = "149",
    year = "2020"
}

@article{Penington:2019kki,
    author = "Penington, Geoff and Shenker, Stephen H. and Stanford, Douglas and Yang, Zhenbin",
    title = "{Replica wormholes and the black hole interior}",
    eprint = "1911.11977",
    archivePrefix = "arXiv",
    primaryClass = "hep-th",
    doi = "10.1007/JHEP03(2022)205",
    journal = "JHEP",
    volume = "03",
    pages = "205",
    year = "2022"
}

@article{Almheiri:2019qdq,
    author = "Almheiri, Ahmed and Hartman, Thomas and Maldacena, Juan and Shaghoulian, Edgar and Tajdini, Amirhossein",
    title = "{Replica Wormholes and the Entropy of Hawking Radiation}",
    eprint = "1911.12333",
    archivePrefix = "arXiv",
    primaryClass = "hep-th",
    doi = "10.1007/JHEP05(2020)013",
    journal = "JHEP",
    volume = "05",
    pages = "013",
    year = "2020"
}

@article{Dyson:2002nt,
    author = "Dyson, Lisa and Lindesay, James and Susskind, Leonard",
    title = "{Is there really a de Sitter/CFT duality?}",
    eprint = "hep-th/0202163",
    archivePrefix = "arXiv",
    reportNumber = "SU-ITP-02-07",
    doi = "10.1088/1126-6708/2002/08/045",
    journal = "JHEP",
    volume = "08",
    pages = "045",
    year = "2002"
}

@article{Dyson:2002pf,
    author = "Dyson, Lisa and Kleban, Matthew and Susskind, Leonard",
    title = "{Disturbing implications of a cosmological constant}",
    eprint = "hep-th/0208013",
    archivePrefix = "arXiv",
    reportNumber = "SU-ITP-02-25, MIT-CTP-3295",
    doi = "10.1088/1126-6708/2002/10/011",
    journal = "JHEP",
    volume = "10",
    pages = "011",
    year = "2002"
}

@article{Goheer:2002vf,
    author = "Goheer, Naureen and Kleban, Matthew and Susskind, Leonard",
    title = "{The Trouble with de Sitter space}",
    eprint = "hep-th/0212209",
    archivePrefix = "arXiv",
    doi = "10.1088/1126-6708/2003/07/056",
    journal = "JHEP",
    volume = "07",
    pages = "056",
    year = "2003"
}

@article{Susskind:2021esx,
    author = "Susskind, Leonard",
    title = "{Entanglement and Chaos in De Sitter Space Holography: An SYK Example}",
    eprint = "2109.14104",
    archivePrefix = "arXiv",
    primaryClass = "hep-th",
    doi = "10.22128/jhap.2021.455.1005",
    journal = "JHAP",
    volume = "1",
    number = "1",
    pages = "1--22",
    year = "2021"
}

@article{Shaghoulian:2021cef,
    author = "Shaghoulian, Edgar",
    title = "{The central dogma and cosmological horizons}",
    eprint = "2110.13210",
    archivePrefix = "arXiv",
    primaryClass = "hep-th",
    doi = "10.1007/JHEP01(2022)132",
    journal = "JHEP",
    volume = "01",
    pages = "132",
    year = "2022"
}

@article{Shaghoulian:2022fop,
    author = "Shaghoulian, Edgar and Susskind, Leonard",
    title = "{Entanglement in De Sitter space}",
    eprint = "2201.03603",
    archivePrefix = "arXiv",
    primaryClass = "hep-th",
    doi = "10.1007/JHEP08(2022)198",
    journal = "JHEP",
    volume = "08",
    pages = "198",
    year = "2022"
}

@article{Cotler:2022weg,
    author = "Cotler, Jordan and Strominger, Andrew",
    title = "{The Universe as a Quantum Encoder}",
    eprint = "2201.11658",
    archivePrefix = "arXiv",
    primaryClass = "hep-th",
    month = "1",
    year = "2022"
}

@article{Cotler:2023xku,
    author = "Cotler, Jordan and Strominger, Andrew",
    title = "{Cosmic ER=EPR in dS/CFT}",
    eprint = "2302.00632",
    archivePrefix = "arXiv",
    primaryClass = "hep-th",
    month = "2",
    year = "2023"
}

@article{Strominger:2003br,
    author = "Strominger, Andrew and Thompson, David Mattoon",
    title = "{A Quantum Bousso bound}",
    eprint = "hep-th/0303067",
    archivePrefix = "arXiv",
    doi = "10.1103/PhysRevD.70.044007",
    journal = "Phys. Rev. D",
    volume = "70",
    pages = "044007",
    year = "2004"
}

@article{Franken:2023ugu,
    author = "Franken, Victor and Rondeau, Fran{\c{c}}ois",
    title = "{On the quantum Bousso bound in JT gravity}",
    eprint = "2311.17152",
    archivePrefix = "arXiv",
    primaryClass = "hep-th",
    reportNumber = "CPHT-RR074.112023, CPHT-RR067.112023",
    doi = "10.1007/JHEP03(2024)178",
    journal = "JHEP",
    volume = "03",
    pages = "178",
    year = "2024"
}

@article{Kawamoto:2023nki,
    author = "Kawamoto, Taishi and Ruan, Shan-Ming and Suzuki, Yu-ki and Takayanagi, Tadashi",
    title = "{A half de Sitter holography}",
    eprint = "2306.07575",
    archivePrefix = "arXiv",
    primaryClass = "hep-th",
    reportNumber = "YITP-23-73",
    doi = "10.1007/JHEP10(2023)137",
    journal = "JHEP",
    volume = "10",
    pages = "137",
    year = "2023"
}

@article{Hao:2024nhd,
    author = "Hao, Peng-Xiang and Kawamoto, Taishi and Ruan, Shan-Ming and Takayanagi, Tadashi",
    title = "{Non-extremal island in de Sitter gravity}",
    eprint = "2407.21617",
    archivePrefix = "arXiv",
    primaryClass = "hep-th",
    reportNumber = "YITP-24-91",
    doi = "10.1007/JHEP03(2025)004",
    journal = "JHEP",
    volume = "03",
    pages = "004",
    year = "2025"
}

@article{Yadav:2024ray,
    author = "Yadav, Gopal",
    title = "{Communicating multiverses in a holographic de Sitter braneworld}",
    eprint = "2404.00763",
    archivePrefix = "arXiv",
    primaryClass = "hep-th",
    doi = "10.1103/PhysRevD.110.026028",
    journal = "Phys. Rev. D",
    volume = "110",
    number = "2",
    pages = "026028",
    year = "2024"
}

@article{Noumi:2025cup,
    author = "Noumi, Toshifumi and Sano, Fumiya and Suzuki, Yu-ki",
    title = "{Holographic entanglement entropy in the FLRW universe}",
    eprint = "2504.10457",
    archivePrefix = "arXiv",
    primaryClass = "hep-th",
    doi = "10.1007/JHEP08(2025)115",
    journal = "JHEP",
    volume = "08",
    pages = "115",
    year = "2025"
}

@article{Liu:2025cml,
    author = "Liu, Hong",
    title = "{Towards a holographic description of closed universes}",
    eprint = "2509.14327",
    archivePrefix = "arXiv",
    primaryClass = "hep-th",
    reportNumber = "MIT-CTP/5923",
    month = "9",
    year = "2025"
}

@article{Espinosa:2018hue,
    author = "Espinosa, J. R.",
    title = "{A Fresh Look at the Calculation of Tunneling Actions}",
    eprint = "1805.03680",
    archivePrefix = "arXiv",
    primaryClass = "hep-th",
    doi = "10.1088/1475-7516/2018/07/036",
    journal = "JCAP",
    volume = "07",
    pages = "036",
    year = "2018"
}

@article{Espinosa:2018voj,
    author = "Espinosa, J. R.",
    title = "{Fresh look at the calculation of tunneling actions including gravitational effects}",
    eprint = "1808.00420",
    archivePrefix = "arXiv",
    primaryClass = "hep-th",
    doi = "10.1103/PhysRevD.100.104007",
    journal = "Phys. Rev. D",
    volume = "100",
    number = "10",
    pages = "104007",
    year = "2019"
}

@article{Susskind:2003kw,
    author = "Susskind, Leonard",
    editor = "Carr, Bernard J.",
    title = "{The Anthropic landscape of string theory}",
    eprint = "hep-th/0302219",
    archivePrefix = "arXiv",
    pages = "247--266",
    month = "2",
    year = "2003"
}

@article{Frey:2003dm,
    author = "Frey, Andrew R. and Lippert, Matthew and Williams, Brook",
    title = "{The Fall of stringy de Sitter}",
    eprint = "hep-th/0305018",
    archivePrefix = "arXiv",
    doi = "10.1103/PhysRevD.68.046008",
    journal = "Phys. Rev. D",
    volume = "68",
    pages = "046008",
    year = "2003"
}

@article{Kachru:2003aw,
    author = "Kachru, Shamit and Kallosh, Renata and Linde, Andrei D. and Trivedi, Sandip P.",
    title = "{De Sitter vacua in string theory}",
    eprint = "hep-th/0301240",
    archivePrefix = "arXiv",
    reportNumber = "SLAC-PUB-9630, SU-ITP-03-01, TIFR-TH-03-03",
    doi = "10.1103/PhysRevD.68.046005",
    journal = "Phys. Rev. D",
    volume = "68",
    pages = "046005",
    year = "2003"
}

@article{Bousso:2000xa,
    author = "Bousso, Raphael and Polchinski, Joseph",
    title = "{Quantization of four form fluxes and dynamical neutralization of the cosmological constant}",
    eprint = "hep-th/0004134",
    archivePrefix = "arXiv",
    reportNumber = "SU-ITP-00-12, NSF-ITP-00-40",
    doi = "10.1088/1126-6708/2000/06/006",
    journal = "JHEP",
    volume = "06",
    pages = "006",
    year = "2000"
}

@inproceedings{Maloney:2002rr,
    author = "Maloney, Alexander and Silverstein, Eva and Strominger, Andrew",
    title = "{De Sitter space in noncritical string theory}",
    booktitle = "{Workshop on Conference on the Future of Theoretical Physics and Cosmology in Honor of Steven Hawking's 60th Birthday}",
    eprint = "hep-th/0205316",
    archivePrefix = "arXiv",
    reportNumber = "SLAC-PUB-9228, HUTP-02-A019",
    pages = "570--591",
    month = "5",
    year = "2002"
}

@article{Chandrasekaran:2022cip,
    author = "Chandrasekaran, Venkatesa and Longo, Roberto and Penington, Geoff and Witten, Edward",
    title = "{An algebra of observables for de Sitter space}",
    eprint = "2206.10780",
    archivePrefix = "arXiv",
    primaryClass = "hep-th",
    doi = "10.1007/JHEP02(2023)082",
    journal = "JHEP",
    volume = "02",
    pages = "082",
    year = "2023"
}

@article{Witten:2023qsv,
    author = "Witten, Edward",
    title = "{Algebras, regions, and observers.}",
    eprint = "2303.02837",
    archivePrefix = "arXiv",
    primaryClass = "hep-th",
    doi = "10.1090/pspum/107/01954",
    journal = "Proc. Symp. Pure Math.",
    volume = "107",
    pages = "247--276",
    year = "2024"
}

@article{Witten:2023xze,
    author = "Witten, Edward",
    title = "{A background-independent algebra in quantum gravity}",
    eprint = "2308.03663",
    archivePrefix = "arXiv",
    primaryClass = "hep-th",
    doi = "10.1007/JHEP03(2024)077",
    journal = "JHEP",
    volume = "03",
    pages = "077",
    year = "2024"
}

@article{Gomez:2023upk,
    author = "Gomez, Cesar",
    title = "{Clocks, Algebras and Cosmology}",
    eprint = "2304.11845",
    archivePrefix = "arXiv",
    primaryClass = "hep-th",
    month = "4",
    year = "2023"
}

@article{Balasubramanian:2023xyd,
    author = "Balasubramanian, Vijay and Nomura, Yasunori and Ugajin, Tomonori",
    title = "{De Sitter space is sometimes not empty}",
    eprint = "2308.09748",
    archivePrefix = "arXiv",
    primaryClass = "hep-th",
    reportNumber = "RIKEN-iTHEMS-Report-23",
    doi = "10.1007/JHEP02(2024)135",
    journal = "JHEP",
    volume = "02",
    pages = "135",
    year = "2024"
}

@article{Mirbabayi:2023vgl,
    author = "Mirbabayi, Mehrdad",
    title = "{An observer\textquoteright{}s measure of de Sitter entropy}",
    eprint = "2311.07724",
    archivePrefix = "arXiv",
    primaryClass = "hep-th",
    doi = "10.1007/JHEP10(2024)077",
    journal = "JHEP",
    volume = "10",
    pages = "077",
    year = "2024"
}

@article{Ryu:2006ef,
    author = "Ryu, Shinsei and Takayanagi, Tadashi",
    title = "{Aspects of Holographic Entanglement Entropy}",
    eprint = "hep-th/0605073",
    archivePrefix = "arXiv",
    reportNumber = "NSF-KITP-06-31, KUNS-2021",
    doi = "10.1088/1126-6708/2006/08/045",
    journal = "JHEP",
    volume = "08",
    pages = "045",
    year = "2006"
}

@article{Engelhardt:2014gca,
    author = "Engelhardt, Netta and Wall, Aron C.",
    title = "{Quantum Extremal Surfaces: Holographic Entanglement Entropy beyond the Classical Regime}",
    eprint = "1408.3203",
    archivePrefix = "arXiv",
    primaryClass = "hep-th",
    doi = "10.1007/JHEP01(2015)073",
    journal = "JHEP",
    volume = "01",
    pages = "073",
    year = "2015"
}

@article{Wall:2012uf,
    author = "Wall, Aron C.",
    title = "{Maximin Surfaces, and the Strong Subadditivity of the Covariant Holographic Entanglement Entropy}",
    eprint = "1211.3494",
    archivePrefix = "arXiv",
    primaryClass = "hep-th",
    doi = "10.1088/0264-9381/31/22/225007",
    journal = "Class. Quant. Grav.",
    volume = "31",
    number = "22",
    pages = "225007",
    year = "2014"
}

@article{Freedman:2016zud,
    author = "Freedman, Michael and Headrick, Matthew",
    title = "{Bit threads and holographic entanglement}",
    eprint = "1604.00354",
    archivePrefix = "arXiv",
    primaryClass = "hep-th",
    reportNumber = "BRX-TH-6302, NSF-KITP-16-051",
    doi = "10.1007/s00220-016-2796-3",
    journal = "Commun. Math. Phys.",
    volume = "352",
    number = "1",
    pages = "407--438",
    year = "2017"
}

@article{Headrick:2022nbe,
    author = "Headrick, Matthew and Hubeny, Veronika E.",
    title = "{Covariant bit threads}",
    eprint = "2208.10507",
    archivePrefix = "arXiv",
    primaryClass = "hep-th",
    doi = "10.1007/JHEP07(2023)180",
    journal = "JHEP",
    volume = "07",
    pages = "180",
    year = "2023"
}

@article{Dong:2016hjy,
    author = "Dong, Xi and Lewkowycz, Aitor and Rangamani, Mukund",
    title = "{Deriving covariant holographic entanglement}",
    eprint = "1607.07506",
    archivePrefix = "arXiv",
    primaryClass = "hep-th",
    doi = "10.1007/JHEP11(2016)028",
    journal = "JHEP",
    volume = "11",
    pages = "028",
    year = "2016"
}

@article{Hartman:2020khs,
    author = "Hartman, Thomas and Jiang, Yikun and Shaghoulian, Edgar",
    title = "{Islands in cosmology}",
    eprint = "2008.01022",
    archivePrefix = "arXiv",
    primaryClass = "hep-th",
    doi = "10.1007/JHEP11(2020)111",
    journal = "JHEP",
    volume = "11",
    pages = "111",
    year = "2020"
}

@article{Chen:2020tes,
    author = "Chen, Yiming and Gorbenko, Victor and Maldacena, Juan",
    title = "{Bra-ket wormholes in gravitationally prepared states}",
    eprint = "2007.16091",
    archivePrefix = "arXiv",
    primaryClass = "hep-th",
    doi = "10.1007/JHEP02(2021)009",
    journal = "JHEP",
    volume = "02",
    pages = "009",
    year = "2021"
}

@article{VanRaamsdonk:2020tlr,
    author = "Van Raamsdonk, Mark",
    title = "{Comments on wormholes, ensembles, and cosmology}",
    eprint = "2008.02259",
    archivePrefix = "arXiv",
    primaryClass = "hep-th",
    doi = "10.1007/JHEP12(2021)156",
    journal = "JHEP",
    volume = "12",
    pages = "156",
    year = "2021"
}

@article{Balasubramanian:2020coy,
    author = "Balasubramanian, Vijay and Kar, Arjun and Ugajin, Tomonori",
    title = "{Entanglement between two disjoint universes}",
    eprint = "2008.05274",
    archivePrefix = "arXiv",
    primaryClass = "hep-th",
    doi = "10.1007/JHEP02(2021)136",
    journal = "JHEP",
    volume = "02",
    pages = "136",
    year = "2021"
}

@article{Balasubramanian:2020xqf,
    author = "Balasubramanian, Vijay and Kar, Arjun and Ugajin, Tomonori",
    title = "{Islands in de Sitter space}",
    eprint = "2008.05275",
    archivePrefix = "arXiv",
    primaryClass = "hep-th",
    doi = "10.1007/JHEP02(2021)072",
    journal = "JHEP",
    volume = "02",
    pages = "072",
    year = "2021"
}

@article{Manu:2020tty,
    author = "Manu, A. and Narayan, K. and Paul, Partha",
    title = "{Cosmological singularities, entanglement and quantum extremal surfaces}",
    eprint = "2012.07351",
    archivePrefix = "arXiv",
    primaryClass = "hep-th",
    doi = "10.1007/JHEP04(2021)200",
    journal = "JHEP",
    volume = "04",
    pages = "200",
    year = "2021"
}

@article{Kames-King:2021etp,
    author = "Kames-King, Joshua and Verheijden, Evita M. H. and Verlinde, Erik P.",
    title = "{No Page curves for the de Sitter horizon}",
    eprint = "2108.09318",
    archivePrefix = "arXiv",
    primaryClass = "hep-th",
    doi = "10.1007/JHEP03(2022)040",
    journal = "JHEP",
    volume = "03",
    pages = "040",
    year = "2022"
}

@article{Azarnia:2021uch,
    author = "Azarnia, Sanam and Fareghbal, Reza and Naseh, Ali and Zolfi, Hamed",
    title = "{Islands in flat-space cosmology}",
    eprint = "2109.04795",
    archivePrefix = "arXiv",
    primaryClass = "hep-th",
    doi = "10.1103/PhysRevD.104.126017",
    journal = "Phys. Rev. D",
    volume = "104",
    number = "12",
    pages = "126017",
    year = "2021"
}

@article{Goswami:2021ksw,
    author = "Goswami, Kaberi and Narayan, K. and Saini, Hitesh K.",
    title = "{Cosmologies, singularities and quantum extremal surfaces}",
    eprint = "2111.14906",
    archivePrefix = "arXiv",
    primaryClass = "hep-th",
    doi = "10.1007/JHEP03(2022)201",
    journal = "JHEP",
    volume = "03",
    pages = "201",
    year = "2022"
}

@article{Bousso:2022gth,
    author = "Bousso, Raphael and Wildenhain, Elizabeth",
    title = "{Islands in closed and open universes}",
    eprint = "2202.05278",
    archivePrefix = "arXiv",
    primaryClass = "hep-th",
    doi = "10.1103/PhysRevD.105.086012",
    journal = "Phys. Rev. D",
    volume = "105",
    number = "8",
    pages = "086012",
    year = "2022"
}

@article{Aguilar-Gutierrez:2021bns,
    author = "Aguilar-Gutierrez, Sergio E. and Chatwin-Davies, Aidan and Hertog, Thomas and Pinzani-Fokeeva, Natalia and Robinson, Brandon",
    title = "{Islands in Multiverse Models}",
    eprint = "2108.01278",
    archivePrefix = "arXiv",
    primaryClass = "hep-th",
    doi = "10.1007/JHEP05(2022)137",
    journal = "JHEP",
    volume = "11",
    pages = "212",
    year = "2021",
    note = "[Addendum: JHEP 05, 137 (2022), Erratum: JHEP 05, 082 (2022)]"
}

@article{Ben-Dayan:2022nmb,
    author = "Ben-Dayan, Ido and Hadad, Merav and Wildenhain, Elizabeth",
    title = "{Islands in the fluid: islands are common in cosmology}",
    eprint = "2211.16600",
    archivePrefix = "arXiv",
    primaryClass = "hep-th",
    doi = "10.1007/JHEP03(2023)077",
    journal = "JHEP",
    volume = "03",
    pages = "077",
    year = "2023"
}

@article{VanRaamsdonk:2009ar,
    author = "Van Raamsdonk, Mark",
    title = "{Comments on quantum gravity and entanglement}",
    eprint = "0907.2939",
    archivePrefix = "arXiv",
    primaryClass = "hep-th",
    month = "7",
    year = "2009"
}

@article{Maldacena:2013xja,
    author = "Maldacena, Juan and Susskind, Leonard",
    title = "{Cool horizons for entangled black holes}",
    eprint = "1306.0533",
    archivePrefix = "arXiv",
    primaryClass = "hep-th",
    doi = "10.1002/prop.201300020",
    journal = "Fortsch. Phys.",
    volume = "61",
    pages = "781--811",
    year = "2013"
}

@article{Engelhardt:2023xer,
    author = "Engelhardt, Netta and Liu, Hong",
    title = "{Algebraic ER=EPR and complexity transfer}",
    eprint = "2311.04281",
    archivePrefix = "arXiv",
    primaryClass = "hep-th",
    reportNumber = "MIT-CTP/5607",
    doi = "10.1007/JHEP07(2024)013",
    journal = "JHEP",
    volume = "07",
    pages = "013",
    year = "2024"
}

@article{Susskind:2022dfz,
    author = "Susskind, Leonard",
    title = "{Scrambling in Double-Scaled SYK and De Sitter Space}",
    eprint = "2205.00315",
    archivePrefix = "arXiv",
    primaryClass = "hep-th",
    month = "4",
    year = "2022"
}

@article{Lin:2022nss,
    author = "Lin, Henry and Susskind, Leonard",
    title = "{Infinite Temperature's Not So Hot}",
    eprint = "2206.01083",
    archivePrefix = "arXiv",
    primaryClass = "hep-th",
    month = "6",
    year = "2022"
}

@article{Susskind:2022bia,
    author = "Susskind, Leonard",
    title = "{De Sitter Space, Double-Scaled SYK, and the Separation of Scales in the Semiclassical Limit}",
    eprint = "2209.09999",
    archivePrefix = "arXiv",
    primaryClass = "hep-th",
    doi = "10.22128/jhap.2024.920.1103",
    journal = "JHAP",
    volume = "5",
    number = "1",
    pages = "1--30",
    year = "2025"
}

@article{Susskind:2023hnj,
    author = "Susskind, Leonard",
    title = "{De Sitter Space has no Chords. Almost Everything is Confined.}",
    eprint = "2303.00792",
    archivePrefix = "arXiv",
    primaryClass = "hep-th",
    doi = "10.22128/jhap.2023.661.1043",
    journal = "JHAP",
    volume = "3",
    number = "1",
    pages = "1--30",
    year = "2023"
}

@article{Narovlansky:2023lfz,
    author = "Narovlansky, Vladimir and Verlinde, Herman",
    title = "{Double-scaled SYK and de Sitter holography}",
    eprint = "2310.16994",
    archivePrefix = "arXiv",
    primaryClass = "hep-th",
    doi = "10.1007/JHEP05(2025)032",
    journal = "JHEP",
    volume = "05",
    pages = "032",
    year = "2025"
}

@article{Verlinde:2024znh,
    author = "Verlinde, Herman",
    title = "{Double-scaled SYK, chords and de Sitter gravity}",
    eprint = "2402.00635",
    archivePrefix = "arXiv",
    primaryClass = "hep-th",
    doi = "10.1007/JHEP03(2025)076",
    journal = "JHEP",
    volume = "03",
    pages = "076",
    year = "2025"
}

@article{Verlinde:2024zrh,
    author = "Verlinde, Herman and Zhang, Mengyang",
    title = "{SYK correlators from 2D Liouville-de Sitter gravity}",
    eprint = "2402.02584",
    archivePrefix = "arXiv",
    primaryClass = "hep-th",
    doi = "10.1007/JHEP05(2025)053",
    journal = "JHEP",
    volume = "05",
    pages = "053",
    year = "2025"
}

@article{Rahman:2022jsf,
    author = "Rahman, Adel A.",
    title = "{dS JT Gravity and Double-Scaled SYK}",
    eprint = "2209.09997",
    archivePrefix = "arXiv",
    primaryClass = "hep-th",
    month = "9",
    year = "2022"
}

@article{Rahman:2023pgt,
    author = "Rahman, Adel A. and Susskind, Leonard",
    title = "{Comments on a Paper by Narovlansky and Verlinde}",
    eprint = "2312.04097",
    archivePrefix = "arXiv",
    primaryClass = "hep-th",
    month = "12",
    year = "2023"
}

@article{Rahman:2024vyg,
    author = "Rahman, Adel and Susskind, Leonard",
    title = "{Infinite Temperature is Not So Infinite: The Many Temperatures of de Sitter Space}",
    eprint = "2401.08555",
    archivePrefix = "arXiv",
    primaryClass = "hep-th",
    month = "1",
    year = "2024"
}

@article{Banks:1996vh,
    author = "Banks, Tom and Fischler, W. and Shenker, S. H. and Susskind, Leonard",
    title = "{M theory as a matrix model: A conjecture}",
    eprint = "hep-th/9610043",
    archivePrefix = "arXiv",
    reportNumber = "RU-96-95, SU-ITP-96-12, UTTG-13-96",
    doi = "10.1201/9781482268737-37",
    journal = "Phys. Rev. D",
    volume = "55",
    pages = "5112--5128",
    year = "1997"
}

@article{Freivogel:2004rd,
    author = "Freivogel, Ben and Susskind, Leonard",
    title = "{A Framework for the landscape}",
    eprint = "hep-th/0408133",
    archivePrefix = "arXiv",
    reportNumber = "SU-ITP-04-33",
    doi = "10.1103/PhysRevD.70.126007",
    journal = "Phys. Rev. D",
    volume = "70",
    pages = "126007",
    year = "2004"
}

@article{Freivogel:2005vv,
    author = "Freivogel, Ben and Kleban, Matthew and Rodriguez Martinez, Maria and Susskind, Leonard",
    title = "{Observational consequences of a landscape}",
    eprint = "hep-th/0505232",
    archivePrefix = "arXiv",
    reportNumber = "SU-ITP-05-19",
    doi = "10.1088/1126-6708/2006/03/039",
    journal = "JHEP",
    volume = "03",
    pages = "039",
    year = "2006"
}

@article{Freivogel:2006xu,
    author = "Freivogel, Ben and Sekino, Yasuhiro and Susskind, Leonard and Yeh, Chen-Pin",
    title = "{A Holographic framework for eternal inflation}",
    eprint = "hep-th/0606204",
    archivePrefix = "arXiv",
    reportNumber = "SU-ITP-06-18, OIQP-06-07, UCB-PTH-06-12, LBNL-60518",
    doi = "10.1103/PhysRevD.74.086003",
    journal = "Phys. Rev. D",
    volume = "74",
    pages = "086003",
    year = "2006"
}

@article{Sekino:2009kv,
    author = "Sekino, Yasuhiro and Susskind, Leonard",
    title = "{Census Taking in the Hat: FRW/CFT Duality}",
    eprint = "0908.3844",
    archivePrefix = "arXiv",
    primaryClass = "hep-th",
    reportNumber = "SU-ITP-09-40, OIQP-09-09",
    doi = "10.1103/PhysRevD.80.083531",
    journal = "Phys. Rev. D",
    volume = "80",
    pages = "083531",
    year = "2009"
}

@phdthesis{Aguilar-Gutierrez:2024yzu,
    author = "Aguilar-Gutierrez, S. E.",
    title = "{De Sitter space, complexity, and the double-scaled SYK model}",
    eprint = "2406.19089",
    archivePrefix = "arXiv",
    primaryClass = "hep-th",
    school = "Leuven U.",
    year = "2024"
}

@article{Langhoff:2021uct,
    author = "Langhoff, Kevin and Murdia, Chitraang and Nomura, Yasunori",
    title = "{Multiverse in an inverted island}",
    eprint = "2106.05271",
    archivePrefix = "arXiv",
    primaryClass = "hep-th",
    doi = "10.1103/PhysRevD.104.086007",
    journal = "Phys. Rev. D",
    volume = "104",
    number = "8",
    pages = "086007",
    year = "2021"
}

@article{Geng:2021wcq,
    author = "Geng, Hao and Nomura, Yasunori and Sun, Hao-Yu",
    title = "{Information paradox and its resolution in de Sitter holography}",
    eprint = "2103.07477",
    archivePrefix = "arXiv",
    primaryClass = "hep-th",
    doi = "10.1103/PhysRevD.103.126004",
    journal = "Phys. Rev. D",
    volume = "103",
    number = "12",
    pages = "126004",
    year = "2021"
}

@article{Nomura:2016ikr,
    author = "Nomura, Yasunori and Salzetta, Nico and Sanches, Fabio and Weinberg, Sean J.",
    title = "{Toward a Holographic Theory for General Spacetimes}",
    eprint = "1611.02702",
    archivePrefix = "arXiv",
    primaryClass = "hep-th",
    doi = "10.1103/PhysRevD.95.086002",
    journal = "Phys. Rev. D",
    volume = "95",
    number = "8",
    pages = "086002",
    year = "2017"
}

@article{Nomura:2017fyh,
    author = "Nomura, Yasunori and Rath, Pratik and Salzetta, Nico",
    title = "{Spacetime from Unentanglement}",
    eprint = "1711.05263",
    archivePrefix = "arXiv",
    primaryClass = "hep-th",
    doi = "10.1103/PhysRevD.97.106010",
    journal = "Phys. Rev. D",
    volume = "97",
    number = "10",
    pages = "106010",
    year = "2018"
}

@article{Murdia:2022giv,
    author = "Murdia, Chitraang and Nomura, Yasunori and Ritchie, Kyle",
    title = "{Black hole and de Sitter microstructures from a semiclassical perspective}",
    eprint = "2207.01625",
    archivePrefix = "arXiv",
    primaryClass = "hep-th",
    doi = "10.1103/PhysRevD.107.026016",
    journal = "Phys. Rev. D",
    volume = "107",
    number = "2",
    pages = "026016",
    year = "2023"
}

@article{Israel:1966rt,
    author = "Israel, W.",
    title = "{Singular hypersurfaces and thin shells in general relativity}",
    doi = "10.1007/BF02710419",
    journal = "Nuovo Cim. B",
    volume = "44S10",
    pages = "1",
    year = "1966",
    note = "[Erratum: Nuovo Cim.B 48, 463 (1967)]"
}

@article{Marolf:2002np,
    author = "Marolf, Donald and Ross, Simon F.",
    title = "{Stringy negative tension branes and the second law of thermodynamics}",
    eprint = "hep-th/0202091",
    archivePrefix = "arXiv",
    reportNumber = "SU-GP-02-2-2",
    doi = "10.1088/1126-6708/2002/04/008",
    journal = "JHEP",
    volume = "04",
    pages = "008",
    year = "2002"
}

\end{document}